\documentclass[11pt,a4paper]{article}

\usepackage[colorlinks=true, linkcolor=black!50!blue, urlcolor=blue, citecolor=blue, anchorcolor=blue]{hyperref}
\usepackage[font=small,labelfont=bf,margin=0mm,labelsep=period,tableposition=top]{caption}
\usepackage[a4paper,top=3cm,bottom=2.5cm,left=2.5cm,right=2.5cm,bindingoffset=0mm]{geometry}

\usepackage{filecontents}

\usepackage{multibib}
\newcites{supp}{Supplementary References}

\usepackage{pdfpages}

\usepackage{graphicx,placeins}
\usepackage{float}
\usepackage{afterpage}
\usepackage{epsfig,cite}
\usepackage{amssymb}
\usepackage{amsmath}
\usepackage{dsfont}
\usepackage{multirow}
\usepackage{url}
\usepackage{xcolor}
\usepackage{float}
\usepackage{afterpage}

\usepackage{url}
\usepackage{hyperref}

\usepackage{booktabs}

\usepackage{tikz}
\usetikzlibrary{arrows}
\usepackage{enumitem}
\usepackage{hyperref}
\usepackage{cite}

\newcommand{\be}{\begin{equation}}
\newcommand{\ee}{\end{equation}}
\newcommand{\bea}{\begin{eqnarray}}
\newcommand{\eea}{\end{eqnarray}}
\newcommand{\bi}{\begin{itemize}}
\newcommand{\ei}{\end{itemize}}
\newcommand{\ben}{\begin{enumerate}}
\newcommand{\een}{\end{enumerate}}

\newcommand{\lc}{\left[}
\newcommand{\rc}{\right]}
\newcommand{\lp}{\left(}
\newcommand{\rp}{\right)}

\def\frac#1#2{{{#1}\over {#2}}}
\def\gsim{\mathrel{\rlap{\lower4pt\hbox{\hskip1pt$\sim$}}
    \raise1pt\hbox{$>$}}}         
\def\lsim{\mathrel{\rlap{\lower4pt\hbox{\hskip1pt$\sim$}}
    \raise1pt\hbox{$<$}}}         

\newcommand{\draft}[1]{}

\def\beq{\begin{equation}}
\def\eeq{\end{equation}}

\def\lapprox{\lower .7ex\hbox{$\;\stackrel{\textstyle <}{\sim}\;$}}
\def\gapprox{\lower .7ex\hbox{$\;\stackrel{\textstyle >}{\sim}\;$}}

\usepackage{tabularx}
\newcolumntype{C}[1]{>{\centering\arraybackslash}p{#1}}

\usepackage{amsmath}
\usepackage{amsfonts}
\usepackage{amssymb}
\usepackage{dsfont}
\usepackage{pifont}
\usepackage{booktabs}
\usepackage{graphicx}
\usepackage{epstopdf}
\usepackage{epsfig}
\usepackage{framed}
\usepackage{makeidx}
\usepackage{hyperref}
\usepackage{placeins}
\usepackage[font=small,labelfont=bf]{caption}

\newcounter{daggerfootnote}
\newcommand*{\daggerfootnote}[1]{%
    \setcounter{daggerfootnote}{\value{footnote}}%
    \renewcommand*{\thefootnote}{\fnsymbol{footnote}}%
    \footnote[2]{#1}%
    \setcounter{footnote}{\value{daggerfootnote}}%
    \renewcommand*{\thefootnote}{\arabic{footnote}}%
}

\usepackage{ulem}

\begin{document}
\newgeometry{top=1.5cm,bottom=1.5cm,left=2.5cm,right=2.5cm,bindingoffset=0mm}
\begin{figure}[h]
  \includegraphics[width=0.32\textwidth]{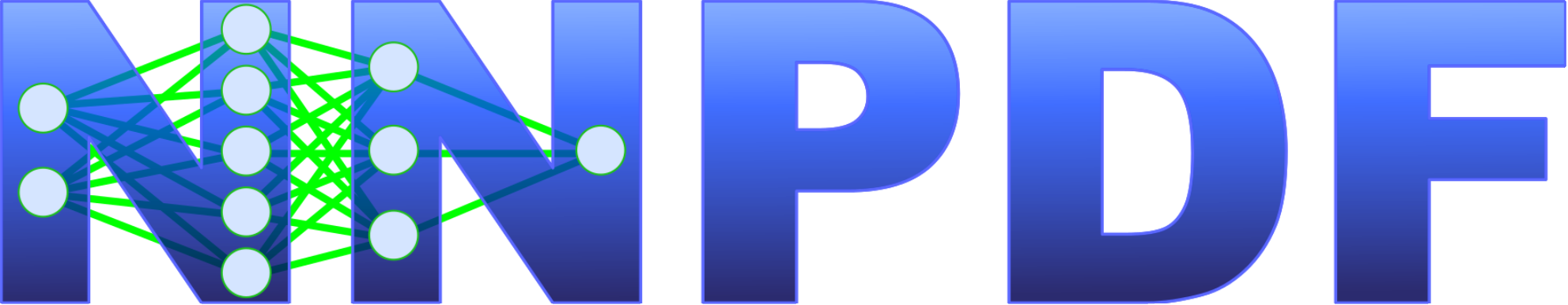}
\end{figure}
\vspace{-2.0cm}
\begin{flushright}
Edinburgh 2021/28\\
Nikhef 2021-032\\
TIF-UNIMI-2021-21
\end{flushright}
\vspace{0.3cm}

\vspace{1.1cm}
\begin{center}
  
 {{\LARGE \bf Evidence for intrinsic charm quarks in the proton}}

  \vspace{1.1cm}

  {\small
  {\bf  The NNPDF Collaboration:} \\[0.2cm]
  Richard D. Ball,$^{1}$
  Alessandro Candido,$^{2}$
Juan Cruz-Martinez,$^{2}$
Stefano Forte,$^{2}$
Tommaso Giani,$^{3,4}$\\[0.1cm]
Felix Hekhorn,$^{2}$
Kirill Kudashkin,$^{2}$
Giacomo Magni,$^{3,4}$ and
Juan Rojo$^{3,4}$\daggerfootnote{Corresponding author (\href{mailto:j.rojo@vu.nl}{j.rojo@vu.nl}).}
}\\

 \vspace{0.7cm}
 
 {\it \small

 ~$^1$The Higgs Centre for Theoretical Physics, University of Edinburgh,\\
   JCMB, KB, Mayfield Rd, Edinburgh EH9 3JZ, Scotland\\[0.1cm]
    ~$^2$Tif Lab, Dipartimento di Fisica, Universit\`a di Milano and\\
   INFN, Sezione di Milano, Via Celoria 16, I-20133 Milano, Italy\\[0.1cm]
    ~$^3$Department of Physics and Astronomy, Vrije Universiteit, NL-1081 HV Amsterdam\\[0.1cm]
~$^4$Nikhef Theory Group, Science Park 105, 1098 XG Amsterdam, The Netherlands\\[0.1cm]

   }

\vspace{1.0cm}

\end{center}

\noindent

The theory of the strong  force, Quantum
Chromodynamics, describes 
the proton in terms of quarks and gluons.
The proton is a state of 
two up quarks and one down quark bound by gluons, but quantum
theory predicts 
that in addition there is an infinite number of
quark-antiquark pairs.
Both light and heavy quarks, whose   mass is respectively smaller or
bigger than the mass of the
proton, are  revealed inside the proton in high-energy collisions.
However, it is unclear whether
heavy quarks also exist as a part of the proton
wave-function, which is  determined by
non-perturbative dynamics and accordingly unknown:
so-called intrinsic heavy quarks~\cite{Brodsky:1980pb}.
It has been
argued for a long time that the proton could have a sizable
intrinsic component of the lightest  heavy
quark, the charm quark.
Innumerable efforts to establish
intrinsic charm in the
proton~\cite{Brodsky:2015fna} have remained inconclusive.
Here we provide 
evidence for intrinsic
charm by exploiting  
a high-precision determination of the quark-gluon content of
the nucleon~\cite{Ball:2021leu} based on 
machine learning and a large experimental dataset.
We disentangle the intrinsic charm component from charm-anticharm pairs 
arising from high-energy radiation~\cite{Ball:2015tna}.
We establish the existence of intrinsic charm at the  $3\sigma$ level,
with a
momentum distribution in remarkable agreement
with model predictions~\cite{Brodsky:1980pb,Hobbs:2013bia}.
We confirm these findings by
comparing to very recent 
data on $Z$-boson production with charm jets  from
the LHCb experiment~\cite{LHCb:2021stx}. 

\section*{Main}

The foundational deep-inelastic scattering experiments at the SLAC linear collider
in the late 60s and early 70s demonstrated the presence inside the
proton of pointlike  constituents, soon identified with quarks, the
elementary particles that interact and are bound inside the proton by
gluons, the carriers of the strong  nuclear force.
It was rapidly clear, and confirmed in detail by subsequent studies,
that these pointlike constituents, collectively called ``partons'' by
Feynman~\cite{Feynman:1969wa}, include the up and down quarks that
carry the proton quantum numbers, but also gluons, as well
as an infinite number of pairs of quarks and their
antimatter counterparts, antiquarks.
The description of electron-proton and proton-proton collisions at high
momentum transfers in terms of collisions between partons is now
rooted in the theory of Quantum  Chromodynamics (QCD), and it provides
the basis
of modern-day precision phenomenology at proton accelerators such as
the Large Hadron Collider
(LHC) of CERN~\cite{Gao:2017yyd} as well
as for future facilities including the
EIC~\cite{AbdulKhalek:2021gbh},
the FPF~\cite{Feng:2022inv},
and  
neutrino telescopes~\cite{IceCube-Gen2:2020qha}.

Knowledge of the structure of the proton, which is necessary in order
to obtain
quantitative prediction for physics processes at the LHC and other
experiments, is encoded in
the distribution of  momentum carried by partons of each type
(gluons, up quarks, down quarks, up antiquarks, etc):
parton distribution functions (PDFs).
These PDFs could be in principle
computed from first principles, but in
practice even their determination from numerical
simulations~\cite{Constantinou:2020hdm} is extremely challenging.
Consequently,  the only 
strategy currently available for obtaining the
reliable determination of the proton PDFs which is required to evaluate LHC
predictions is empirical, through the global analysis of
data for which precise theoretical predictions and experimental
measurements are available, so that the PDFs are the only
unknown~\cite{Gao:2017yyd}.

While this successful framework has by now been worked through in great detail, several key open questions remain open.
One of the most controversial of these concerns the treatment of
so-called heavy quarks, i.e.\ those whose mass is greater than that of
the proton ($m_p=0.94$~GeV). Indeed, virtual quantum effects and
energy-mass considerations suggest that the three light quarks and
antiquarks (up, 
down, and strange) should all be present in the proton
wave-function.
Their PDFs are therefore surely determined by the low-energy
dynamics that controls the nature of the proton as a bound
state.
However, it is a well-known fact~\cite{DeRoeck:2011na,
  Kovarik:2019xvh,Gao:2017yyd,Rojo:2019uip}
that in high enough energy collisions all species of quarks can be
excited and hence observed
inside the proton, so their PDFs are nonzero.
This excitation
follows from standard QCD radiation and it can be computed accurately
in perturbation theory.

But then the question arises: do heavy quarks also contribute to the
proton wave-function? Such a contribution is called ``intrinsic'', to
distinguish it from that computable in
perturbation theory, which originates from QCD radiation.
Already since the dawn of QCD, it
was argued that all kinds of intrinsic heavy quarks must be
present in the
proton wave-function~\cite{Brodsky:1984nx}.
In particular, it was
suggested~\cite{Brodsky:1980pb}
that the intrinsic component could be non-negligible for the
charm quark, whose mass ($m_c\simeq 1.51$ GeV) is of the same order of
magnitude as the mass of the proton.

This question has remained highly controversial, and indeed recent
dedicated studies have resulted in disparate claims,
from excluding momentum fractions carried by intrinsic  charm larger than 0.5\% at the 4$\sigma$
level~\cite{Jimenez-Delgado:2014zga} to allowing up to a 2\% charm momentum
fraction~\cite{Hou:2017khm}.
A particularly delicate issue in this context is that of
separating the radiative component: finding that the charm PDF
is nonzero at a low scale is not sufficient to argue that intrinsic charm
has been identified.

Here we present a resolution of this four-decades-long conundrum
by providing unambiguous evidence for intrinsic charm  in the proton.
This is achieved by means of a determination of the charm
PDF~\cite{Ball:2021leu} from the most extensive hard-scattering 
global dataset analyzed to date, using state-of-the-art perturbative
QCD calculations~\cite{Heinrich:2020ybq}, adapted to accommodate the possibility of massive quarks inside the proton~\cite{Forte:2010ta,Ball:2015dpa,Ball:2015tna}, and sophisticated machine 
learning (ML)
techniques~\cite{Ball:2016neh,Ball:2017nwa,Ball:2021leu}. This
determination is performed at next-to-next-to-leading-order (NNLO) in an
expansion in powers of the strong coupling, $\alpha_s$, which
represents the precision frontier for collider phenomenology.

The charm PDF determined in this manner includes a 
radiative component, and
indeed it depends on the resolution scale: it is 
given in a four-flavor-number scheme (4FNS), in which up, 
down, strange and charm quarks are subject to  perturbative
radiative corrections and mix with each other and the gluon as the
resolution is increased.
The
intrinsic charm component can be disentangled from it as follows.
First, we
note that in the absence of an intrinsic component, the initial
condition for the charm PDF is determined using perturbative matching
conditions~\cite{Collins:1986mp}, computed  up to NNLO in~\cite{pdfnnlo},
and recently (partly) extended up to N$^3$LO~\cite{Bierenbaum:2009zt,Bierenbaum:2009mv,Ablinger:2010ty,Ablinger:2014vwa,Ablinger:2014uka,Behring:2014eya,Ablinger_2014,Ablinger:2014nga,Blumlein:2017wxd}.
These matching conditions 
determine the charm PDF in terms of the PDFs of the
three-flavor-number-scheme (3FNS), in which only the three lightest quark 
flavors are radiatively corrected.
Hence this perturbative charm PDF is
entirely determined in terms of the three light quarks and antiquarks
and the gluon.
However, the 3FNS charm quark PDF needs not
vanish: in fact, if the charm quark PDF in the 4FNS is freely
parametrized and thus determined from the data~\cite{Ball:2015tna},
the matching conditions can be inverted.
The 3FNS charm PDF
thus obtained is then by definition the intrinsic charm PDF: indeed, in
the absence of intrinsic charm it would vanish~\cite{Ball:2015dpa}. 
Thus unlike the 4FNS charm PDF, that
includes both an intrinsic and a radiative
component, the 3FNS charm
PDF is purely intrinsic.
In this work we have performed this inversion at
NNLO~\cite{pdfnnlo} as well as at N$^3$LO~\cite{Bierenbaum:2009zt,Bierenbaum:2009mv,Ablinger:2010ty,Ablinger:2014vwa,Ablinger:2014uka,Behring:2014eya,Ablinger_2014,Ablinger:2014nga,Blumlein:2017wxd},
which as we shall see provides a handle on the perturbative uncertainty of the NNLO result.

\begin{figure}[h]
  \begin{center}
    \includegraphics[width=0.99\linewidth]{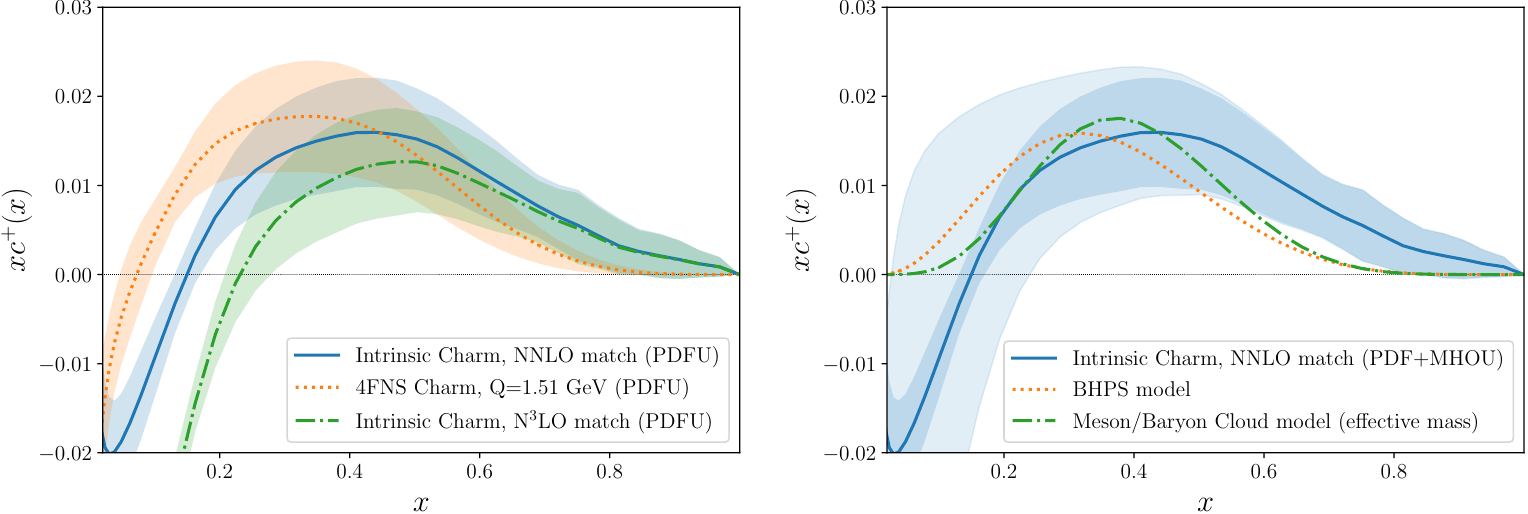}
    \caption{\small  {\bf The intrinsic charm PDF
      and comparison with models}.
      Left: the purely
      intrinsic (3FNS) result (blue)
      with PDF uncertainties only, compared to the 4FNS PDF, that
      includes both an intrinsic and radiative
      component,   at
      $Q=m_c=1.51$ GeV (orange). The purely intrinsic (3FNS)
      result obtained using N$^3$LO matching is also shown (green).
      Right: the purely
      intrinsic (3FNS)
      final result with total uncertainty (PDF+MHOU), with the PDF
      uncertainty indicated as a dark shaded band;
the predictions from the original 
BHPS model~\cite{Brodsky:1980pb} and from the more recent meson/baryon
      cloud model~\cite{Hobbs:2013bia} are also shown for comparison
      (dotted and dot-dashed curves respectively).
         \label{fig:charm_content_3fns} }
\end{center}
\end{figure}

Our starting point is the NNPDF4.0 global
analysis~\cite{Ball:2021leu}, which provides a determination of
the sum of the charm and
anticharm PDFs, namely  $c^+(x,Q)\equiv c(x,Q)+\bar  c(x,Q)$, in the
4FNS. 
This can be viewed 
as a probability density in $x$, the fraction of the proton momentum
carried by charm, in the sense that the integral over all 
values of $0\le x\le1$ of 
$xc^+(x)$ is equal to  the fraction of
the proton momentum carried by charm quarks, though note that PDFs are
generally not necessarily positive-definite. 
Our result for  the 4FNS $xc^+(x,Q)$  at
the charm mass scale, $Q=m_c$ with $m_c=1.51$ GeV, 
is displayed in Fig.~\ref{fig:charm_content_3fns}~(left).
The ensuing intrinsic charm is determined from it
by transforming to the 3FNS using
NNLO matching.
This result is also shown 
in Fig.~\ref{fig:charm_content_3fns}~(left).
The bands  indicate the 68\% confidence level (CL) interval
associated with the PDF uncertainties  (PDFU) in each case.  Henceforth, we will refer to
the  3FNS $xc^+(x,Q)$ PDF as the
intrinsic charm PDF. 

The intrinsic (3FNS) charm PDF
displays a characteristic valence-like
 structure at large-$x$ peaking at $x\simeq 0.4$.
 While intrinsic charm is found to be small in absolute terms
 (it contributes less than 1\% to the proton  total momentum),
 it is significantly different from zero.
 Note that the transformation to the 3FNS has little effect on the peak region,
 because there is almost no charm radiatively generated at such large values of $x$: in
 fact, a very similar valence-like peak is already found in the 4FNS calculation.

Because at the charm mass scale the strong coupling $\alpha_s$ is rather
large, the perturbative expansion converges slowly.
In order to
estimate the effect of missing higher order uncertainties (MHOU), we
have also performed the transformation from the 4FNS NNLO charm PDF
determined from the data to the 3FNS (intrinsic) charm PDF at one
order higher, namely at N$^3$LO. 
The result is also shown
Fig.~\ref{fig:charm_content_3fns} (left). Reassuringly, the intrinsic
valence-like structure is unchanged.
On the other hand, it is clear that for
$x\lsim 0.2$ perturbative uncertainties become very large.
We can estimate  the total uncertainty on our determination
of intrinsic charm by adding in quadrature the PDF uncertainty and a
MHOU estimated from the shift between the result found using NNLO
and N$^3$LO matching.

This procedure leads to our final result for intrinsic charm and its total
uncertainty, shown in Fig.~\ref{fig:charm_content_3fns} (right).
The intrinsic charm PDF is found to be compatible with zero for
$x\lsim 0.2$: the negative trend 
seen  in Fig.~\ref{fig:charm_content_3fns} with PDF uncertainties only 
becomes 
compatible with zero upon inclusion of  theoretical
uncertainties. However, at
larger $x$ even with theoretical uncertainties
the intrinsic charm PDF
 differs 
from zero by about 2.5 standard deviations ($2.5\sigma$) in the peak region.
This result  is stable upon variations of dataset, methodology (in
particular the PDF  parametrization basis) and Standard Model
parameters (specifically the charm mass),
as demonstrated in the Supplementary Information (SI) Sects.~\ref{app:charm_stability_4fns}
and~\ref{app:charm_stability_3fns}. 

Our determination of intrinsic charm can be compared to theoretical expectations.
Subsequent to the
original intrinsic charm model of~\cite{Brodsky:1980pb} (BHPS
model),
a variety of other models were  
proposed~\cite{Hoffmann:1983ah,Pumplin:2005yf,Paiva:1996dd,Steffens:1999hx,Hobbs:2013bia},
see~\cite{Brodsky:2015fna} for a review.
Irrespective of their specific details, most models predict a valence-like
structure at large $x$ 
with a maximum located  between $x\simeq 0.2$ and $x\simeq 0.5$, and a
vanishing intrinsic component for
$x\lsim 0.1$.
In Fig.~\ref{fig:charm_content_3fns}~(right) we compare our result to
the original BHPS model and to the more recent meson/baryon cloud model of~\cite{Hobbs:2013bia}.

As these models predict only the shape of the
intrinsic charm distribution, but not 
its overall normalization, we have normalized them by requiring
that they reproduce the same 
charm momentum fraction as our determination.
We find remarkable agreement between the shape of our 
determination and the model predictions.
In particular, we reproduce  the presence and location of the large-$x$ valence-like peak
structure (with  better agreement, of marginal statistical significance, with
the meson/baryon cloud calculation),  and the vanishing of 
intrinsic charm at small-$x$.
The fraction of the proton momentum carried by charm quarks that we
obtain from our analysis, 
used in this comparison to models,  is $\lp 0.62 \pm 0.28\rp \%$
including PDF uncertainties only (see
SI Sect.~\ref{app:charm_mom_frac} for details).
However, the uncertainty
upon inclusion of MHOU greatly increases, and we obtain
$\lp 0.62 \pm 0.61\rp \%$, due to the contribution from the small-$x$
region, $x\lsim 0.2$, where the MHOU is very large, see
Fig.~\ref{fig:charm_content_3fns}~(right).
Note that in most previous
analyses~\cite{Hou:2017khm} (see SI Sect.~\ref{app:ct}) intrinsic charm models (such as the BHPS
model) are fitted to the data, with only the momentum fraction left as
a free parameter.

We emphasize that in our analysis the charm PDF is entirely
determined by the experimental data included in the PDF determination.
The data with the most impact on charm are from recently measured LHC
processes, which are both accurate and precise.
Since these measurements are made at high scales, the corresponding
hard cross-sections can be reliably computed in QCD perturbation theory.

Independent evidence for intrinsic charm
is provided by the very recent LHCb measurements of $Z$-boson production
in association with charm-tagged jets in the forward
region~\cite{LHCb:2021stx}, which were not included in our baseline dataset.
This process, and specifically the ratio $\mathcal{R}_j^c$
of charm-tagged jets normalized to flavor-inclusive jets,
is directly sensitive to the charm PDF~\cite{Boettcher:2015sqn}, and
with LHCb kinematics also
in the kinematic region  where the  intrinsic component is relevant.
Following~\cite{Boettcher:2015sqn,LHCb:2021stx}, we have 
evaluated $\mathcal{R}_j^c$ at NLO~\cite{Alioli:2010xd,Sjostrand:2007gs}  
(see  SI Sect.~\ref{sec:zcharm} for details), both with our default PDFs
that include intrinsic charm, and also with an independent PDF determination in
which intrinsic charm is constrained to vanish 
identically, so charm is determined by perturbative matching
(see SI Sect.~\ref{app:consistency}).

\begin{figure}[htbp]
  \begin{center}
    \includegraphics[width=0.99\linewidth]{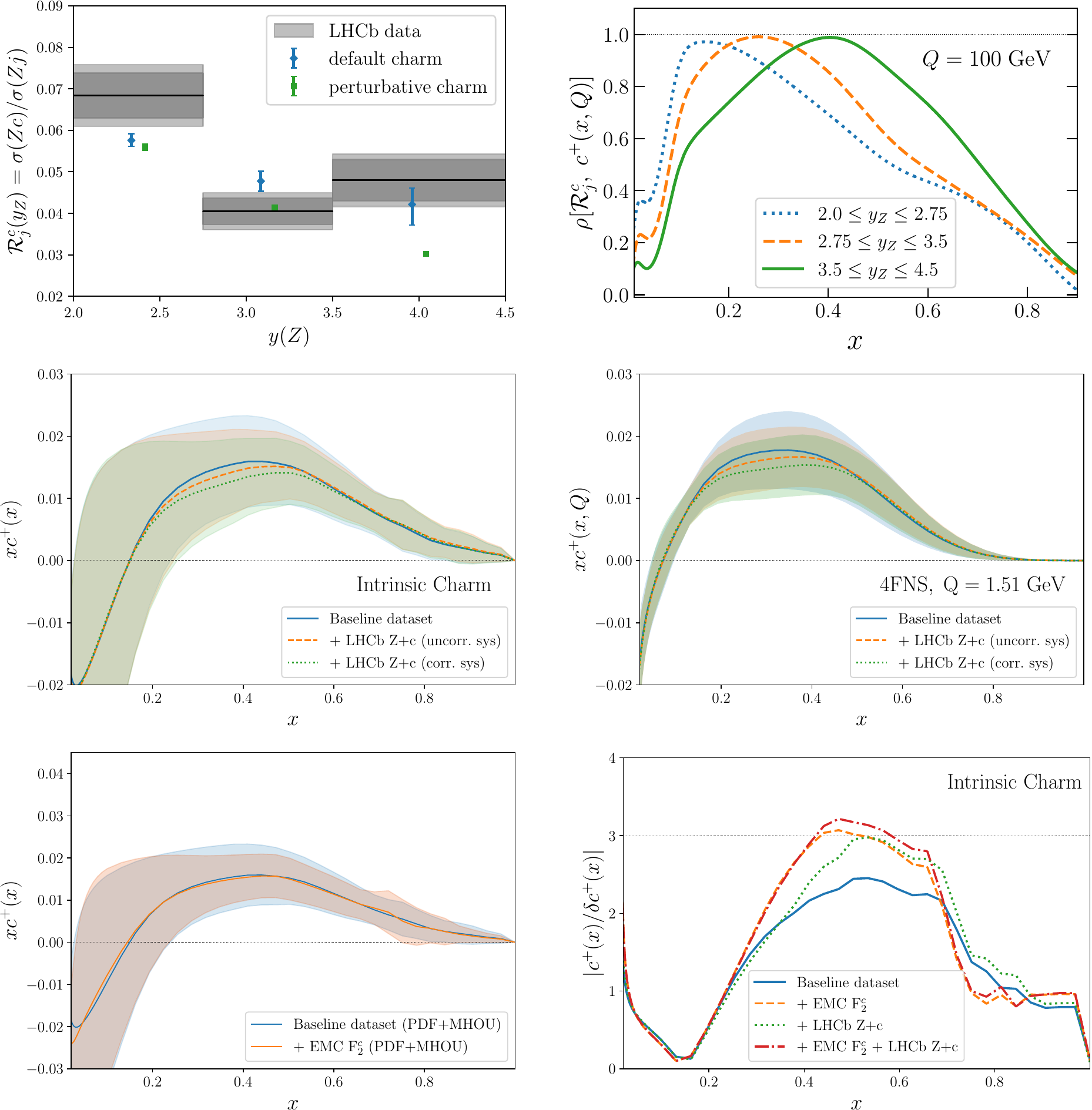}
     \caption{\small
       {\bf Intrinsic charm and $Z+$charm production at LHCb.}
       Top left: the LHCb measurements of $Z$ boson production
      in association with charm-tagged jets, $\mathcal{R}_j^c$, at $\sqrt{s}=13$ TeV,  compared with
      our default prediction which includes an intrinsic charm component,
      as well as with a variant in which we impose the
      vanishing of the intrinsic charm component.
       The thicker (thinner) bands in the LHCb data indicate the statistical
      (total) uncertainty, while the theory predictions include both PDF and MHO uncertainties.
      Top right: the correlation coefficient between
     the  charm PDF at $Q=100$ GeV in NNPDF4.0
      and the LHCb measurements of $\mathcal{R}_j^c$ 
     for the three $y_Z$ bins.
     Center: the charm PDF
     in the 4FNS (right) and the intrinsic (3FNS) charm PDF (left)
     before and after inclusion of the LHCb $Z$+charm data.
     Results are shown
     for both experimental correlation models discussed in the text.
     Bottom left: the intrinsic charm PDF before and after inclusion
     of the EMC charm structure function data.
     Bottom right: the statistical significance of the
     intrinsic charm PDF in our baseline analysis, compared to the results
     obtained also including either the LHCb $Z$+charm (with uncorrelated
     systematics) or the EMC
     structure function data, or both.
  \label{fig:Zc} }
\end{center}
\end{figure}

In Fig.~\ref{fig:Zc}~(top left) we compare the LHCb measurements of $\mathcal{R}_j^c$, provided
in three bins of the $Z$-boson rapidity
$y_Z$, with the theoretical predictions
 based on both our default PDFs as well as the PDF set in
 which we impose the vanishing of intrinsic charm.
 In Fig.~\ref{fig:Zc}~(top right)
we also display the  correlation coefficient between
 the  charm PDF at $Q=100$ GeV 
 and the observable  $\mathcal{R}_j^c$, demonstrating how this observable
 is highly
 correlated to charm in a localized
 $x$ region that depends on the rapidity bin.
 It is clear that
 our prediction is in excellent agreement with the LHCb measurements, while in the
 highest rapidity bin, which is highly correlated to the charm PDF in
 the region of the observed valence peak $x\simeq 0.45$, the prediction
 obtained by imposing the vanishing of intrinsic charm undershoots the
 data at the $3\sigma$ level.
 Hence this measurement provides
 independent direct evidence in support of our result.

 We have also determined the impact of these LHCb $Z$+charm measurements on the
charm PDF.
Since the experimental covariance matrix is not available,
we have considered two limiting scenarios in which the total
systematic uncertainty is either completely uncorrelated 
($\rho_{\rm sys}=0$) or fully correlated  ($\rho_{\rm sys}=1$) between
 rapidity bins. The charm PDF in the 4FNS before and after
inclusion of the LHCb data (with either correlation model), and the intrinsic
charm PDF obtained from it, are displayed in
Fig.~\ref{fig:Zc}~(center left and right respectively).
The bands account for both PDF and MHO uncertainties.
The results show full consistency: inclusion of the LHCb  $\mathcal{R}_j^c$ data leaves
the intrinsic charm PDF unchanged, while moderately reducing the
uncertainty on it.

In the past, the main indication for  intrinsic charm came from EMC data~\cite{Aubert:1982tt} on deep inelastic scattering with charm in the final state~\cite{Harris:1995jx}.
These data are relatively imprecise, their accuracy has often been questioned,
and they were taken at relatively low scales where radiative corrections are large.
For these reasons, we have not included them in our baseline
analysis.
However, it is interesting to assess the impact of
their inclusion.
Results are shown in 
Fig.~\ref{fig:Zc}~(bottom left), where we display the
intrinsic charm PDF before and after inclusion of the EMC data.
Just
like in the case of the LHCb data we find full consistency: unchanged
shape and a moderate reduction of uncertainties.

We can summarize our results  through their so-called local statistical
significance, namely, the size of the intrinsic charm PDF
in units of its total uncertainty.
This displayed  in Fig.~\ref{fig:Zc}~(bottom right) for our default determination of
intrinsic charm, as well as after inclusion of either the LHCb $Z$+charm or the
EMC data, or both.
We find a local significance for intrinsic charm at the $2.5\sigma$ level
in the region $0.3 \lsim x \lsim 0.6$.
This is increased to about
$3\sigma$ by the inclusion of either the EMC or the LHCb
data, and above if they are both included.
The similarity of the impact of the EMC and LHCb measurements is
especially remarkable in view of the fact that they involve very
different physical processes and energies.

In summary, in this work we have presented
long-sought evidence for intrinsic charm quarks in the proton.
Our findings
close a fundamental open question
in the understanding of nucleon structure that has been hotly
debated by  particle and nuclear physicists for the last 40 years.
By carefully disentangling the perturbative component,
we obtain unambiguous evidence for intrinsic charm, 
which turns out to be in qualitative agreement with
the expectations from model calculations.
Our determination of the charm PDF, driven by indirect constraints from the 
latest high-precision LHC data, is perfectly
consistent with direct constraints both from EMC charm production
data taken forty years  
ago, and with very recent  $Z$+charm production data in the
forward region from LHCb.
Combining all data, we find
local significance for intrinsic charm in the large-$x$
region just above the  $3\sigma$ level.
Our results motivate
further dedicated studies of intrinsic charm through a wide range
of nuclear, particle and astro-particle physics experiments,
from the High-Luminosity LHC~\cite{Azzi:2019yne}
and the fixed-target programs of LHCb~\cite{LHCb:2018jry} and
ALICE~\cite{QCDWorkingGroup:2019dyv}, to
the  
Electron Ion Collider, AFTER~\cite{Hadjidakis:2018ifr},
the Forward Physics Facility~\cite{Anchordoqui:2021ghd},
and neutrino telescopes~\cite{Halzen:2016thi}.


\subsection*{Acknowledgments}

We thank our colleagues of the NNPDF Collaboration
for many illuminating discussions concerning the charm PDF.
We are grateful to Johannes Bl\"umlein for communicating  {\sc\small Mathematica}
code with the results
of~\cite{Bierenbaum:2009zt,Bierenbaum:2009mv,Ablinger:2010ty,Ablinger:2014vwa,Ablinger:2014uka,Behring:2014eya,Ablinger_2014,Ablinger:2014nga,Blumlein:2017wxd}, to
Jakob Ablinger for assistance
in the implementation of the $\mathcal{O}\lp \alpha_s^3 \rp$ calculation
of the heavy quark matching conditions, and to Silvia Zanoli
for sharing her {\sc\small Mathematica} implementation with us.
We are grateful to Rhorry Gauld for discussions, assistance and sharing his {\sc\small Pythia8} 
implementation for the calculation $Z$+charm production.
We thank Marco Guzzi and Pavel Nadolsky for discussions concerning intrinsic charm
in the CT family of global PDF fits, and Tim Hobbs and Wally Melnitchouk for providing
us with their predictions of the meson/baryon cloud model.
We are grateful to Tom Boettcher, Philip Ilten, and Michael Williams to assistance
with the LHCb $Z$+charm measurements.

\subsection*{Funding information}

S.~F., J.~C.-M., F.~H., A.~C., and K.~K. are supported by
the European Research Council under 
the European Union's Horizon 2020 research and innovation Programme
(grant agreement n.740006).
R.~D.~B. is supported by the U.K.
Science and Technology Facility Council (STFC) grant ST/P000630/1. 
J.~R. and G.~M. are partially supported by NWO (Dutch Research Council).
T.~G. is supported by NWO (Dutch Research Council) via an ENW-KLEIN-2 project.

\subsection*{Code and data availability}

The analysis presented in this work has been carried out using two
open-source software frameworks, {\sc\small NNPDF} for the global
PDF determination and {\sc\small EKO} for the calculation of the 3FNS charm.
These codes are publicly available from \url{https://docs.nnpdf.science/}
and \url{https://eko.readthedocs.io/} respectively.
Both the {\sc\small LHAPDF} grids produced in this work and the version
of {\sc\small EKO} with the respective run cards used are made available from
\url{http://nnpdf.mi.infn.it/nnpdf4-0-charm-study/}.

\subsection*{Author contributions}

As customary in high-energy physics, authors are listed
in alphabetical order.

J.~C.~M. is the main author of the new algorithm used in the NNPDF4.0
PDF determination.
A.~C., F.~H., and G.~M. developed the {\sc\small EKO} code used
to evaluate the 3FNS charm PDF, and specifically G.~M. implemented the
matching conditions, with the help of K.~K. for the implementation of
some harmonic sums.
T.~G. performed the analysis of the LHCb $Z$+charm data.
R.~D.~B. and S.~F. designed the general procedure. J.~R. coordinated
the intrinsic charm determination and S.~F. supervised the whole
project. J.~R.~ and S.~F. wrote the paper and R.~D.~B. revised it.
All authors discussed the results and their implications.

\clearpage
\section*{Methods}
\label{sec:methods}

The strategy adopted in this work in order to
determine the intrinsic charm content of the proton is 
based on the following
observation.
The assumption that there is no intrinsic charm
amounts to the assumption
that all 4FNS PDFs are determined~\cite{Collins:1986mp} using
perturbative matching conditions~\cite{pdfnnlo} in terms of 
3FNS PDFs
that do not include
a charm PDF.
However, these perturbative matching conditions are
actually given by a square matrix that also includes a 3FNS charm
PDF.
So the assumption of no intrinsic charm amounts to the assumption
that if the 4FNS PDFs are transformed back to the 3FNS, the 3FNS charm
PDF is found to vanish. Hence, intrinsic charm is by definition the
deviation from zero of the 3FNS charm PDF~\cite{Ball:2015dpa}. Note
that whereas the 3FNS charm PDF is purely intrinsic, while the 4FNS
charm PDF includes both an intrinsic and a perturbative
 radiative component, the
4FNS intrinsic component is not equal to the 3FNS charm PDF, since
matching conditions reshuffle all PDFs among each other. 

\begin{figure}[h]
\begin{center}
  \includegraphics[width=0.8\textwidth]{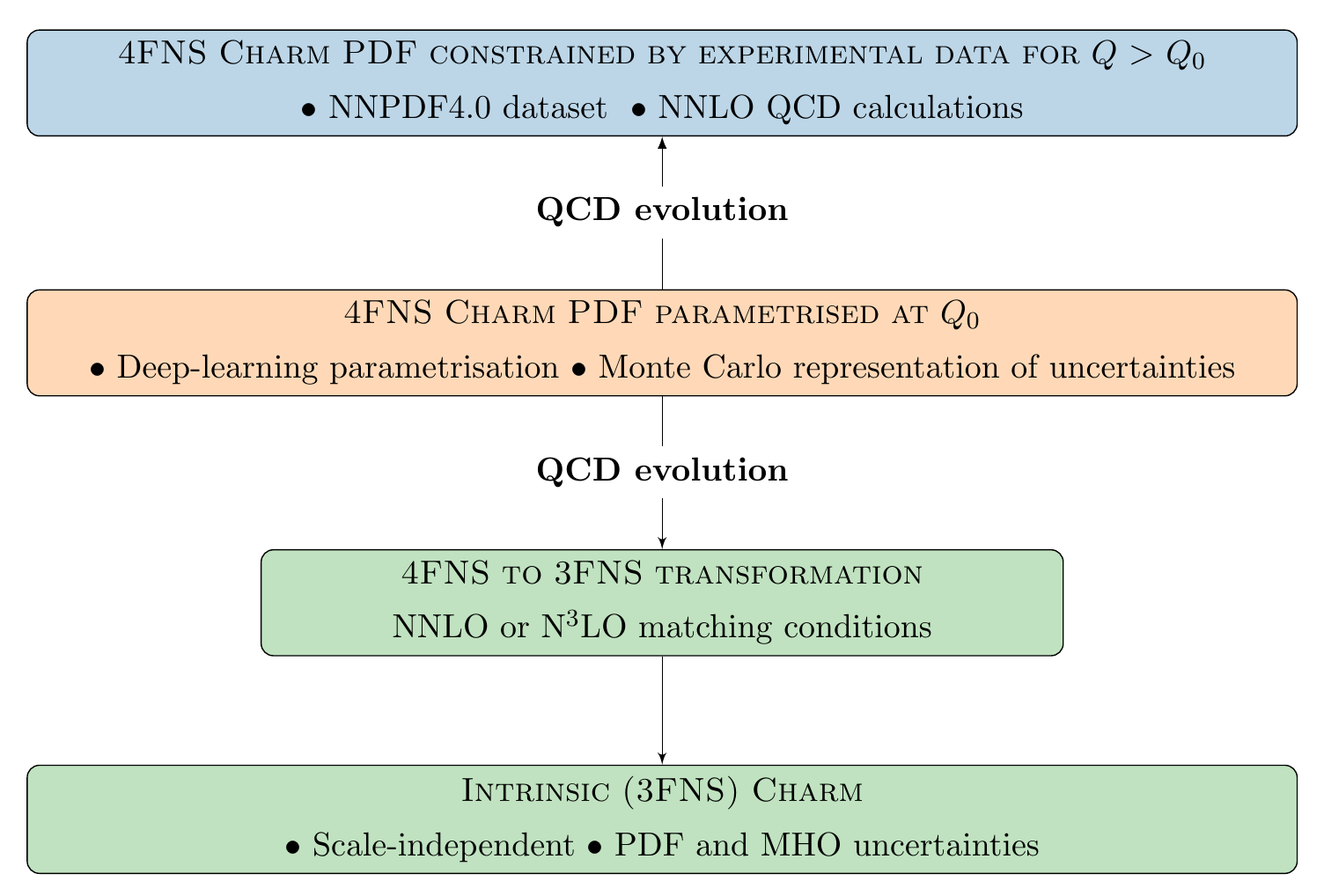}
 \end{center}
\vspace{-0.2cm}
\caption{The 4FNS charm PDF is parametrized  at $Q_0$
  and evolved to all  $Q$, where it is  constrained by the NNPDF4.0
  global dataset. 
 Subsequently, it is transformed to the 3FNS where (if nonzero) it
 provides the intrinsic charm component.
  \label{fig:strategy}
}
\end{figure}

Intrinsic charm can then be determined through the following two steps,
summarized in Fig.~\ref{fig:strategy}. 
First, all the PDFs, including the charm PDF, are parametrized 
in the 4FNS at an input scale $Q_0$ and evolved 
using NNLO perturbative QCD to   $Q \not = Q_0$.
These evolved PDFs can be used to 
compute physical cross-sections, also at NNLO, which then are
compared to a global dataset of experimental measurements.
The result of this first step in our procedure is 
a Monte Carlo (MC) representation
of the probability distribution for the 4FNS PDFs at the input
parametrization scale $Q_0$.

Next, this 4FNS charm PDF is transformed to the 3FNS at some scale matching scale
$Q_c$.
Note that the choice of both $Q_0$ and $Q_c$ are immaterial. The former
because perturbative evolution is invertible, so
results for the PDFs do not depend on the choice of
parametrization scale $Q_0$. The latter because 
the 3FNS charm is scale independent, so it does not depend on the
value of $Q_c$.
Both statements of course hold up to fixed perturbative accuracy, and
are violated by MHO corrections.
In practice, we parametrize PDFs at the scale
$Q_0=$~1.65~GeV and perform the inversion at a scale
chosen equal to the charm mass $Q_c=m_c=1.51$~GeV.

The scale-independent 3FNS charm PDF is then the sought-for intrinsic
charm.

\paragraph{Global QCD analysis.}
The 4FNS charm PDF and its associated
uncertainties is determined by means of a global QCD analysis
within the NNPDF4.0 framework.
All PDFs, including the charm PDF, are  parametrized at $Q_0=1.65$ GeV in 
a model-independent manner using a neural network, which is fitted to data using 
supervised machine learning techniques.
The Monte Carlo replica method
is deployed to ensure a faithful uncertainty estimate.
Specifically, we express the 4FNS total charm PDF ($c^+=c+\bar{c}$)  in terms of the output neurons associated to the quark singlet $\Sigma$ and non-singlet $T_{15}$
distributions, see Sect.~3.1 of~\cite{Ball:2021leu}, as
\be
\label{eq:fitted_charm_param}
xc^+(x,Q_0;{\boldsymbol \theta}) =
\lp x^{\alpha_{\Sigma}}(1-x)^{\beta_{\Sigma}} {\rm NN}_{\Sigma}(x,{\boldsymbol \theta})-
x^{\alpha_{T_{15}}}(1-x)^{\beta_{T_{15}}} {\rm NN}_{T_{15}}(x,{\boldsymbol \theta})
\rp/4 \, ,
\ee
where ${\rm
  NN}_{i}(x,{\boldsymbol \theta})$ is the $i$-th output neuron of a
neural network with input $x$ and  parameters ${\boldsymbol \theta}$,
and 
$\lp \alpha_i,\beta_i\rp $ are
preprocessing exponents.
A crucial feature of Eq.~(\ref{eq:fitted_charm_param}) is that no {\it ad hoc} specific model assumptions are used: the shape and size of $xc^+(x,Q_0)$ are entirely determined from experimental data.
Hence, our determination of the 4FNS fitted charm PDF, and thus of the intrinsic charm, is unbiased.
%

\begin{figure}[t]
\begin{center}
  \includegraphics[width=1.0\textwidth]{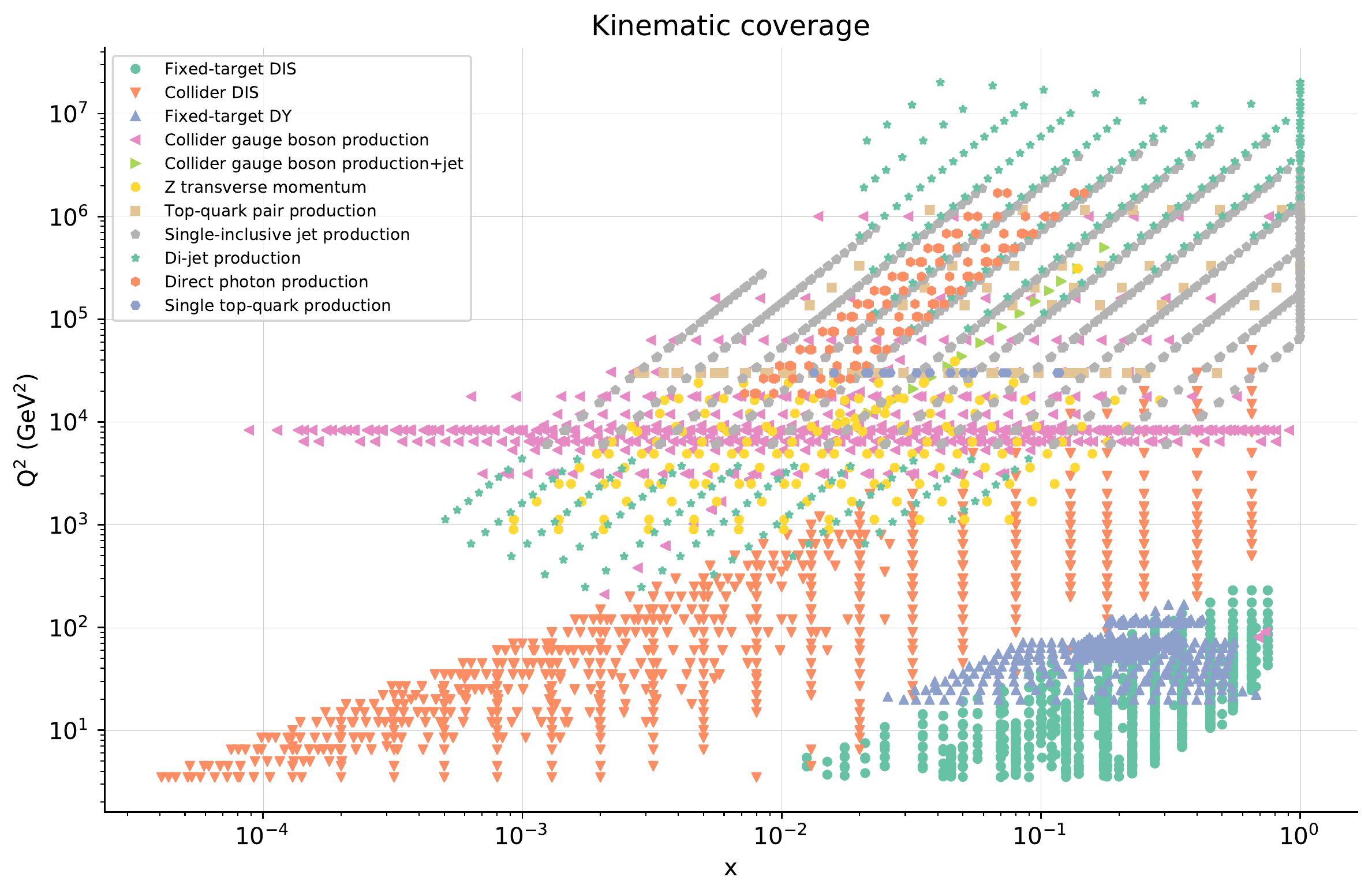}
 \end{center}
\vspace{-0.8cm}
\caption{The kinematic coverage in the $(x,Q)$ plane
  covered by the 4618 cross-sections used for the
  determination of the charm PDF in the present work.
  These cross-sections have been classified into the main different
  types of processes entering the global analysis.
  \label{fig:kinplot}
}
\end{figure}

The neural network parameters ${\boldsymbol \theta}$ in
Eq.~(\ref{eq:fitted_charm_param})
are determined by fitting an extensive global dataset that consists of 4618 
cross-sections from a wide range of different processes, measured over
the years in a variety of fixed-target and collider experiments  (see~\cite{Ball:2021leu} for a complete list).
Fig.~\ref{fig:kinplot} displays the kinematic coverage in the $(x,Q)$ plane
covered by these cross-sections, where $Q$ is
the  scale, and  $x$ is
the parton momentum fraction that correspond to leading-order kinematics.
Many of these processes provide direct or indirect sensitivity 
to the charm content of the proton.
Particularly important constraints come from $W$ and $Z$ production from 
ATLAS, CMS, and LHCb as well as from
neutral and charged current deep-inelastic 
scattering (DIS) structure functions from HERA.
The 4FNS  PDFs at the input scale $Q_0$ are related
to experimental measurements at $Q \not =Q_0$ by means of NNLO QCD calculations, including
the FONLL-C general-mass scheme for DIS~\cite{Forte:2010ta} generalized to 
allow for fitted charm~\cite{Ball:2015tna}.

We have verified (see
SI Sects.~\ref{app:charm_stability_4fns} and~\ref{app:charm_stability_3fns}) that the
determination of 4FNS charm PDF Eq.~(\ref{eq:fitted_charm_param}) and
the ensuing 3FNS intrinsic charm PDF are  stable upon variations
of methodology (PDF parametrization basis), input dataset, and values
of Standard Model parameters (the charm mass).
We have also studied the stability of our results upon replacing the
current NNPDF4.0 methodology~\cite{Ball:2021leu} with the previous
NNPDF3.1 methodology~\cite{NNPDF:2017mvq}. It turns out that results
are  perfectly consistent. Indeed, the old methodology leads to somewhat larger
uncertainties, corresponding to a moderate reduction of the local statistical
significance for intrinsic charm, and to a central value which is
within the smaller  error band of our current result.

A determination in which the vanishing of intrinsic charm is
imposed has also been performed.
In this case, the fit quality significantly
deteriorates: the values of the $\chi^2$ per data point of 1.162,
1.26, and 1.22 for total, Drell-Yan, 
and neutral-current DIS data respectively, found when fitting charm, are 
increased to 1.198, 1.31, 1.28 when the vanishing of intrinsic charm
is imposed.
The absolute worsening of the total $\chi^2$ when the vanishing of intrinsic charm is imposed is therefore
of 166 units, corresponding to
a $2\sigma$ effects in units of $\sigma_{\chi^2}= \sqrt{2n_{\rm dat}}$.

\paragraph{Calculation of the 3FNS charm PDF.}
The Monte Carlo representation of the probability distribution associated to
the 4FNS charm PDF determined by the global
analysis contains an intrinsic component mixed with a perturbatively
generated contribution, with the latter
becoming larger in the $x\lsim 0.1$ region as the scale $Q$ is increased.
In order to extract the intrinsic component, 
we transform PDFs to the 3FNS at the scale $Q_c=m_c=1.51$~GeV using
{\sc\small EKO}, a novel {\sc\small Python} open source
PDF evolution framework (see  SI Sect.~\ref{app:eko}).
In its current implementation, {\sc\small EKO} performs  QCD 
evolution of PDFs to any scale
up to NNLO. For
the sake of the current analysis, N$^3$LO matching conditions have also
been implemented, by 
using  the results
of~\cite{Bierenbaum:2009zt,Bierenbaum:2009mv,Ablinger:2010ty,Ablinger:2014vwa,Ablinger:2014uka,Behring:2014eya,Ablinger_2014,Ablinger:2014nga,Blumlein:2017wxd}
for $\mathcal{O}(\alpha_s^3)$ operator matrix elements~
so that the direct and inverse transformations from the 3FNS to the
4FNS can be performed at one order
higher.
The N$^3$LO contributions to the matching conditions are a subset of
the full N$^3$LO terms that would be required to perform a PDF determination
 to one higher perturbative order, and would
also include currently unknown
N$^3$LO contributions to QCD evolution. Therefore, our results have 
NNLO accuracy and we can only use the  N$^3$LO contributions to the
 $\mathcal{O}(\alpha_s^3)$ corrections to the
heavy quark matching
matching conditions as a way to estimate the 
the size of the missing higher orders. 
Indeed, these corrections have a very 
significant impact on the
perturbatively generated component, see SI Sect.~\ref{app:consistency}.
They become large for $x \lsim 0.1$, which coincides with the region
dominated by the perturbative component of the charm PDF,
  and are relatively small for the valence region
  where intrinsic charm dominates.
  
\paragraph{$Z$ production in association with charm-tagged jets.}
The production of $Z$ bosons in association with charm-tagged jets (or alternatively,
with identified $D$ mesons) at the LHC is directly sensitive to the charm content
of the proton via the dominant $gc \to Zc$ partonic scattering process.
Measurements of this process at  the forward rapidities covered by the
LHCb acceptance provide access to the large-$x$ region where the intrinsic 
contribution is expected to dominate.
This is in contrast with the corresponding measurements from ATLAS and CMS,
which only become sensitive to intrinsic charm
at rather larger values of $p_T^Z$ than those
currently accessible experimentally.

We have obtained  theoretical predictions for $Z$+charm production
at LHCb with NNPDF4.0, based on
NLO QCD calculations using
{\sc\small POWHEG-BOX} 
interfaced to {\sc\small Pythia8}
with the Monash 2013 tune for showering,
hadronization, and underlying event.
Acceptance requirements and event selection follow the LHCb analysis,
where in particular charm jets are defined as those anti-$k_T$ $R=0.5$ jets
containing a reconstructed charmed hadron.
The ratio between $c$-tagged and untagged $Z$+jet events can then
be compared with the LHCb measurements
\begin{equation}
  \mathcal{R}_j^c(y_Z) \equiv \frac{N(c~{\rm tagged~jets};y_Z)}{ 
    N({\rm jets};y_Z)} =
  \frac{\sigma(pp\to Z+{\rm charm~ jet};y_Z)}{\sigma(pp \to Z+{\rm jet};y_Z)} \, ,
\end{equation}
as a function of the $Z$ boson rapidity $y_Z$ (see SI Sect.~\ref{sec:zcharm} for details).
The more forward the rapidity $ y_{Z}$, the higher the values
of the charm momentum $x$ being probed.
Furthermore, we have also included  the LHCb measurements in the global PDF determination  
by means of the 
Bayesian reweighting (see SI Sect.~\ref{sec:zcharm}).

\clearpage

\appendix

\setcounter{page}{1}

\numberwithin{equation}{section}
\numberwithin{figure}{section}
\numberwithin{table}{section}


\begin{center}

{\bf \LARGE Supplementary Information }\\[0.8cm]

  {{\LARGE \bf Evidence for intrinsic charm quarks in the proton}}

  \vspace{1.1cm}

  {\small
    {\bf  The NNPDF Collaboration:} \\[0.2cm]
     Richard D. Ball,$^{1}$
  Alessandro Candido,$^{2}$
Juan Cruz-Martinez,$^{2}$
Stefano Forte,$^{2}$
Tommaso Giani,$^{3,4}$\\[0.1cm]
Felix Hekhorn,$^{2}$
Kirill Kudashkin,$^{2}$
Giacomo Magni,$^{3,4}$ and
Juan Rojo$^{3,4}$\daggerfootnote{Corresponding author (\href{mailto:j.rojo@vu.nl}{j.rojo@vu.nl}).}
  }\\

 \vspace{0.7cm}
 
 {\it \small

 ~$^1$The Higgs Centre for Theoretical Physics, University of Edinburgh,\\
   JCMB, KB, Mayfield Rd, Edinburgh EH9 3JZ, Scotland\\[0.1cm]
    ~$^2$Tif Lab, Dipartimento di Fisica, Universit\`a di Milano and\\
   INFN, Sezione di Milano, Via Celoria 16, I-20133 Milano, Italy\\[0.1cm]
    ~$^3$Department of Physics and Astronomy, Vrije Universiteit, NL-1081 HV Amsterdam\\[0.1cm]
~$^4$Nikhef Theory Group, Science Park 105, 1098 XG Amsterdam, The Netherlands\\[0.1cm]
}

 \vspace{1cm}

\end{center}


\tableofcontents

\clearpage

\clearpage
\section{The EKO evolution framework}
\label{app:eko}

A crucial ingredient in the derivation of our results
is the determination of the 3FNS intrinsic
charm PDF starting from the 4FNS, which requires the inversion of the matching
conditions implementing the 3FNS to 
the 4FNS transformation.
This inversion is not available in the open-source NNPDF
code~\citesupp{NNPDF:2021uiq} and it is performed here by means of a novel code
for QCD evolution, 
 {\sc\small EKO} (Evolution Kernel Operators), that we use to take the
 PDF set determined at the reference scale $Q_0$ and evolve it to a
 matching scale $Q_c$ where it is transformed to the 3FNS.
Here we provide a brief summary of {\sc\small
EKO}, some details of the way the direct and inverse matching
conditions are implemented, and some benchmarks between {\sc\small
EKO} and other existing QCD evolution codes, including the  {\sc\small
APFEL}~\citesupp{Bertone:2013vaa} evolution code that is used by the
NNPDF code.
{\sc\small EKO} is written in {\sc\small Python} and is available
open source from its {\sc\small GitHub} repository
\begin{center}
\url{https://github.com/N3PDF/eko}
\end{center}
A more detailed description of the code can be found
in its online documentation
\begin{center}
\url{https://eko.readthedocs.io/en/latest/}
\end{center}
as well as in a dedicated publication~\citesupp{Candido:2022tld}.

The scale dependence of PDFs in QCD is determined by solving a set
of coupled integro-differential equations (evolution equations) in two
variables $x$ (momentum fraction) and $Q$ (scale) on which PDFs depend.
Two families of approaches are commonly used in order to this purpose.
One possibility is to treat the (integral) dependence on $x$ of the
PDFs and evolution kernels by
sampling it on a grid of points.
This is the strategy adopted by, among others, the  {\sc\small APFEL}~\citesupp{Bertone:2013vaa},
{\sc\small HOPPET}~\citesupp{Salam:2008qg},
and {\sc\small QCDNUM}~\citesupp{Botje:2010ay} evolution codes.
An alternative possibility is to perform an integral transform (Mellin
transform) with
respect to $x$ thereby turning the integro-differential equations into
coupled ordinary differential equations. These can then be solved
analytically, but the integral transform has to be inverted numerically
to arrive at a final result.
This approach is adopted by {\sc\small PEGASUS}~\citesupp{pegasus} as well as by the
internal PDF evolution code {\sc\small FastKernel} used in earlier NNPDF analyses
and described in~\citesupp{DelDebbio:2007ee,Ball:2008by,Ball:2010de}.
One limitation of {\sc\small PEGASUS} is that it requires the analytic
computation of the Mellin transforms of
the PDFs, which is generally not possible, specifically if PDFs are
parametrized as neural networks.

This restriction is bypassed in {\sc\small FastKernel} by transforming only the
evolution kernel (i.e. the evolution operator, solution of differential
equations, evaluated on a given interpolation basis), which can be then
convoluted with the $x$-space PDFs at the input evolution scale $Q_0$.
Following a similar strategy, {\sc\small EKO}
solves evolution equations in Mellin space and then produces PDF-independent
evolution kernel operators (EKO) which can be convoluted with input 
PDFs.
Variable flavor number scheme evolution (VFNS) is implemented in {\sc\small EKO}, with
the possibility of freely choosing the value of the matching scales
$Q_{h}$ between the $N$-flavor number scheme (NFNS) in which heavy quark $h$
is not included in QCD evolution, and the $(N+1)$-flavor number
scheme in which it is included.
Schematically, for evolution between $Q_0$ and $Q_1$ if no matching
scales are crossed $\left(Q_{h}^2 < Q_0^2 , Q_1^2 < Q_{h'}^2\right)$ one has:
\be
\label{eq:EKO1}
{\mathbf{f}}^{(n_f)}(Q^2_1)= {\mathbf{E}}^{(n_f)}(Q^2_1\leftarrow Q^2_0) \otimes {\mathbf{f}}^{(n_f)}(Q^2_0) \, ,
\ee
where
${\mathbf{E}}^{(n_f)}$ is the  NFNS {\sc\small EKO},
and $\otimes$ is the Mellin convolution operation. Note that {\sc\small EKO}
can perform both ``forward'' ($Q_0< Q_1$)  and ``backward'' ($Q_1< Q_0$)
evolution.
Bold quantities indicate either vectors or matrices
in the $(2n_f+1)$-dimensional flavor space.
If instead a single matching scale $Q_h$ is crossed, assuming for definiteness $Q_0< Q_1$,  
$Q_{h'}^2 < Q_0^2 < Q_{h}^2 < Q_1^2 < Q_{h''}^2$,
one has
\be
\label{eq:EKO2}
{\mathbf{f}}^{(n_f+1)}(Q^2_1)= \lc {\mathbf{E}}^{(n_f+1)}(Q^2_1\leftarrow Q_{h}^2)  
{\mathbf{A}}^{(n_f)}(Q_{h}^2) {\mathbf{E}}^{(n_f)}(Q_{h}^2\leftarrow Q^2_0) \rc \otimes {\mathbf{f}}^{(n_f)}(Q^2_0) \, ,
\ee
where $\mathbf{A}^{(n_f)}(Q_{h}^2)$ is the scheme transformation
between the NFNS and (N+1)FNS, given as a perturbatively computable
series expansion in $\alpha_s$. 
The quantity in square parenthesis is evaluated in Mellin space and then transformed to $x$-space.
This procedure can be  extended to the case in which  more than
one threshold is crossed.
Also, the scales $Q_0$ and $Q_1$ can be ordered in any way, because both direct
and inverse scheme transformations are implemented in
{\sc\small EKO}. Furthermore, the inverse scheme change  is implemented both
perturbatively (i.e.\ as a series expansion in $\alpha_s$ to the same
accuracy as the direct scheme change) or exactly (i.e.\ as the numerical
inverse, completely equivalent to the analytic one within the numeral accuracy
of the rest of the calculation).

If the heavy quark $h$ has no intrinsic component, then below $Q_h$
its PDF is identically zero, and above $Q_h$ it is determined by
$\mathbf{A}^{(n_f)}(Q_{h}^2)$. If it does have an intrinsic
component, then  below $Q_h$ its PDF is scale-independent, but nonzero.
While  {\sc\small EKO} is currently an NNLO code, on
top of the standard~\cite{pdfnnlo} NNLO scheme change for this work
an N$^3$LO scheme change has been implemented, based on recent higher-order computations
of the relevant operator matrix elements~\cite{Bierenbaum:2009zt,Bierenbaum:2009mv,Ablinger:2010ty,Ablinger:2014vwa,Ablinger:2014uka,Behring:2014eya,Blumlein:2017wxd,Ablinger_2014,Ablinger:2014nga}
and the work of~\citesupp{zanoli}.

\begin{figure}[t]
    \begin{center}
        \includegraphics[width=0.47\linewidth]{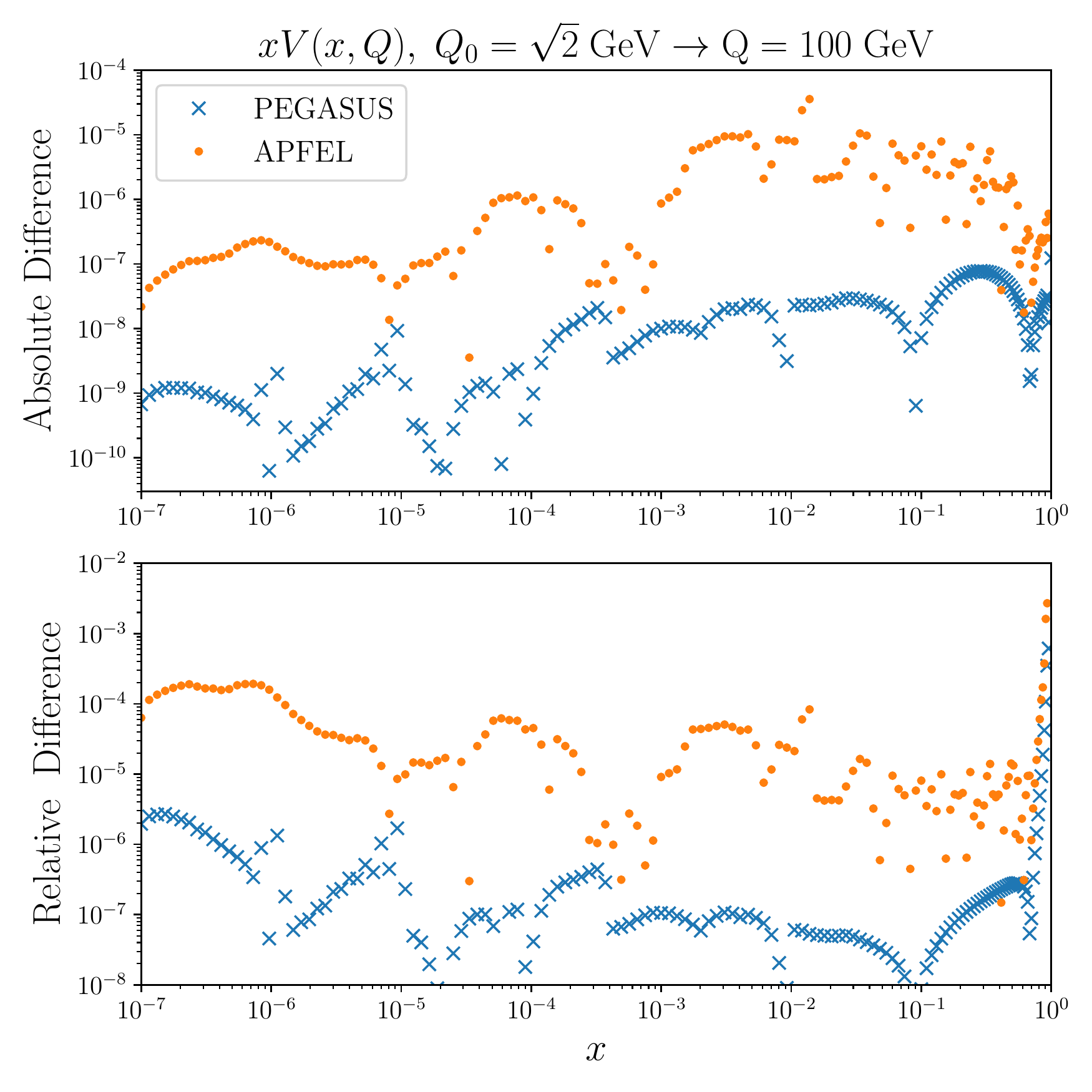}
        \includegraphics[width=0.47\linewidth]{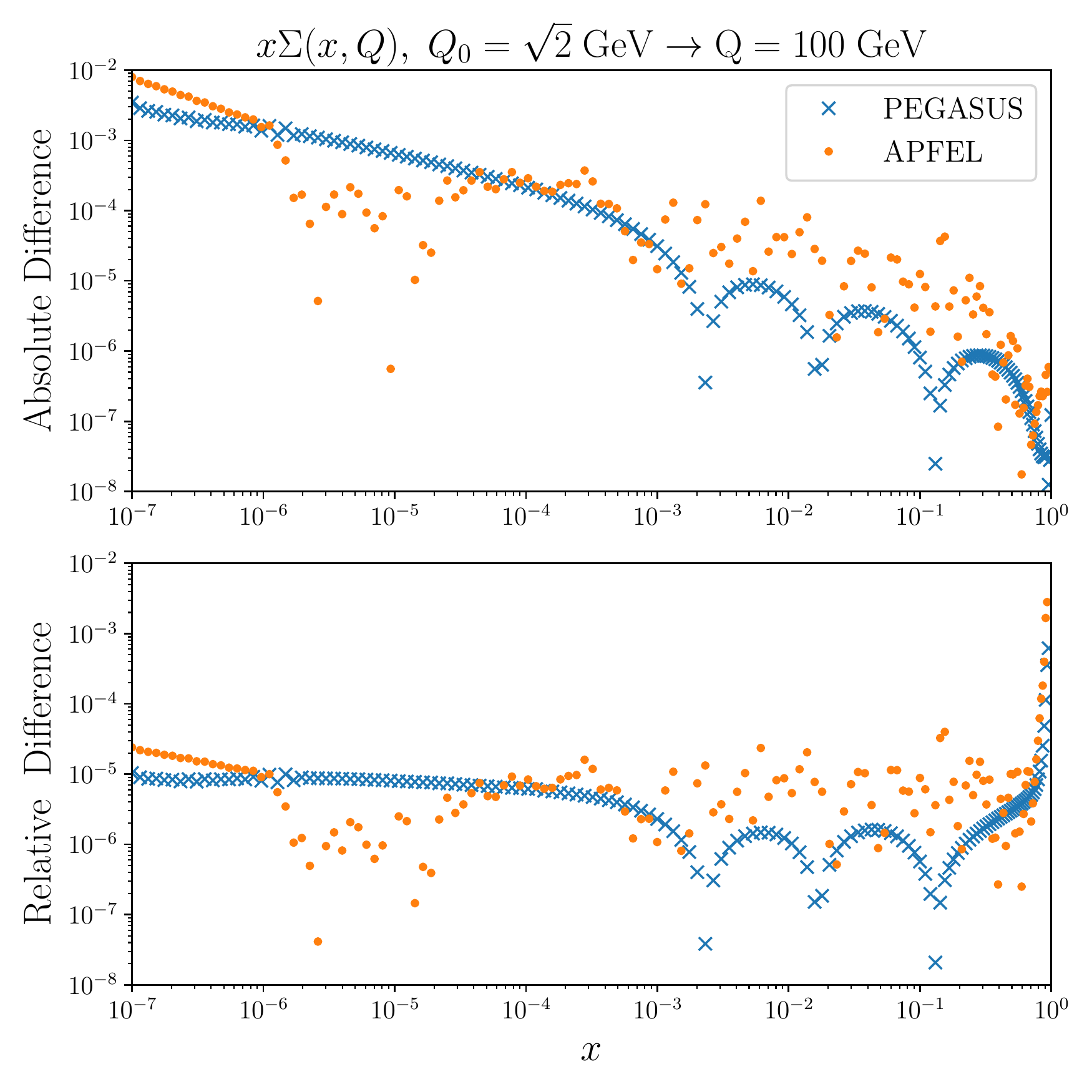}
        \caption{\small Absolute (upper) and relative (bottom) differences between 
        the outcome of NNLO QCD evolution
        as implemented in
        {\sc\small EKO} and the
        corresponding results from {\sc\small APFEL} and {\sc\small PEGASUS}.
We adopt the settings of the Les Houches PDF evolution benchmarks: we
consider  VFNS evolution from $Q_0=\sqrt{2}$ GeV up to $Q=100$ GeV,
and we show  results for the total valence quark distribution $V$ (left)
and the quark singlet distribution $\Sigma$ (right).
      \label{fig:EKObench} }
    \end{center}
\end{figure}

The  {\sc\small EKO} implementation of QCD evolution has been
benchmarked against the Les Houches PDF evolution
benchmarks~\citesupp{Dittmar:2005ed,Giele:2002hx} and with
{\sc\small APFEL} and {\sc\small PEGASUS},
finding excellent agreement beyond the per-mille level.
The implementation of the  matching conditions
has been benchmarked up to N$^3$LO against the independent {\sc\small Mathematica}-based calculation 
presented in~\citesupp{zanoli} finding also good agreement.
To illustrate some of these benchmarks, Fig.~\ref{fig:EKObench} displays
the absolute and relative difference between {\sc\small EKO},
{\sc\small APFEL}, and {\sc\small PEGASUS}
for NNLO VFNS evolution
carried out
following the settings of~\citesupp{Dittmar:2005ed,Giele:2002hx}.
A toy PDF set at $Q_0=\sqrt{2}$ GeV is evolved up to $Q=100$ GeV
for equal values of the factorization and renormalization scales, $Q_f=Q_r=Q$.
We show as representative results those corresponding to
the  total valence quark  $V$ 
and the quark singlet $\Sigma$ distributions.
Excellent agreement is found, in particular
with {\sc\small PEGASUS} which also perform QCD
evolution in Mellin space, with relatively differences
at most at the $\mathcal{O}\lp 10^{-4}\rp$ level.
A similar level of agreement is found
for the gluon and for the other quark PDF combinations.

\clearpage
\section{The perturbative charm PDF}
\label{app:consistency}

In the absence of intrinsic charm, the charm PDF is fully determined by
perturbative matching conditions, i.e.\ by the matrix
$\mathbf{A}^{(n_f)}(Q_{c}^2)$ in Eq.~(\ref{eq:EKO2}).
We will denote the
charm PDF thus obtained
``perturbative charm PDF'', for short. The PDF
uncertainty on the perturbative charm PDF is directly related to that 
of the light quarks and especially the gluon, and is typically much smaller
than  the  uncertainty on our default charm PDF, that includes
intrinsic charm. Here and in the following we will refer to our final
result, as shown in Fig.~\ref{fig:charm_content_3fns} (right) as ``default''.
It should be noticed that the matching conditions for charm are 
nontrivial starting
at NNLO: at NLO the perturbative charm PDF vanishes at threshold.
Hence, having implemented in EKO also the N$^3$LO matching conditions,
we are able to assess the MHOU of the perturbative charm at the
matching scale $Q_c$, by comparing
results obtained at the first two nonvanishing perturbative
orders.

\begin{figure}[h]
  \begin{center}
    \includegraphics[width=0.49\linewidth]{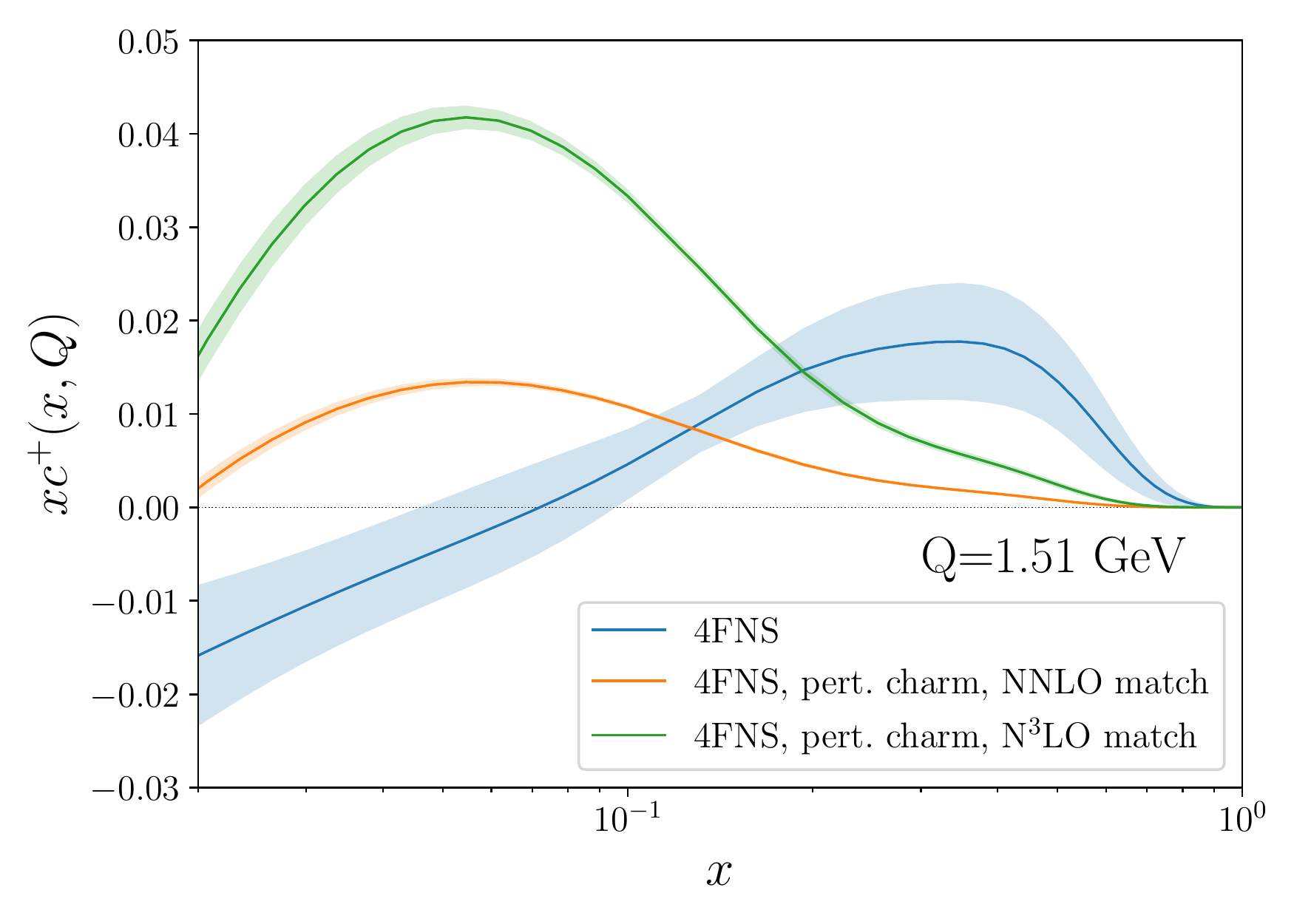}
    \includegraphics[width=0.49\linewidth]{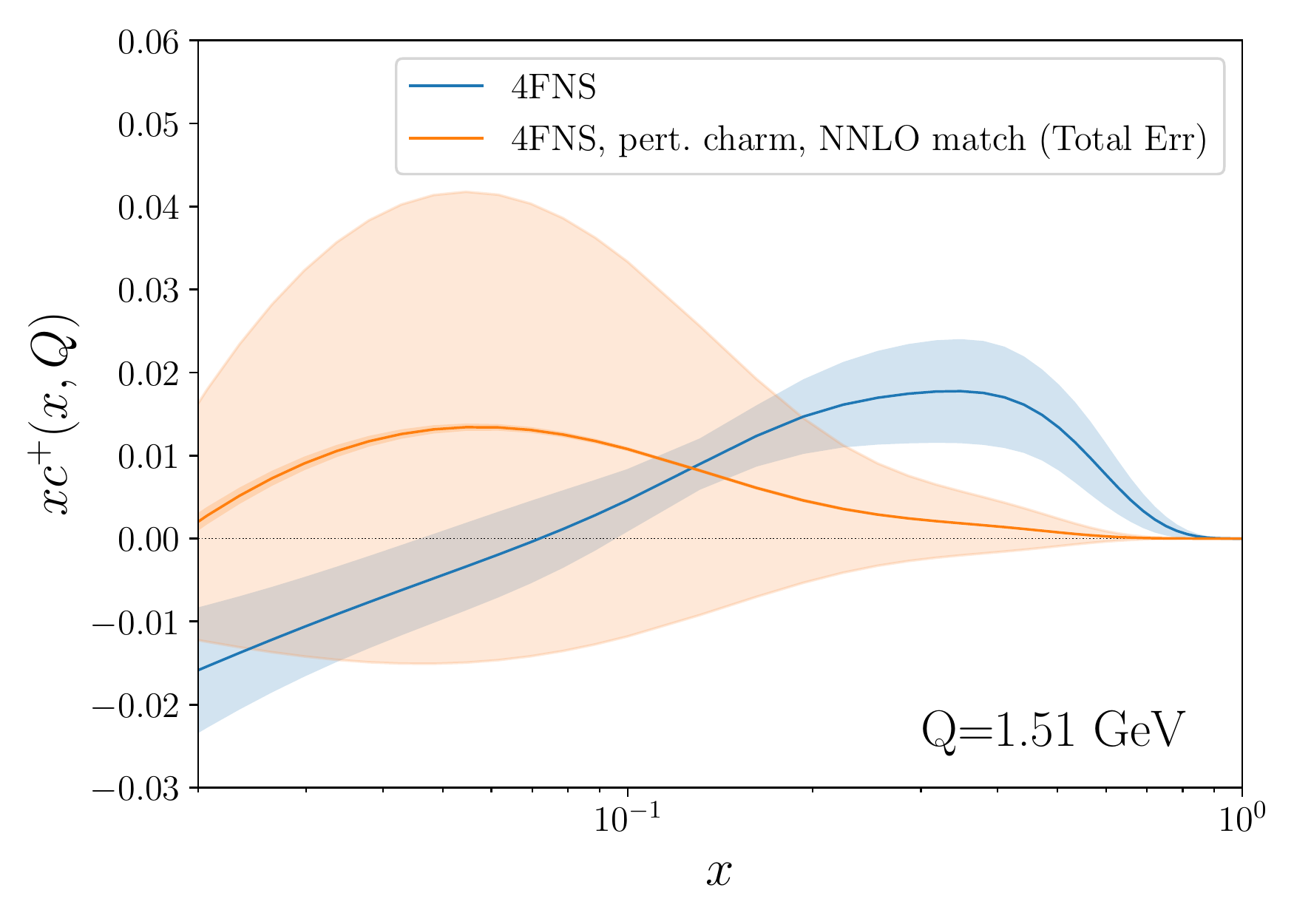}
    \caption{\small Left: the perturbative charm PDF at $Q=1.51$~GeV
  obtained from NNLO PDFs using NNLO and N$^3$LO matching
    conditions.
      Right: the NNLO perturbative charm PDF including the MHOU
    computed as the difference between NNLO and N$^3$LO matching.
In both plots our default (intrinsic) charm PDF is also shown for comparison.  
  \label{fig:charm_fitted_vs_perturbative_mhous} }
\end{center}
\end{figure}

As already mentioned, see also Fig.~\ref{fig:Zc}~(top left) in the main manuscript, we have
constructed a PDF set with perturbative charm, in which the full PDF
determination from the global dataset leading to the NNPDF4.0 PDF set
is repeated, but now with the assumption of vanishing intrinsic charm,
i.e.\ with a perturbative charm PDF.
This perturbative charm PDF is compared to our default result
in Fig.~\ref{fig:charm_fitted_vs_perturbative_mhous}~(left), where the 4FNS
perturbative 
charm PDF at scale  $Q_c=m_c$ obtained using either NNLO or N$^3$LO
under the assumption of no intrinsic charm are shown, together with
our  result allowing for intrinsic charm.
It is clear that while on the one hand, the PDF uncertainty on the
perturbative charm PDF is indeed tiny, on the other
hand the difference between the result for perturbative charm
obtained using NNLO or N$^3$LO matching is large, and in
fact larger at small $x$ than the difference between perturbative charm and our
default (intrinsic) result.

In the same manner as we used the difference between the results obtained from
inversion of NNLO and N$^3$LO  matching as an estimate of the MHOU on
intrinsic charm, we may use the difference between the 4FNS
 perturbative charm obtained from NNLO and N$^3$LO matching as an
 estimate of the MHOU on perturbative charm at the scale $Q_c$.
 The total uncertainty is found by adding
 this in quadrature to the PDF uncertainty (which however in practice
 is negligible).
The result is shown in 
Fig.~\ref{fig:charm_fitted_vs_perturbative_mhous}~(right).
Within this total uncertainty there is now good agreement between our
intrinsic charm result and perturbative charm for all
$x\lsim 0.2$. On the other hand, there is a clear deviation for larger
$x$. We may view the difference between the 4FNS default result
and the 4FNS perturbative  charm as the intrinsic component in the
4FNS, and indeed it is clear from
Fig.~\ref{fig:charm_fitted_vs_perturbative_mhous} that the 4FNS
intrinsic component is sizable and positive at large $x$.
This is of course consistent with our main finding that we
only see evidence of intrinsic charm for large $x\gsim 0.2$, while for
smaller $x$ our result for the charm PDF is compatible with zero, as demonstrated by
Fig.~\ref{fig:charm_content_3fns}~(right) in the main manuscript.

\clearpage
\section{Stability of the 4FNS charm PDF}
\label{app:charm_stability_4fns}

The main input to our determination of intrinsic charm is the 4FNS
charm PDF extracted from high-energy data. While this
determination comes with an uncertainty estimate, it is important to
verify that this adequately reflects the various sources
of uncertainty, and that there are no further sources of uncertainty
that may be unaccounted for.
To this purpose, here we assess the
stability of our results first, upon the choice of underlying dataset,
next upon changes in methodology, and finally, upon variation of
standard model parameters.
In each case we verify stability upon the
most important possible source of instability: respectively, the use
of collider vs.\ fixed target and deep-inelastic vs.\ hadronic data
(dataset); the choice of parametrization basis (methodology); and the
value of the charm quark mass (standard model parameters).
As a final consistency check, we compare our result with that which we
would have obtained by using the same input dataset, but the previous
NNPDF3.1 fitting methodology.
Because we
are interested in intrinsic charm, in all comparisons we focus on
the large-$x$ region in which the intrinsic valence-like peak is found.
In this section, the 4FNS
charm PDF is displayed at the scale $Q = 1.65$ GeV so that
results for all fit variants, including
those with with different $m_c$ values, can be shown at a common scale.

\begin{figure}[t!]
  \begin{center}
    \includegraphics[width=0.99\linewidth]{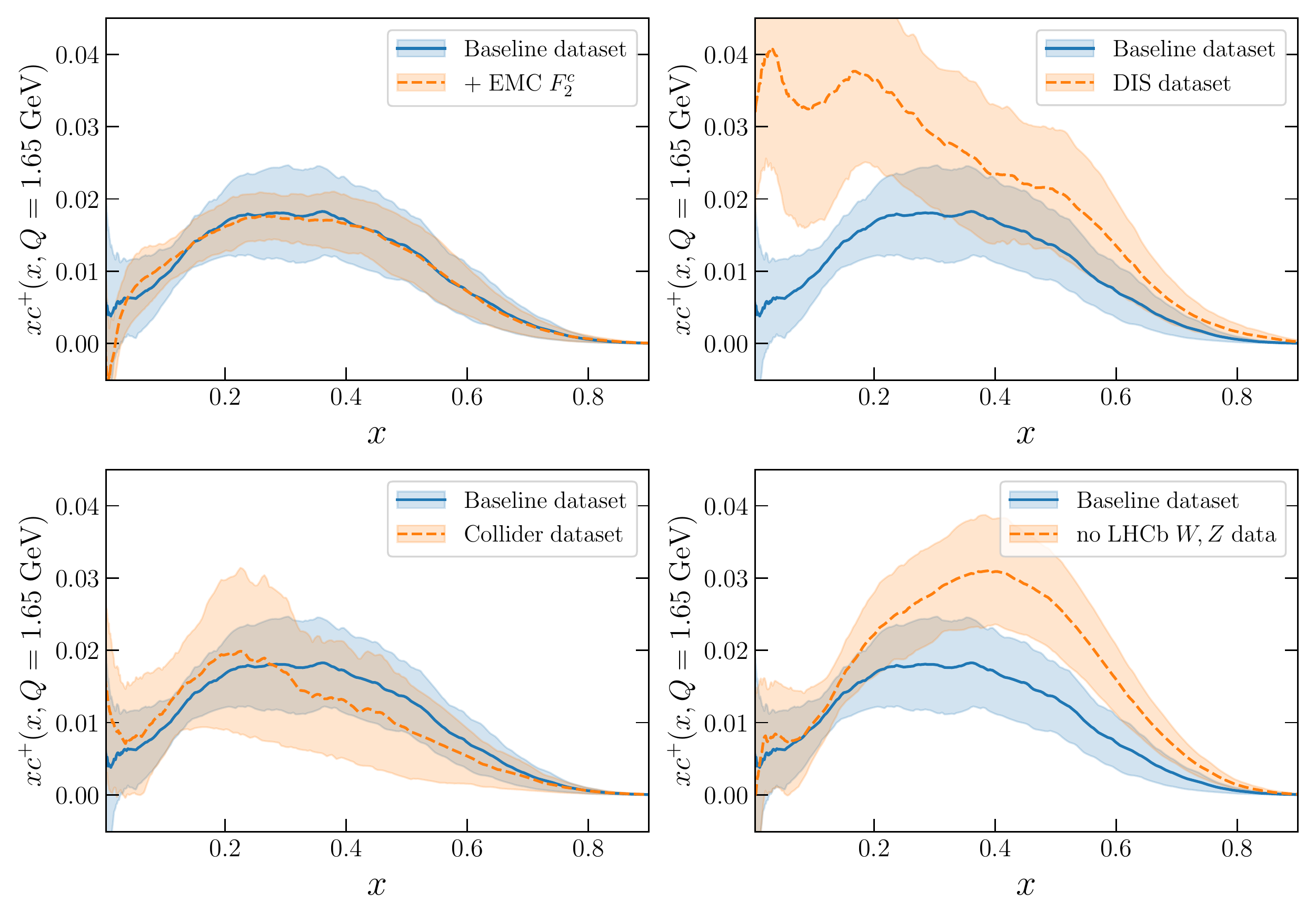}
    \caption{\small The dependence of the 4FNS charm PDF at $Q=1.65$ GeV on
      the input dataset.
      We compare the
    baseline result with that obtained by also including 
      EMC $F_2^c$ data (top left), only including DIS data (top
    right), only including collider data (bottom left) and removing
    LHCb gauge boson production data (bottom right). 
  \label{fig:charm_dataset_dep} }
\end{center}
\end{figure}

\paragraph{Dependence on the choice of dataset.}
We now study the stability of the  4FNS charm determination upon
variation of the
underlying data, which also allows us to
identify the datasets or groups of processes that provide
the leading constraints on intrinsic charm.
To this purpose, we have repeated our PDF
determination using a  variety of subsets of the global dataset used for
our default determination. Results are shown in
Fig.~\ref{fig:charm_dataset_dep}, where we compare the result using
the 
baseline dataset to determinations performed by adding to the baseline
the  EMC charm
structure function data (already discussed in the main text); by only
including  DIS data; by only including collider data (HERA,
Tevatron and LHC); and by removing the LHCb  $W$ and $Z$ production data.

As already noted in the main  text in the case of the 3FNS
result, we find that the extra information provided by the  EMC
$F_2^c$ data is subdominant in comparison to that from the global
dataset. The result is stable and only a moderate
uncertainty reduction at the peak is observed. It is interesting to
contrast this with the previous NNPDF study~\cite{Ball:2016neh}, in
which the global fit provided only very loose constraints on the charm
PDF, which was then determined mostly by the EMC data.
Indeed, a DIS-only fit (for which most data were already available at the time
of the previous determination) determines charm with very large
uncertainties. On the other hand, both the central value and
uncertainty found in the collider-only fit are quite similar to the
baseline result.
This shows that the dominant constraint is now coming from
collider, and specifically hadron collider data (indeed, as we have
seen constraints from DIS data are quite loose). Among these, LHCb
data (which are taken at large rapidity and thus impact PDFs at large
and small $x$) are especially important, as demonstrated by the
increase in uncertainty when they are removed.

In all these determinations, the charm
PDF at $x\simeq 0.4$ remains consistently nonzero and positive, thus
emphasizing the stability of our results.

\paragraph{Dependence on the parametrization basis.}
Among the various methodological choices, a possibly critical one is
the choice of basis functions. Specifically, in our default analysis,
the output of the neural network does not provide the individual
quark flavor and antiflavor PDFs, but rather linear combinations
corresponding to the so-called evolution
basis~\cite{Ball:2021leu}. Indeed, our charm PDF is given in
Eq.~(\ref{eq:fitted_charm_param})  as the linear combination of the
two basis PDFs $\Sigma$ and $T_{15}$.
One may thus ask whether this choice may influence the final results
for individual quark flavors, specifically charm.
Given that physical results are basis
independent, the outcome of a PDF determination should not depend
on the basis choice.

\begin{figure}[t!]
  \begin{center}
    \includegraphics[width=0.60\linewidth]{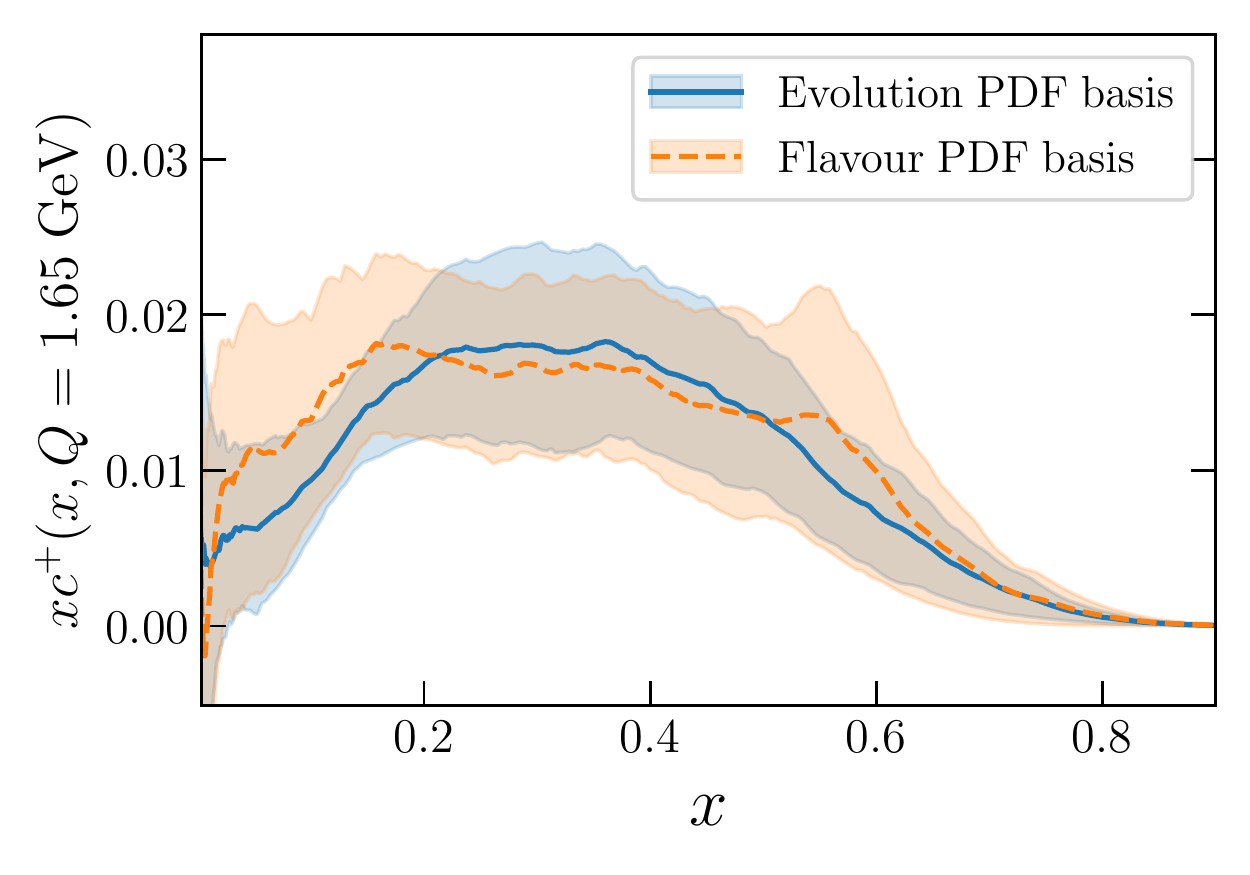}
    \caption{\small The default 4FNS charm PDF at $Q=1.65$ GeV
    compared to a result obtained by parametrizing PDFs in the flavor
    basis instead of the evolution basis. 
  \label{fig:charm_basisdep} }
\end{center}
\end{figure}

In order to check this, we have repeated the PDF determination, but
now using the flavor basis, see Sect.~3.1 of~\cite{Ball:2021leu}, in which
each of the  neural network output neurons now correspond to individual quark
flavors, so in particular,
instead of Eq.~(\ref{eq:fitted_charm_param}),  one has
\be
\label{eq:fitted_charm_param_flavour}
xc^+(x,Q_0;{\boldsymbol \theta}) =
 (1-x)^{\beta_{c^+}} {\rm NN}_{c^+}(x,{\boldsymbol \theta}) \, ,
\ee
where ${\rm NN}_{c^+}(x,{\boldsymbol \theta})$
indicates the value of the output neuron associated to the charm PDF $c^+$.
The 4FNS charm PDFs determined using either basis are compared 
in Fig.~\ref{fig:charm_basisdep}  at $Q=1.65$ GeV.
We find excellent consistency, and in particular 
the valence-like structure at high-$x$ is independent of the choice
of parametrization basis.

\paragraph{Dependence on the charm mass.}
The kinematic threshold for producing charm perturbatively depends on
the value of the charm mass. Therefore the perturbative contribution
to the 4FNS charm PDF, and thus the whole charm PDF if one assumes
perturbative charm, depends strongly on the value of the charm
mass.
On the other hand, the intrinsic charm PDF is of nonperturbative
origin, so it should be essentially independent of the numerical value of the
charm mass that is used in  perturbative computations employed in  its 
determination (though it will of course depend on the true underlying 
physical value of the charm mass).

\begin{figure}[t]
  \begin{center}
    \includegraphics[width=0.99\linewidth]{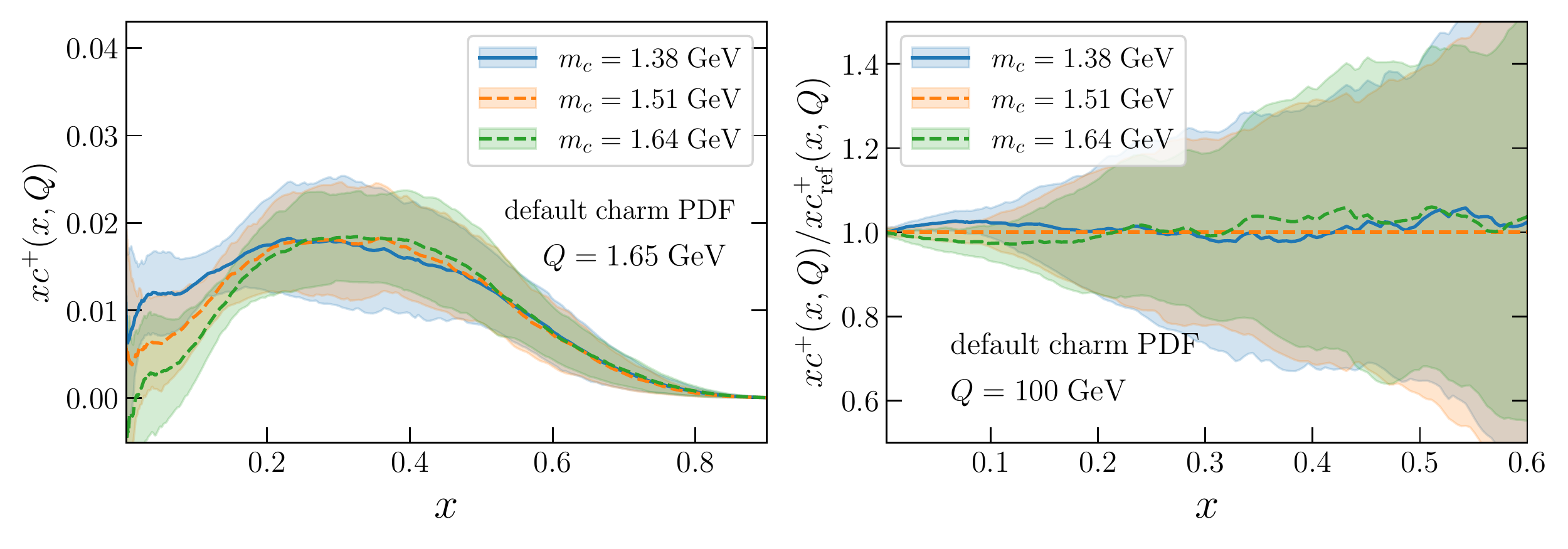}
    \caption{\small The 4FNS charm PDF determined
    using three different values of the charm mass. The absolute
    result (left) is shown at $Q=1.65$ GeV, while the ratio to the 
       default value
       $m_c=1.51$~GeV (right) used elsewhere in this paper is shown at $Q=100$ GeV.
   \label{fig:charm_fitted_mcdep} }
\end{center}
\end{figure}

In order to study this mass dependence, we have repeated our determination using different values for the charm mass.
The definition of the charm mass which is relevant for kinematic
thresholds is the pole mass, for which we adopt the value recommended
by the Higgs cross-section working group~\citesupp{deFlorian:2016spz}
based on the study of~\citesupp{Bauer:2004ve}, namely
 $m_c = 1.51 \pm 0.13$~GeV. 
Results are shown in Fig.~\ref{fig:charm_fitted_mcdep}, where
our default charm PDF determination with  $m_c = 1.51$~GeV is
repeated with $m_c=1.38$~GeV and
$m_c=1.64$~GeV.
In order to understand these results note that this is
the 4FNS PDF, so it includes 
both a nonperturbative and a perturbative component. The latter is
strongly dependent on the charm mass, but of course the data
correspond to the unique true value of the mass and the mass
dependence of the perturbative component is present only due to our
ignorance of the actual true value. When determining the PDF from the
data, as we do, we expect this spurious dependence to be to some extent
reabsorbed into the fitted PDF. So we expect results to display a
moderate dependence on the charm mass --- full independence should
hold for the intrinsic (3FNS) PDF and will be investigated in
SI Sect.~\ref{app:charm_stability_3fns}. 

\begin{figure}[t]
  \begin{center}
    \includegraphics[width=0.99\linewidth]{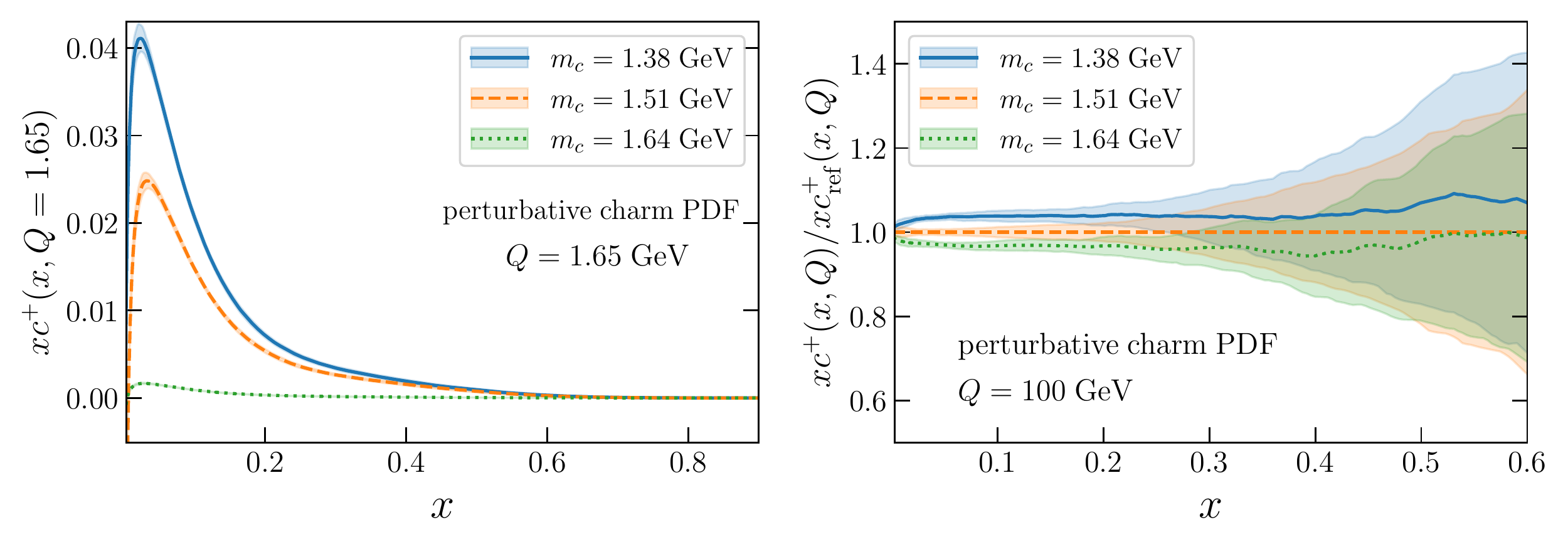}
    \caption{\small The same as Fig.~\ref{fig:charm_fitted_mcdep} but now
    for the perturbative charm PDF.
  \label{fig:charm_pert_mcdep} }
\end{center}
\end{figure}

In Fig.~\ref{fig:charm_pert_mcdep} the same result is shown, but now
for the perturbative charm PDF discussed in
SI Sect.~\ref{app:consistency},  so the charm PDF is of
purely perturbative origin and fully determined by the strongly
mass-dependent matching condition. This dependence is clearly seen in
the plots. Furthermore, comparison with
Fig.~\ref{fig:charm_fitted_mcdep} shows that indeed this spurious
dependence is partly reabsorbed in the fit when the charm PDF is
determined from the data, so that  the residual mass dependence is moderate.
In particular, the large-$x$ valence peak, which is dominated by the
intrinsic component, is very stable.

\begin{figure}[t]
  \begin{center}
    \includegraphics[width=0.65\linewidth]{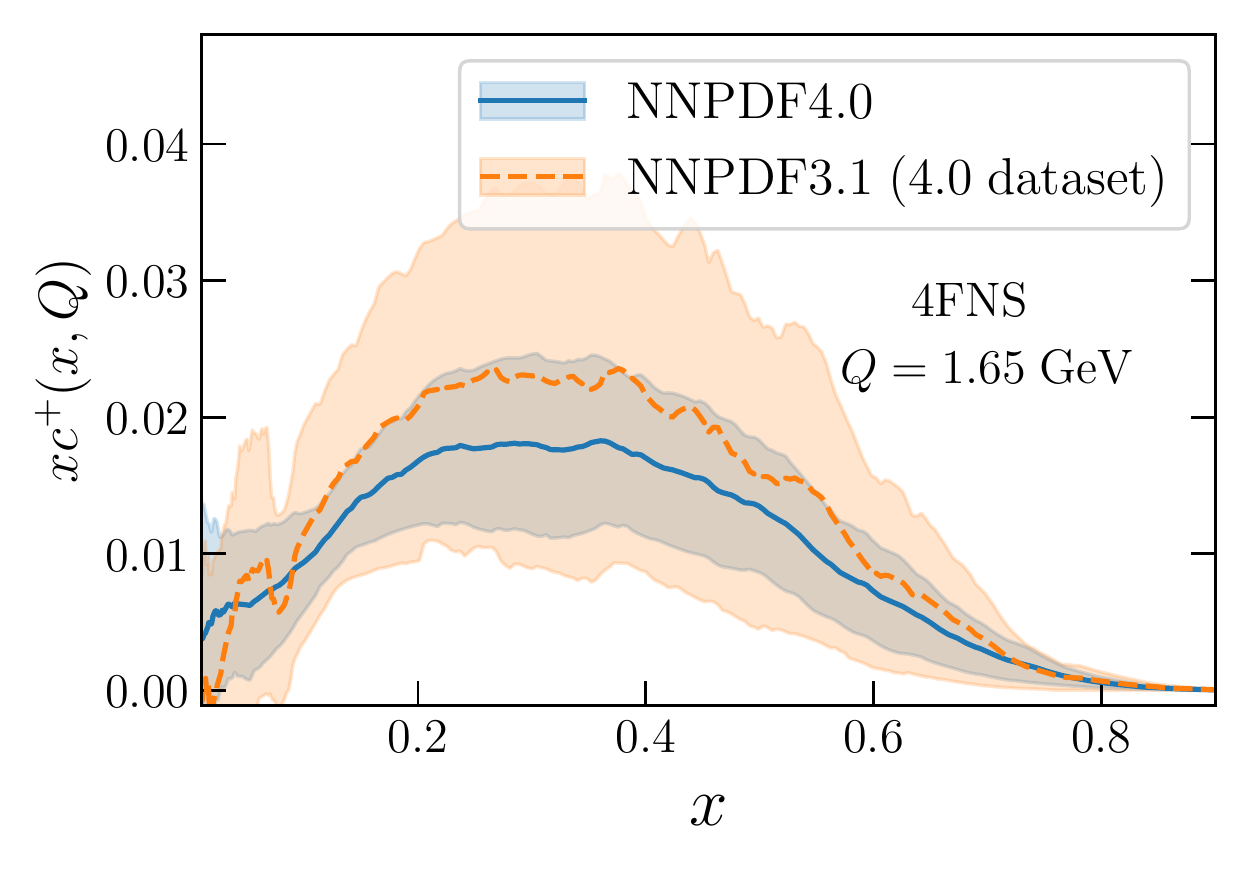}
    \caption{\small Same as Fig.~\ref{fig:charm_basisdep}, comparing
      the baseline determination of the 4FNS charm PDF, based
      on NNPDF4.0, with that obtained
      from the same dataset using the NNPDF3.1 fitting methodology.
  \label{fig:pdfplot-abscharm-40-vs-31meth} }
\end{center}
\end{figure}

  \paragraph{Comparison with NNPDF3.1.}
  Fig.~\ref{fig:pdfplot-abscharm-40-vs-31meth} compares
  the baseline determination of the 4FNS charm PDF based
  on NNPDF4.0 with that obtained
  from the same input dataset but using instead
  the NNPDF3.1 fitting methodology and related settings such those related to positivity
  and integrability.
  Results are fully consistent between the two methodologies, with our default
  determination exhibiting reduced uncertainties due to
  the various improvements implemented
  in the NNPDF4.0 analysis framework.
  
\clearpage
\section{Stability of the 3FNS charm calculation}
\label{app:charm_stability_3fns}

We now repeat the stability and uncertainty
study of the previous section, but for our final
result, namely the intrinsic charm PDF. The main difference to be kept
in mind is that the uncertainty now also includes the dominant MHOU,
due to the matching condition required in order to determine the 3FNS
PDF from the 4FNS result. In order to get a complete picture, we now
add a further set of dataset variations.

\begin{figure}[h]
  \begin{center}
    \includegraphics[width=0.49\linewidth]{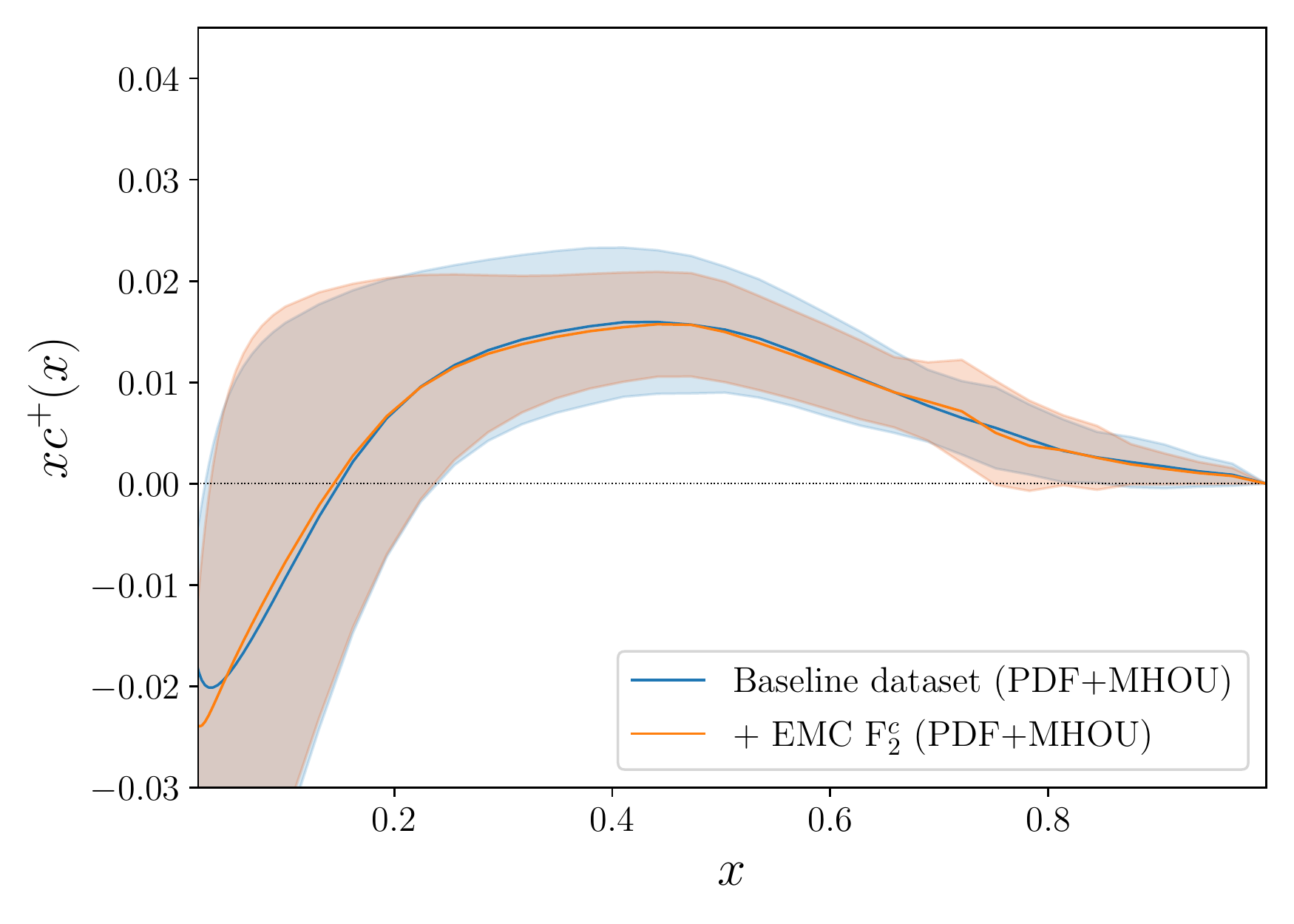}
    \includegraphics[width=0.49\linewidth]{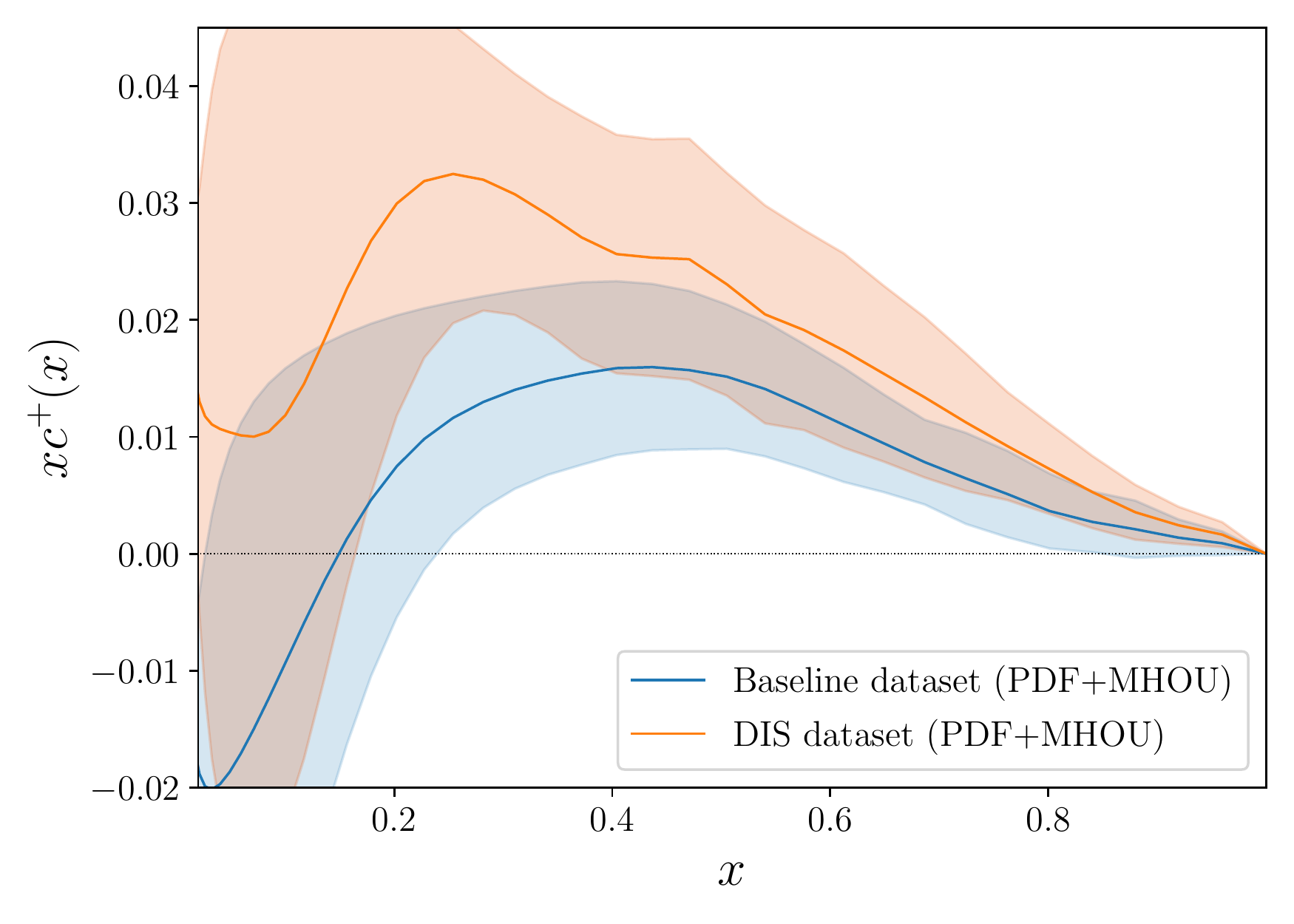}
    \includegraphics[width=0.49\linewidth]{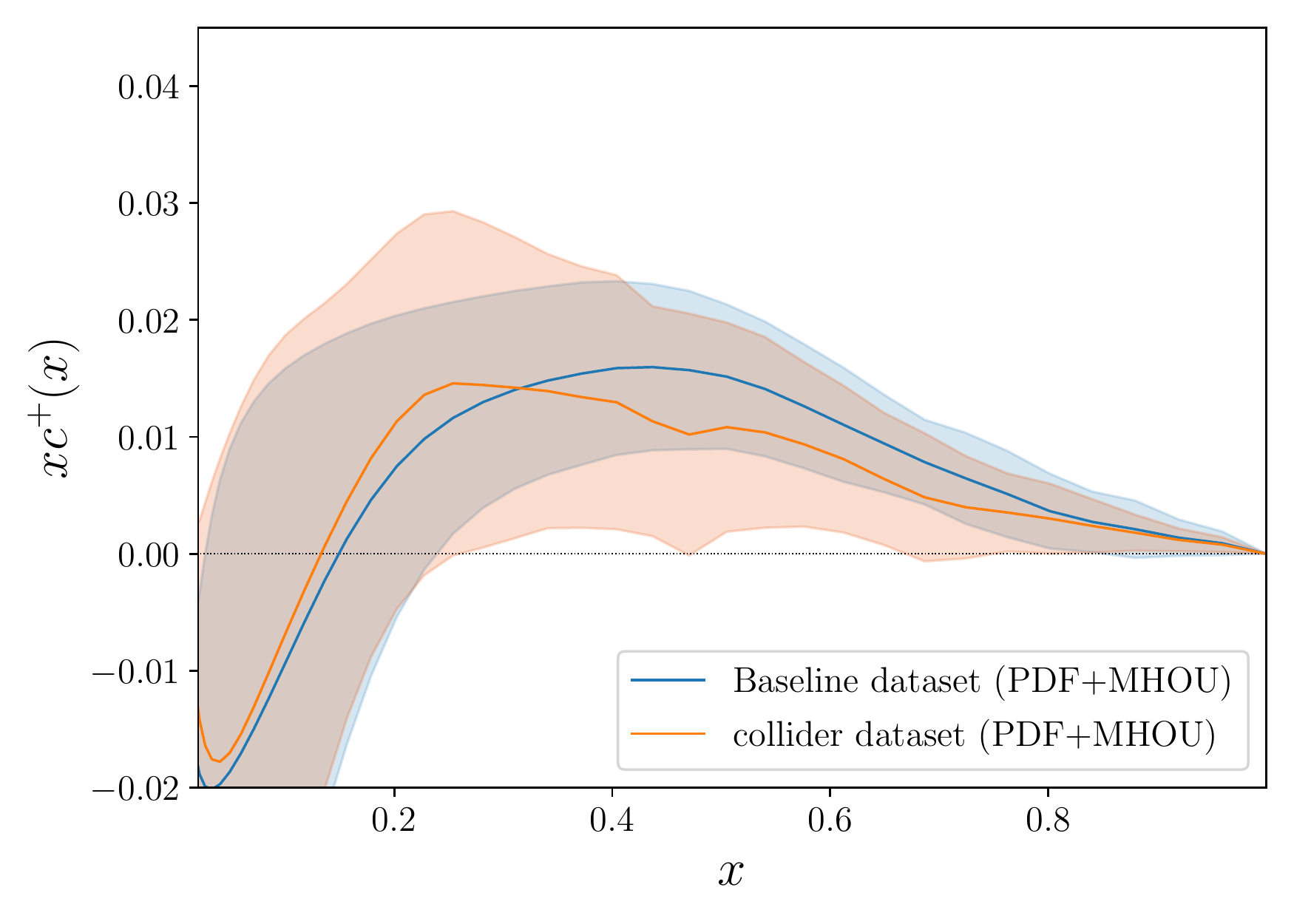}
    \includegraphics[width=0.49\linewidth]{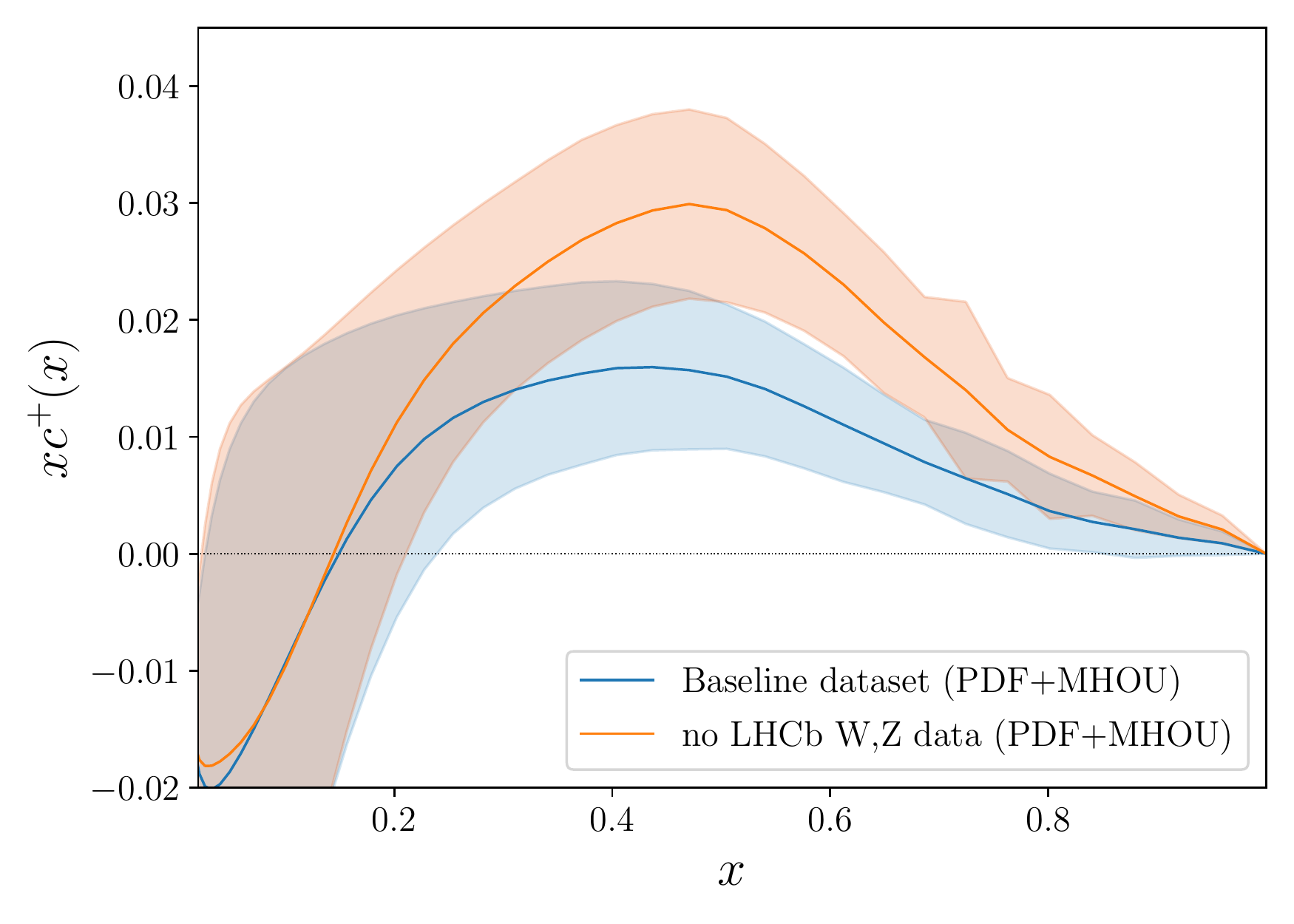}
    \includegraphics[width=0.49\linewidth]{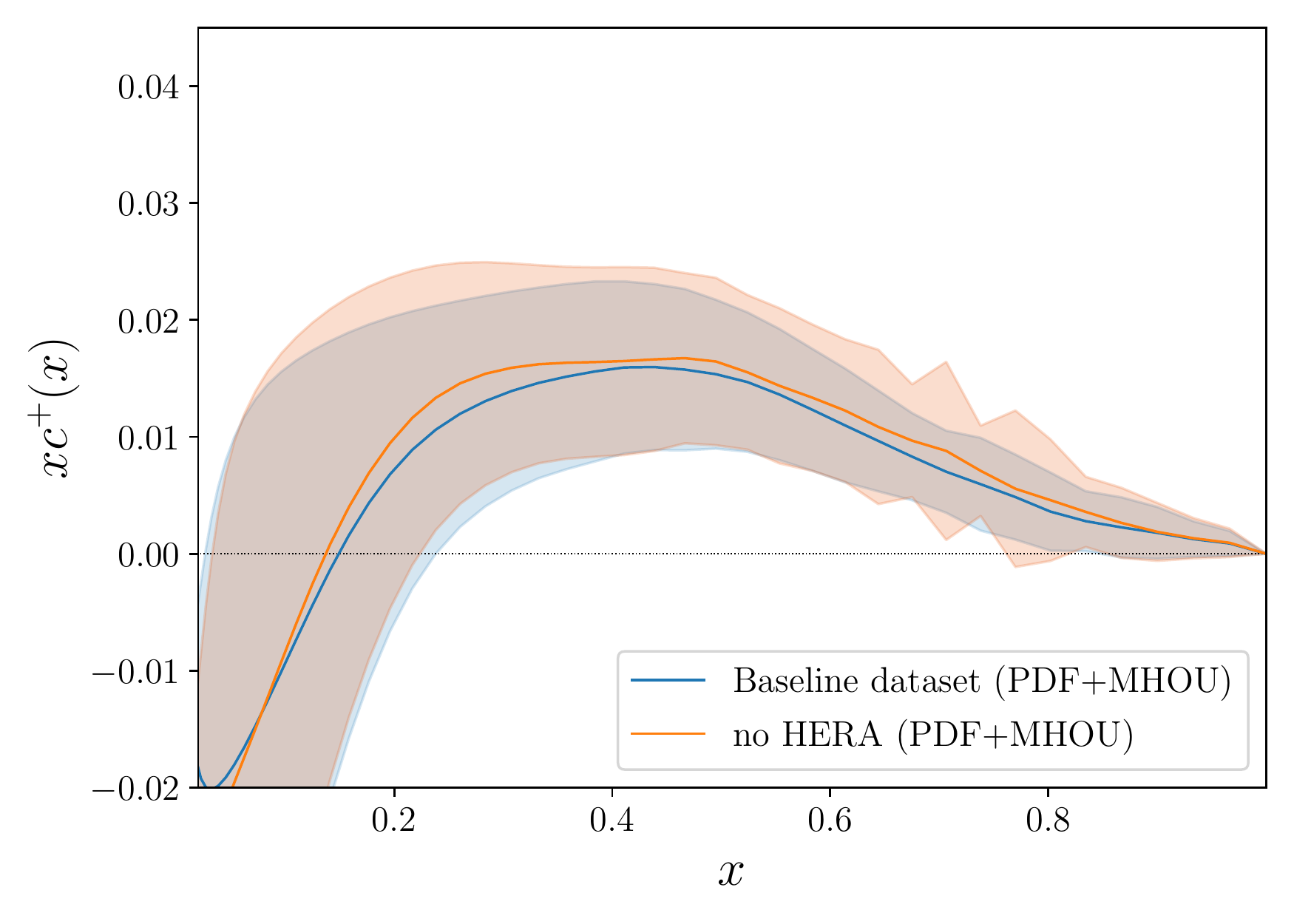}
    \includegraphics[width=0.49\linewidth]{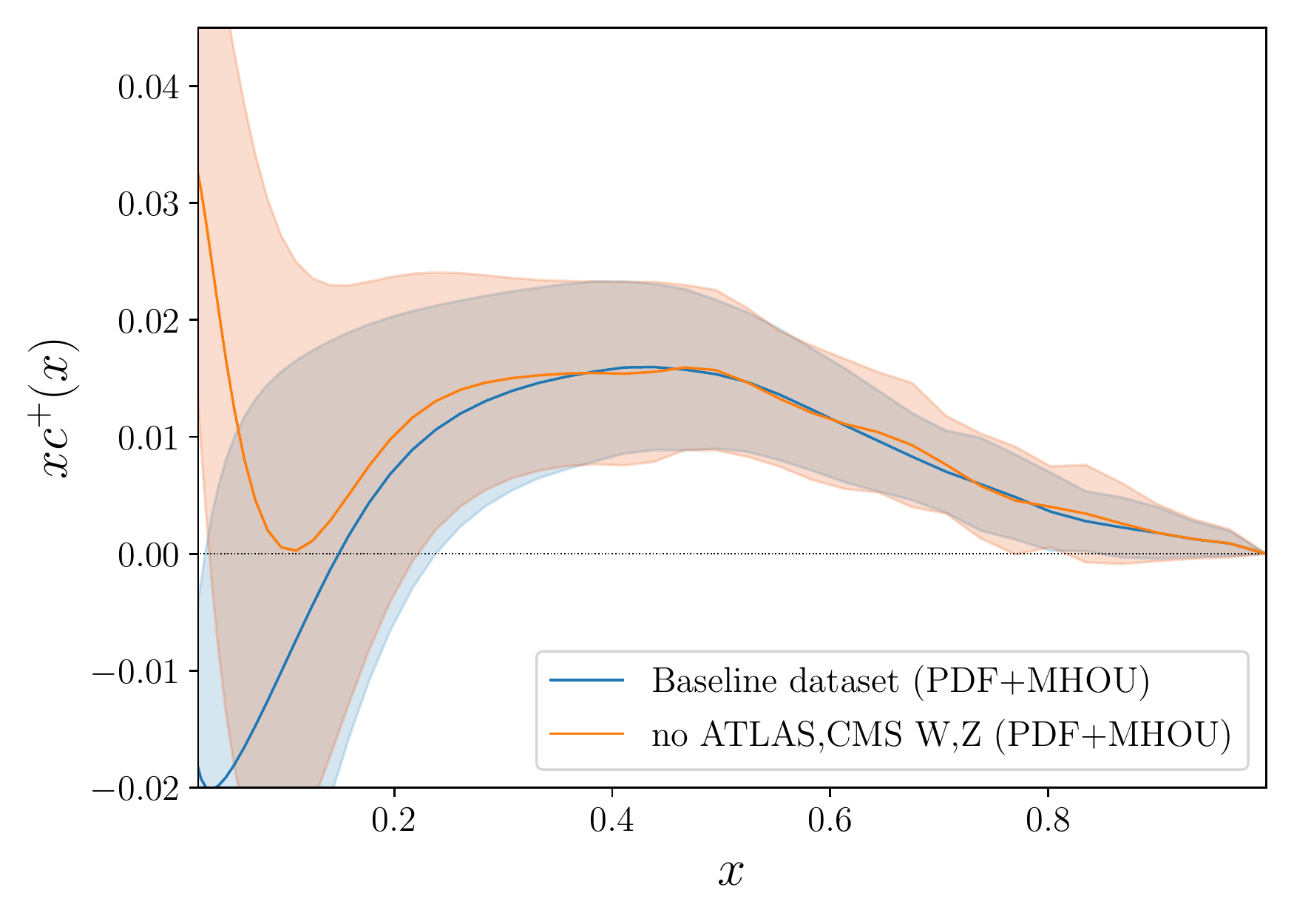}
    \includegraphics[width=0.49\linewidth]{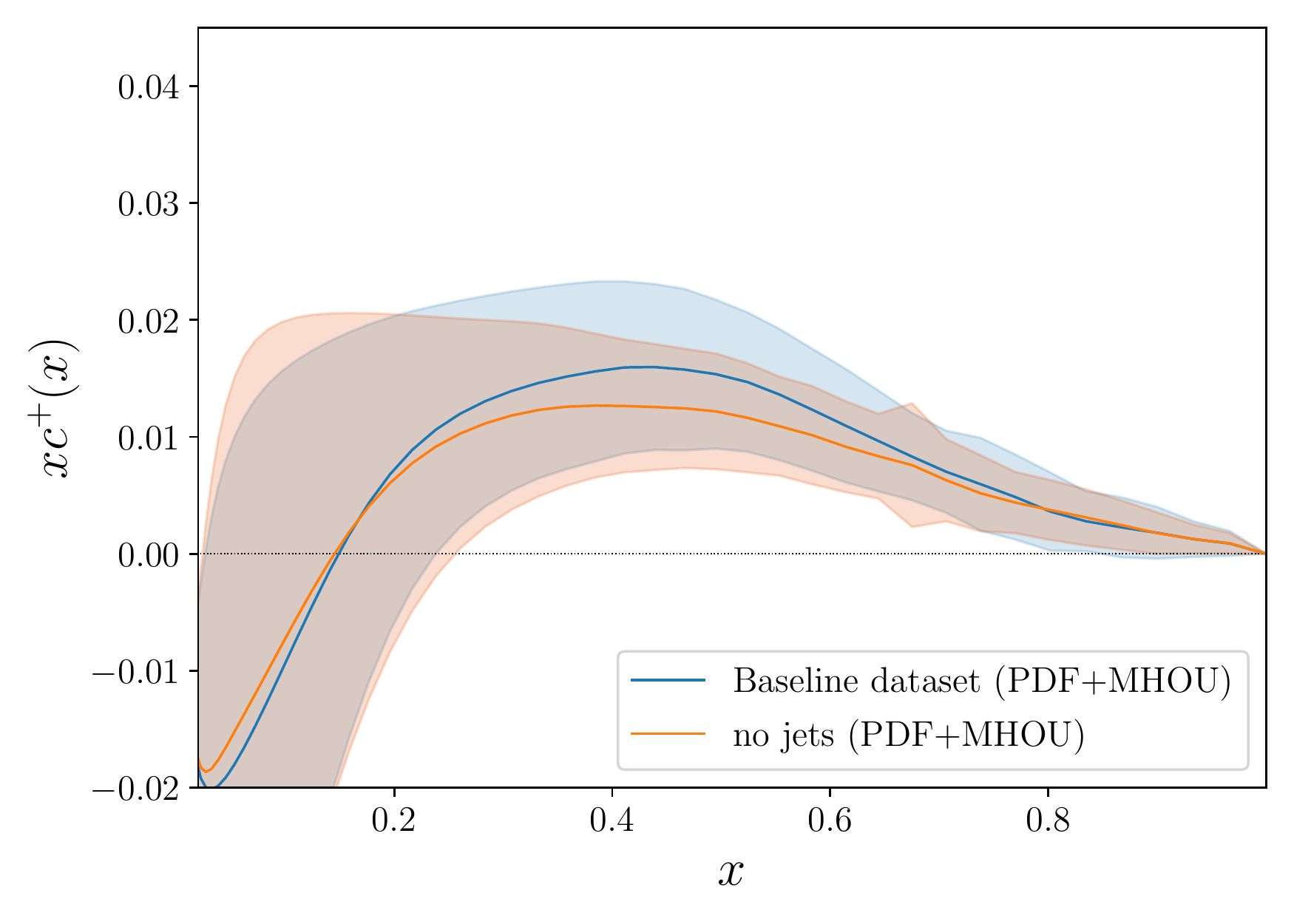}
    \includegraphics[width=0.49\linewidth]{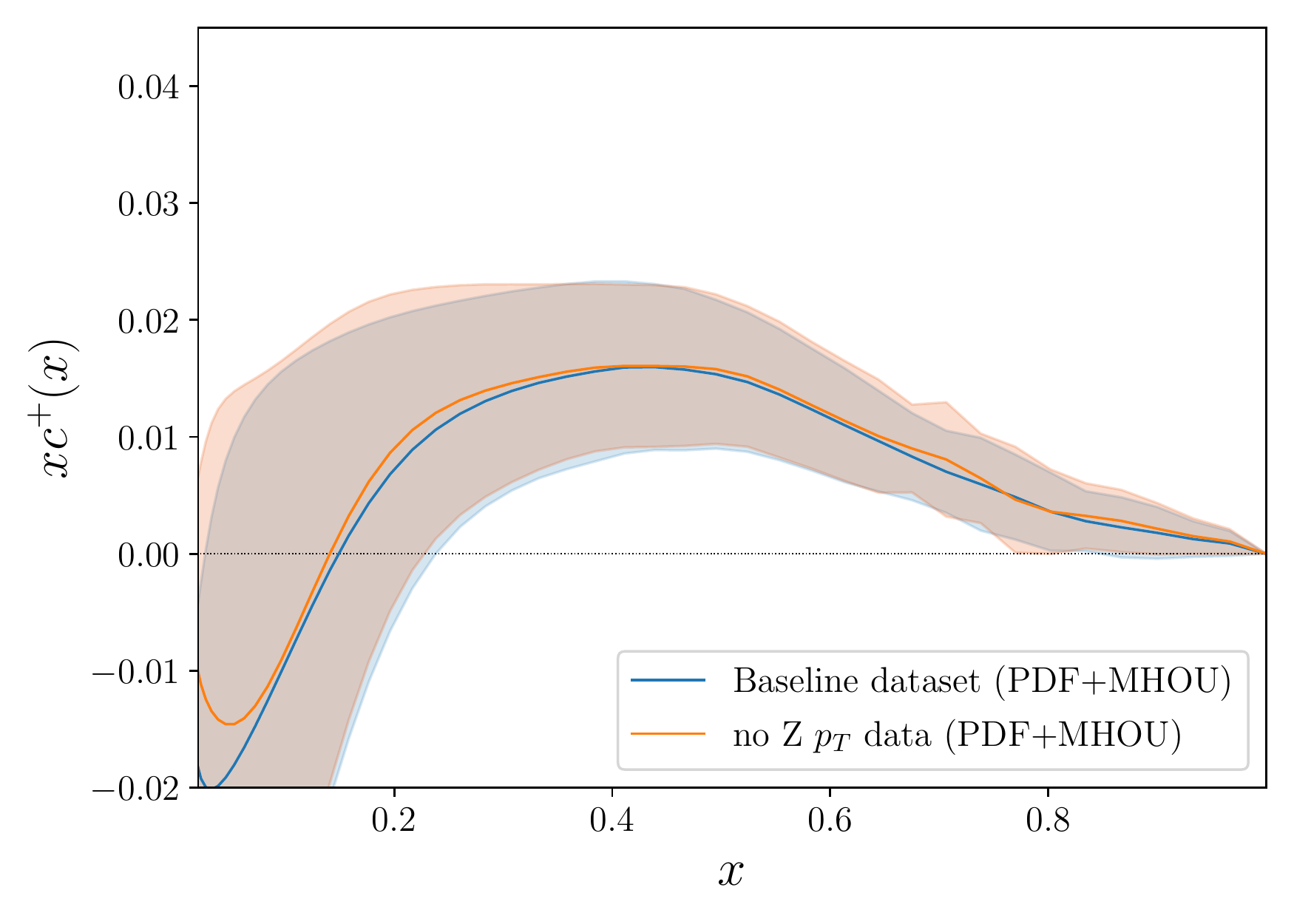}
    \caption{\small Same as Fig.~\ref{fig:charm_dataset_dep}
      for the intrinsic charm (3FNS) PDF (top four plots), now also including
      four additional dataset variations:  no ATLAS and CMS $W, Z$
      production data   (third row left),
      no jet data (third row right), no $Z$ $p_T$
      measurements (bottom row left), no HERA
      DIS data (bottom row right).
The error band indicates the PDF uncertainties combined in quadrature with the MHOUs.
\label{fig:charm_dataset_dep_nf3} }
\end{center}
\end{figure}

\paragraph{Dependence on the input dataset.}
Fig.~\ref{fig:charm_dataset_dep_nf3} displays the dataset variations shown in
Fig.~\ref{fig:charm_dataset_dep}, now for the intrinsic (3FNS) charm
PDF, but with the total uncertainty now being the sum in quadrature of
the PDF and MHO uncertainties, with the latter determined as the difference between
results obtained using NNLO and N$^3$LO matching.
Additionally, we also performed a few extra  dataset
variations: a fit without any $W, Z$ production data from ATLAS and CMS,
a fit without jet data, a fit without $Z$ $p_T$ measurements, and a fit without
HERA structure function data.
Note that the collider-only dataset includes both HERA electron-proton
collider data and Tevatron and LHC hadron collider data, but not
fixed-target deep-inelastic scattering and Drell-Yan production data.

Results are qualitatively very similar to those seen in the 4FNS, a
consequence of the fact that we are focusing on the large-$x$ region where the
effect of the matching is moderate, though now the presence of a
valence-like peak in all determinations is even clearer, specifically
for the DIS-only fit where it was less pronounced in the 4FNS.
 Note however that the DIS-only determination
  exhibits larger uncertainties
  (up to factor 2) and point-by-point fluctuations,
  and is dominated by relatively old fixed-target measurements.
Comparison of all the dataset variations shows that, in terms of their
impact on intrinsic charm,
hadron collider data are generally more important
that deep-inelastic data, that among the former the
LHCb inclusive $W,Z$ data are playing a dominant role,
and that jet observables also play a non-negligible role.

It should be stressed that the agreement between results found using
DIS data and hadron collider data is highly nontrivial, since in the region
relevant for intrinsic charm these determinations are based on disjoint datasets
and are  affected by
very different theoretical and experimental uncertainties:
in particular, potential higher-twist
effects in the DIS observables are highly suppressed for collider observables.
 this respect, a DIS-only determination of intrinsic charm
  is potentially affected by sources of theory uncertainties, such as higher twists,
which are not accounted for in global PDF determinations.

We conclude that
the characteristic valence-like peak structure at large-$x$
predicted by non-perturbative intrinsic charm models (Fig.~\ref{fig:charm_content_3fns}
in the main manuscript)
is always present even under very significant changes of the dataset.

\begin{figure}[t!]
  \begin{center}
    \includegraphics[width=0.54\linewidth]{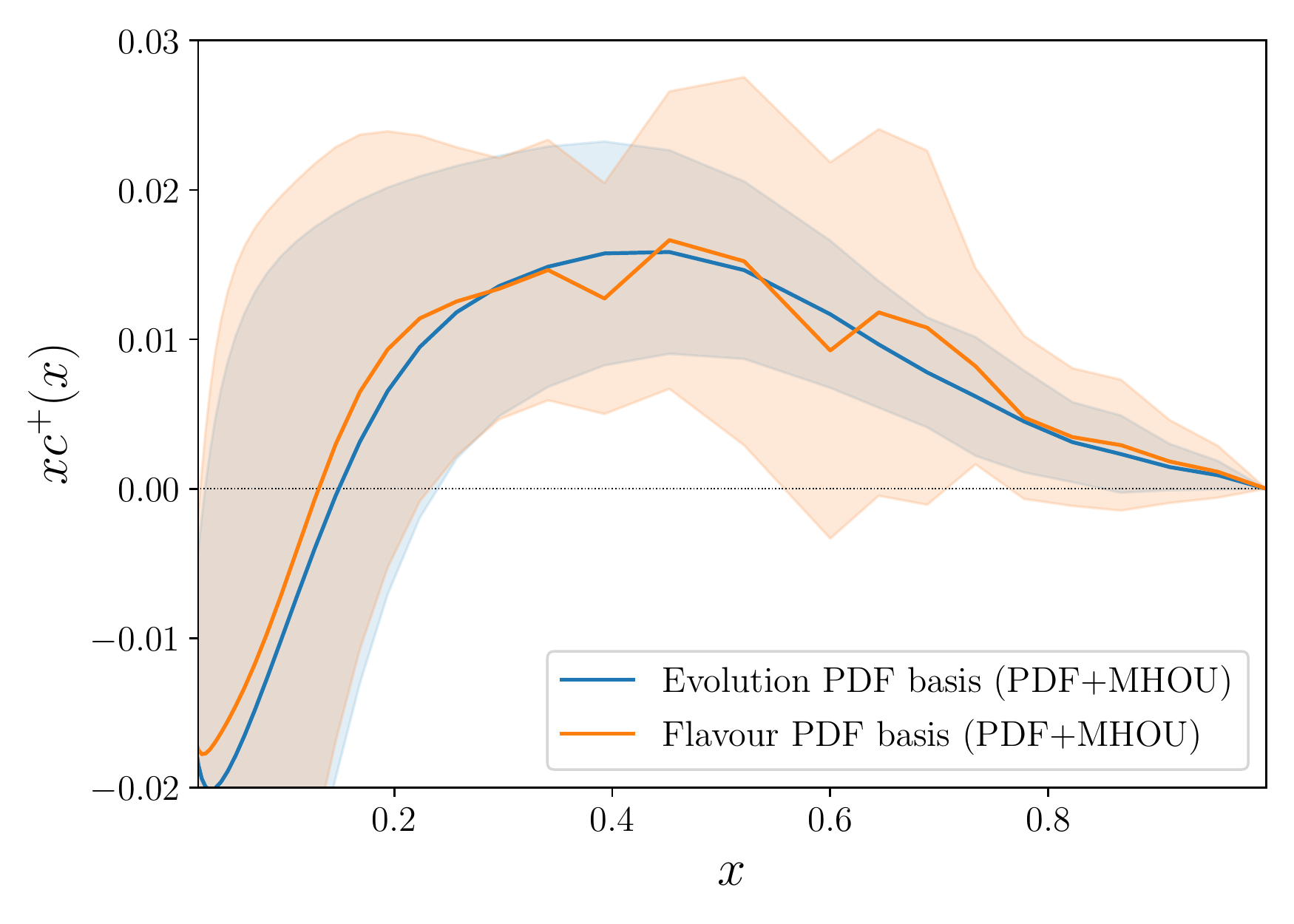}
    \caption{\small Same as Fig.~\ref{fig:charm_basisdep}
    for the intrinsic (3FNS) charm.
  \label{fig:charm_basisdep_3FNS} }
\end{center}
\end{figure}

\paragraph{Dependence on the parametrization basis.}
Fig.~\ref{fig:charm_basisdep_3FNS} displays
the comparison between the intrinsic charm
PDF determined with the default evolution basis choice, and the flavor
basis. Complete consistency of central values is found, with somewhat
larger uncertainties in the case of the flavor basis, due to the more 
challenging fitting environment in this basis (see the discussion in~\cite{Ball:2021leu}).

\begin{figure}[h]
  \begin{center}
\includegraphics[width=0.49\linewidth]{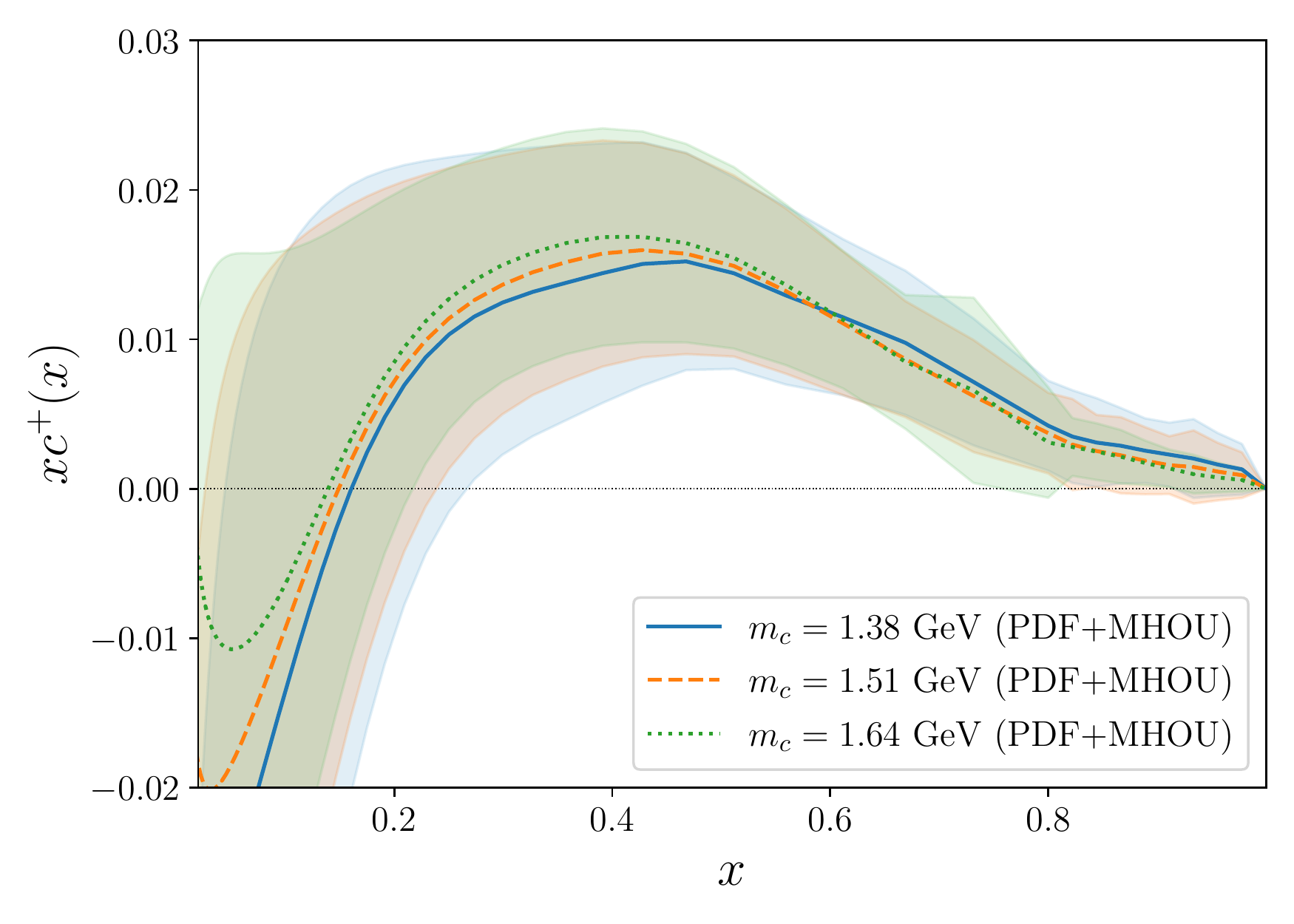}
    \includegraphics[width=0.49\linewidth]{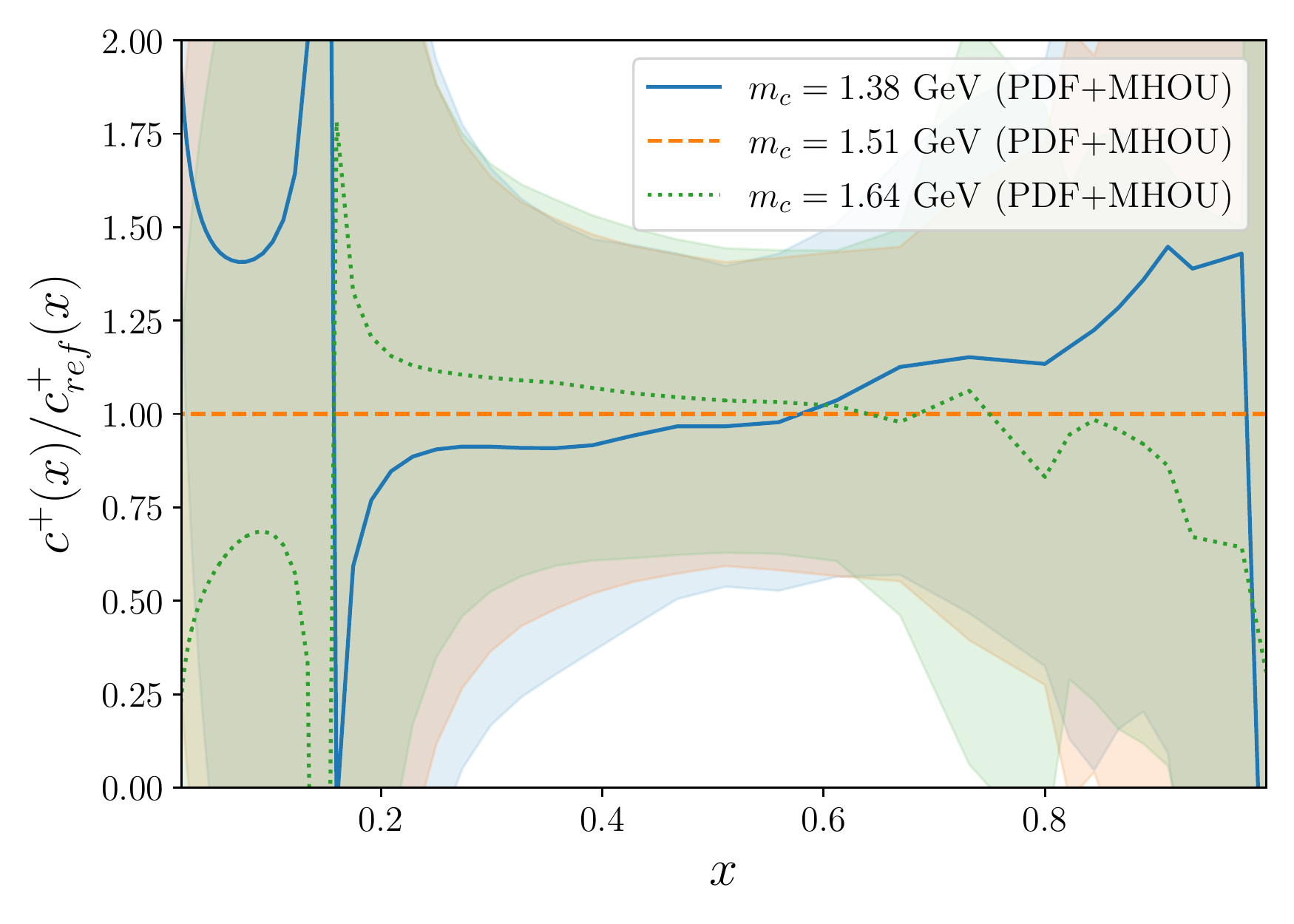}
\caption{\small      
 Same as Fig.~\ref{fig:charm_fitted_mcdep}, now for
      the intrinsic (3FNS) charm PDF. Note that the intrinsic charm
      PDF is scale independent.
  \label{fig:mass_variations_Quad_MHOU} }
\end{center}
\end{figure}

\paragraph{Dependence on the charm mass value.}
The study of the charm mass dependence is particularly interesting,
because the intrinsic component should be independent of it, hence the
residual dependence seen in Fig.~\ref{fig:charm_fitted_mcdep} in the
4FNS, due to the mass dependence of the perturbative component that
could not be reabsorbed in the fitting, should no longer be present. 
Results are shown in
Fig.~\ref{fig:mass_variations_Quad_MHOU}, and it is apparent that
indeed the dependence on the charm mass has all but disappeared.

\clearpage
\section{The charm momentum fraction}
\label{app:charm_mom_frac}

The fraction of the proton momentum carried by charm quarks
is given by
\be
\label{eq:charm_momentum_fraction}
\lc c\rc = \int_0^1dx\, x c^+(x,Q^2) \, .
\ee
Model predictions, as mentioned, are typically provided up to an
overall normalization, which in turn determines the charm momentum fraction.
Consequently, the momentum fraction is often cited as a characteristic
parameter of intrinsic charm.
It should however be borne in mind that,
even in the absence of intrinsic charm, this charm momentum fraction is nonzero due
to the perturbative contribution.

In Table~\ref{tab:momfrac_lowQ} we indicate
the values of the charm momentum fraction
 in the 3FNS for our default charm
  determination and in the 4FNS  (at $Q=1.65$ GeV) both for the
  default result and for perturbative charm PDF (see SI Sect.~\ref{app:consistency}).
We provide results for  three different values of the charm mass $m_c$ and
indicate separately the PDF and the MHO uncertainties.
The 3FNS result is scale-independent, it corresponds to the
momentum fraction carried by intrinsic charm, and it vanishes identically
by assumption in the perturbative charm case.
The 4FNS result corresponds to
the scale-dependent momentum fraction that combines the intrinsic and
perturbative contribution, while of course it contains only the
perturbative contribution in the case of perturbative charm.
As
discussed in SI Sect.~\ref{app:consistency}, the large uncertainty
associated to higher order corrections to the matching conditions
affects the 3FNS result (intrinsic charm) in the default case, in
which the charm PDF is determined from data in the 4FNS scheme, while
it affects the 4FNS result for perturbative charm, that is determined
assuming the vanishing of the 3FNS result.

 For our default determination, the charm
momentum fraction in the 4FNS at low scale
differs from zero at the $3\sigma$
level.
However, it is not possible to tell whether this is of
perturbative or intrinsic origin, because, due to  the large MHOU in
the matching condition, the intrinsic (3FNS) charm momentum fraction
is compatible with zero. This large uncertainty is entirely due to the
small $x\lsim 0.2$ region, see see
Fig.~\ref{fig:charm_content_3fns}~(right).
Accordingly, for perturbative charm the
low-scale 4FNS
momentum fraction is compatible with zero.
Consistently with the results of SI Sect.~\ref{app:charm_stability_4fns},
the 4FNS result is essentially independent of the value of the charm
mass, but it becomes strongly dependent on it if one assumes
perturbative charm.

\begin{table}[t]
  \footnotesize
  \centering
    \renewcommand{\arraystretch}{1.30}
\begin{tabularx}{\textwidth}{C{2.0cm}C{2.3cm}C{2.2cm}C{2.2cm}C{5.6cm}}
  \toprule
  Scheme  & $Q$ & Charm PDF & $m_c$  &  $\lc c\rc~\lp\%\rp$ \\
  \midrule
  \midrule
 3FNS  & --  &default  &  1.51 GeV  &   $ 0.62\pm0.28_{\rm pdf}\pm 0.54_{\rm mhou} $ \\
 3FNS  & --  &default  &  1.38 GeV  &   $ 0.47\pm0.27_{\rm pdf}\pm 0.62_{\rm mhou} $ \\
 3FNS  & --  &default  &  1.64 GeV  &    $ 0.77\pm0.28_{\rm
  pdf}\pm 0.48_{\rm mhou} $ \\
  \midrule

 4FNS  & 1.65 GeV  & default  &  1.51 GeV  &   $0.87 \pm 0.23_{\rm pdf}$  \\
 4FNS  & 1.65 GeV  & default &  1.38 GeV  &   $0.94 \pm 0.22_{\rm pdf}$  \\
 4FNS  & 1.65 GeV  & default   &  1.64 GeV  &  $0.84 \pm 0.24_{\rm pdf}$  \\
  \midrule
 \midrule
 4FNS  & 1.65 GeV   & perturbative  &  1.51 GeV  &   $0.346\pm 0.005_{\rm pdf}\pm 0.44_{\rm mhou}$ \\
 4FNS  & 1.65 GeV   & perturbative  &  1.38 GeV  &    $0.536\pm 0.006_{\rm pdf}\pm 0.49_{\rm mhou}$ \\
 4FNS  & 1.65 GeV   & perturbative  &  1.64 GeV  &    $0.172\pm 0.003_{\rm pdf}\pm 0.41_{\rm mhou}$ \\
\bottomrule
\end{tabularx}
\vspace{0.3cm}
\caption{\label{tab:momfrac_lowQ}
  The charm momentum fraction, Eq.~(\ref{eq:charm_momentum_fraction}).
  We show  results both in the 3FNS and the 4FNS (at $Q=1.65$ GeV)
  for our default charm, and also in the 4FNS for perturbative charm.
We provide results for  three different values of the charm mass $m_c$ and
indicate separately the PDF and the MHO uncertainties.
}
\end{table}

\begin{figure}[h]
  \begin{center}
     \includegraphics[width=0.60\linewidth]{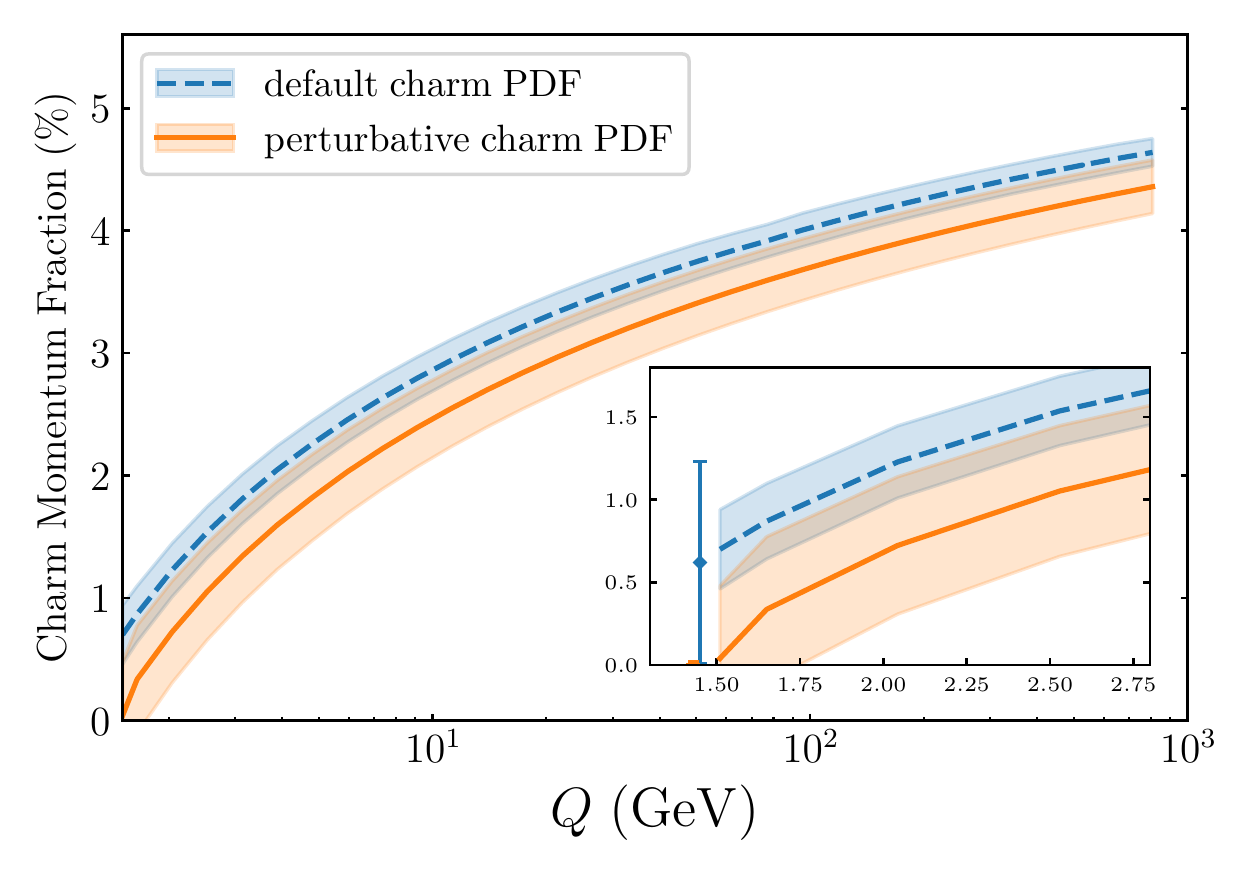}
    \caption{\small 
      The 4FNS charm momentum fraction in NNPDF4.0 as a function of scale $Q$,
      both for the default and perturbative charm cases,
      for a charm mass value of $m_c=1.51$ GeV.
     The inset zooms on the low-$Q$ region and includes the 3FNS
     (default) result
     from Table~\ref{tab:momfrac_lowQ}. 
     Note that the uncertainty includes the MHOU for the 3FNS default
     and 4FNS perturbative charm cases, while it is the PDF
     uncertainty for the 4FNS default charm case.
  \label{fig:comparison_IC_models} }
\end{center}
\end{figure}

\begin{figure}[t]
  \begin{center}
    \includegraphics[width=0.49\linewidth]{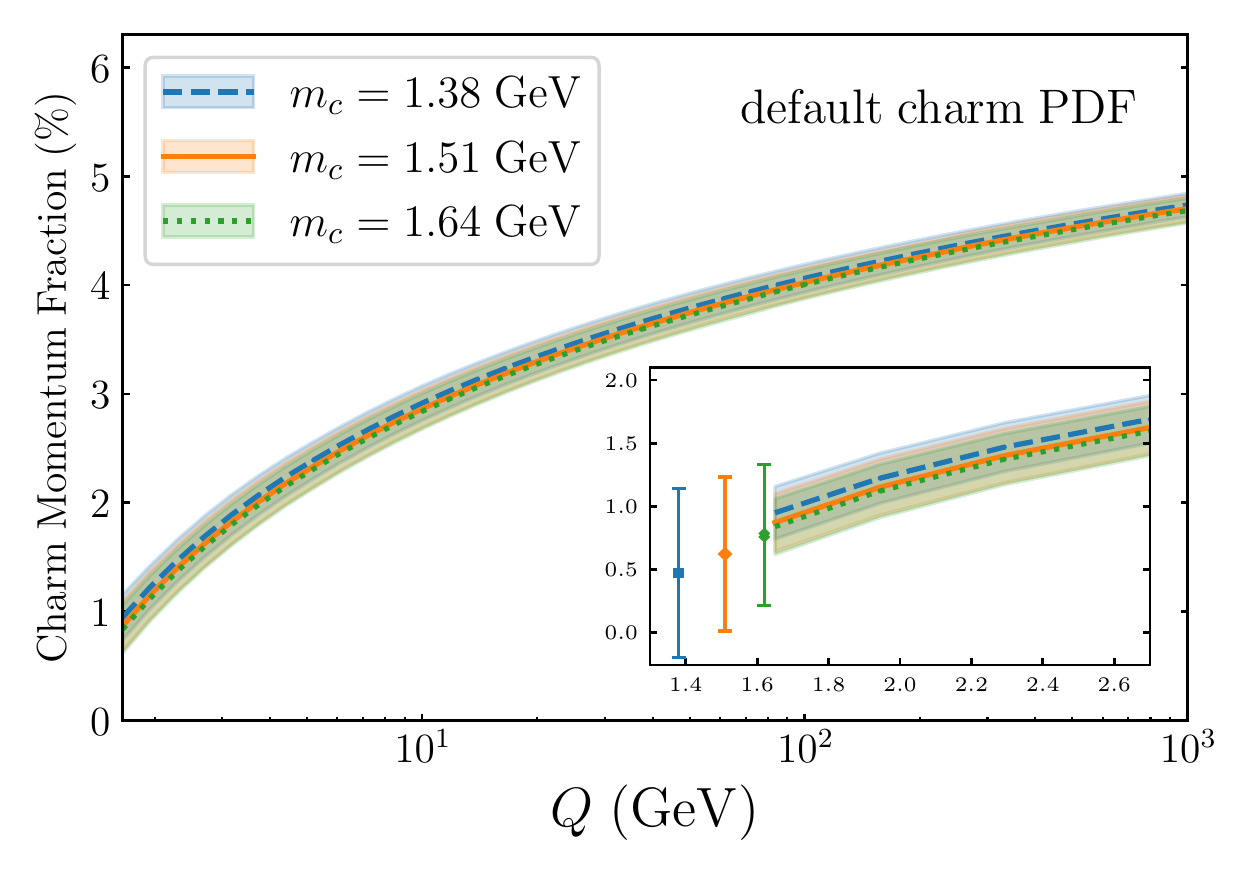}
    \includegraphics[width=0.49\linewidth]{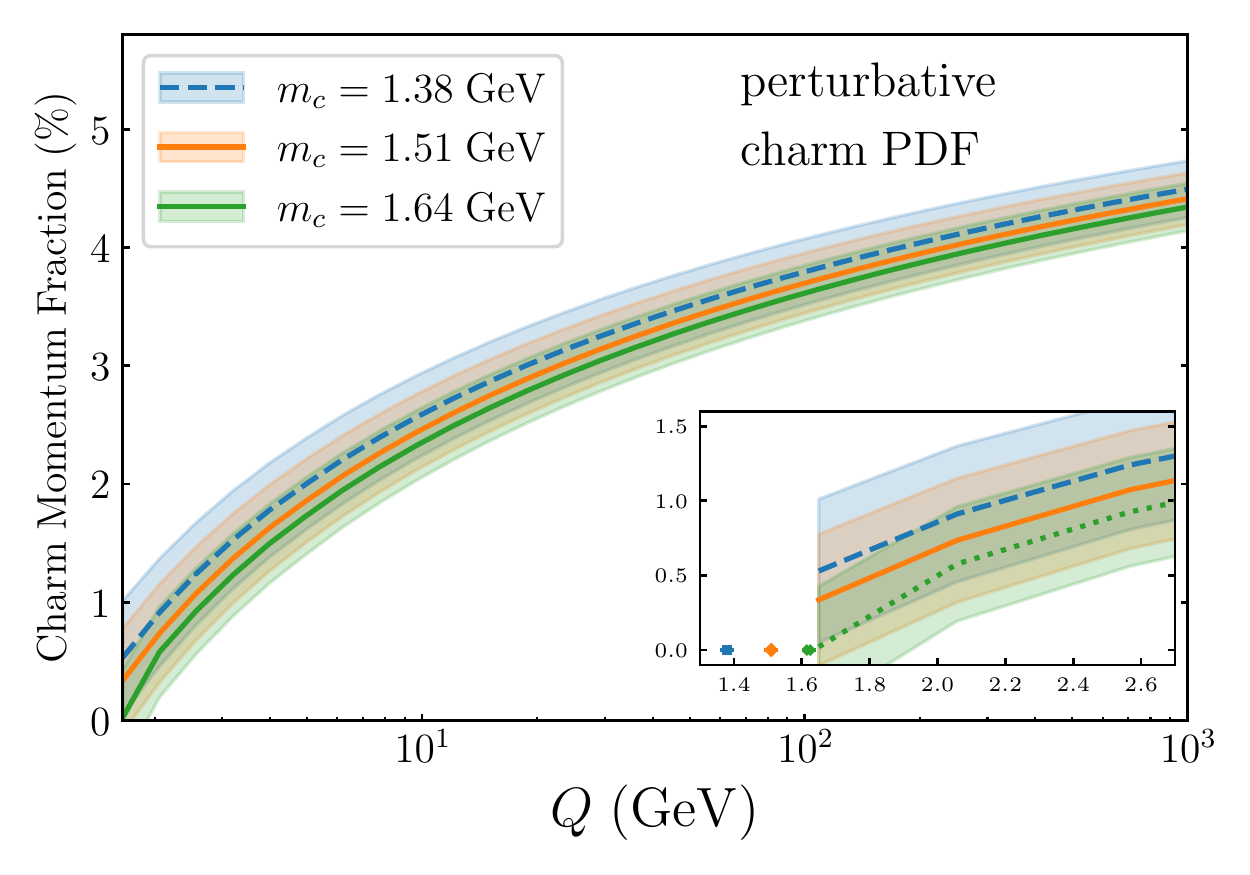}
    \caption{\small
    Same as Fig.~\ref{fig:comparison_IC_models} for different values
    of the charm mass. Note that the 3FNS momentum fraction for
     perturbative charm vanishes identically by assumption.
   \label{fig:charm_momfrac_qdep_mc} }
\end{center}
\end{figure}

The 4FNS charm momentum fraction is plotted as a function of scale
in Fig.~\ref{fig:comparison_IC_models}, both in the default case and
for perturbative charm, with the 3FNS values and the detail of the low-$Q$ 
4FNS results shown in an inset.
The dependence on the value of the charm mass
is shown in Fig.~\ref{fig:charm_momfrac_qdep_mc}.
The large MHOUs on the 3FNS result, and on the 
4FNS result in the case of perturbative charm, are apparent.
The stability of the default result upon variation of  the value of
$m_c$, and the strong dependence of the perturbative charm result on
$m_c$, are  also clear.
Both the large MHOU uncertainty, and the strong dependence on
the value of $m_c$
for perturbative charm are seen to persist up to large scales.

\begin{figure}[h!]
  \begin{center}
    \includegraphics[width=0.65\linewidth]{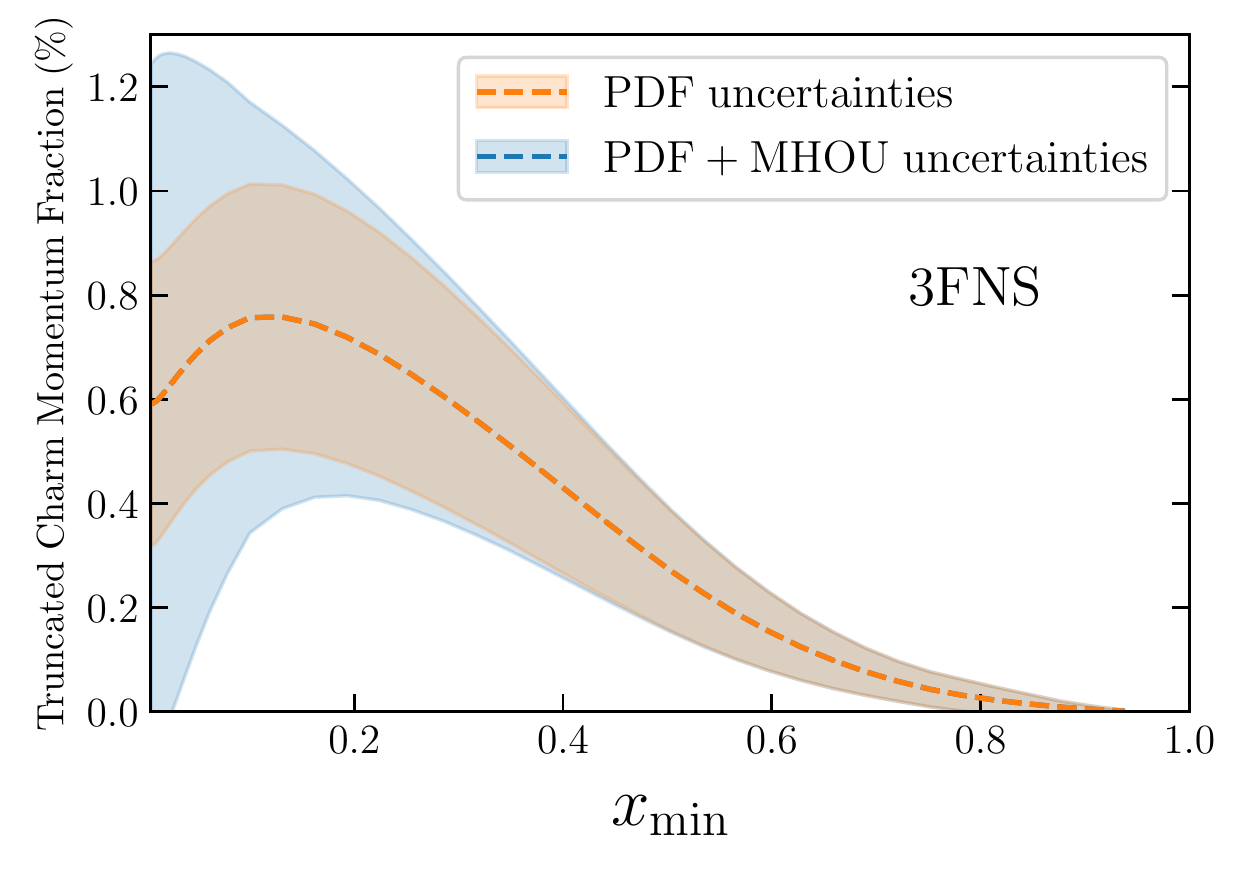}
    \caption{\small The value of the truncated charm momentum integral,
      Eq.~(\ref{eq:charm_momentum_fraction_truncated}), as a function
      of the lower integration limit $x_{\rm min}$
      for our baseline determination of the 3FNS intrinsic charm PDF.
      We display separately the PDF and the total (PDF+MHOU) uncertainties.
  \label{fig:charm_momfrac_xmin_dep} }
\end{center}
\end{figure}

It is interesting to understand in detail the impact of the MHOU on
the momentum fraction carried by intrinsic charm. To this purpose, we
have computed  the truncated momentum integral, i.e. 
  Eq.~(\ref{eq:charm_momentum_fraction}) but only integrated down to
  some  lower
  integration limit $x_{\rm min}$:
  \be
\label{eq:charm_momentum_fraction_truncated}
\lc c\rc_{\rm tr}(x_{\rm min}) \equiv \int_{x_{\rm min}}^1dx\, x c^+(x,Q^2) \, .
\ee
Note than in the 3FNS   $x
c^+(x)$ does not depend on scale, so  this becomes
a scale-independent quantity.
The result for our default intrinsic charm determination is displayed
in Fig.~\ref{fig:charm_momfrac_xmin_dep}, as a function of
of the lower integration limit $x_{\rm min}$.
It is clear that for $x_{\rm min} \gtrsim 0.2$ the truncated momentum
fraction  differs significantly from zero, thereby providing evidence
for intrinsic charm with similar statistical  significance as the
local pull shown in Fig.~\ref{fig:Zc} bottom left.
For $x \lsim 0.2$
this  significance is then washed out
by the large MHOUs.

Hence, while the total momentum fraction has been traditionally adopted
as a measure of intrinsic charm, 
our analysis shows that, once MHOUs are accounted for, the information
provided by the total momentum fraction is limited, at least with
current data and theory.

\clearpage
\section{Comparison with  CT14IC}
\label{app:ct}

The possibility of an intrinsic charm component was recently studied
in~\cite{Hou:2017khm}, by modifying the CT14 PDF set, with the
initial 4FNS charm PDF taken equal to the BHPS
model~\cite{Brodsky:1980pb} form with the normalization fitted as a
free parameter.
A 4FNS  charm PDF with uncertainties at $Q=1.3$~GeV was then
constructed by taking 
the BHPS model with best-fit normalization as central value (called
the `BHPS1 model' in~\cite{Hou:2017khm}); the lower
edge of the uncertainty band was taken to coincide with the standard
CT14 charm PDF  (i.e. the charm PDF determined by perturbative
matching from the 3FNS to the 4FNS); the upper edge of the uncertainty
band was taken as 
the BHPS model but with  normalization fixed to the upper  90\% CL limit (called the
`BHPS2 model' in~\cite{Hou:2017khm}).

\begin{figure}[h]
  \begin{center}
    \includegraphics[width=0.49\linewidth]{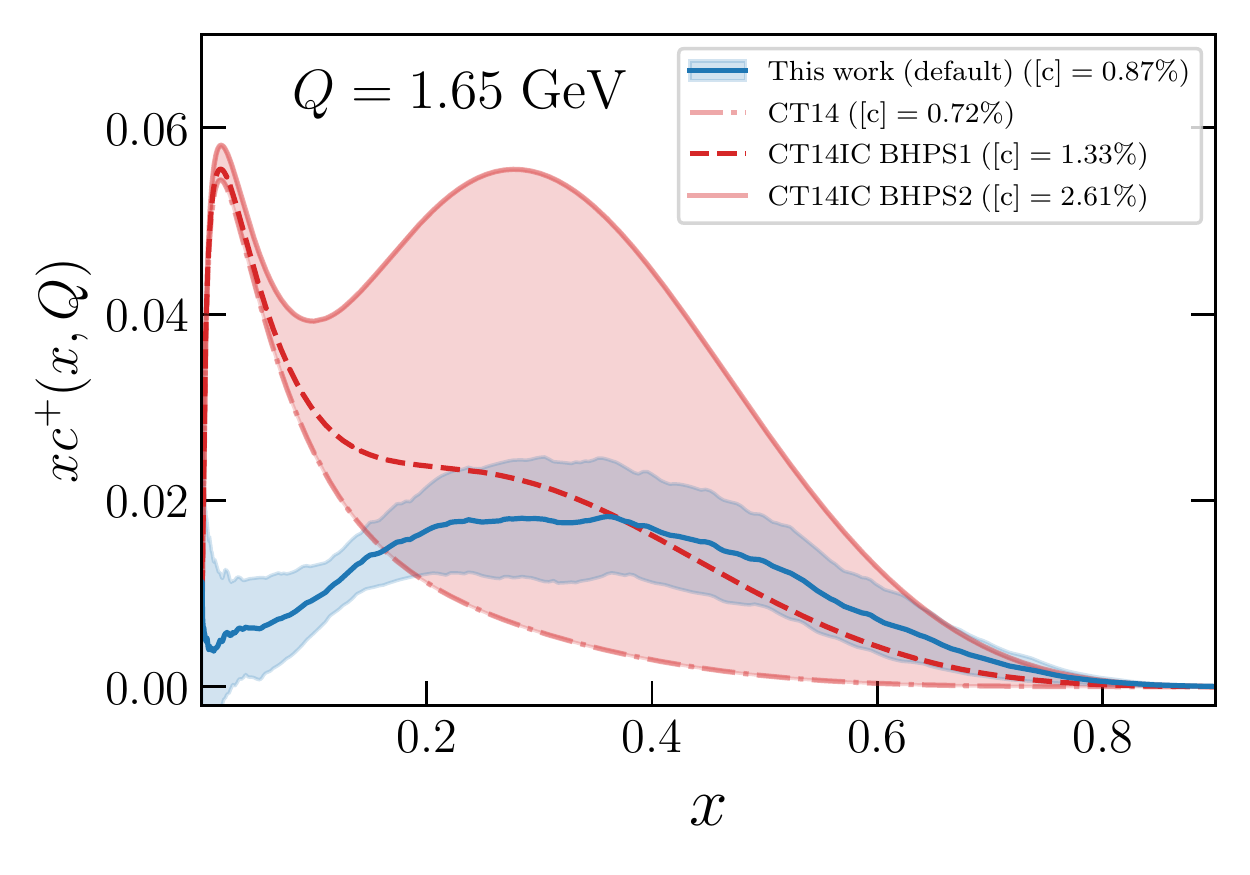}
    \includegraphics[width=0.49\linewidth]{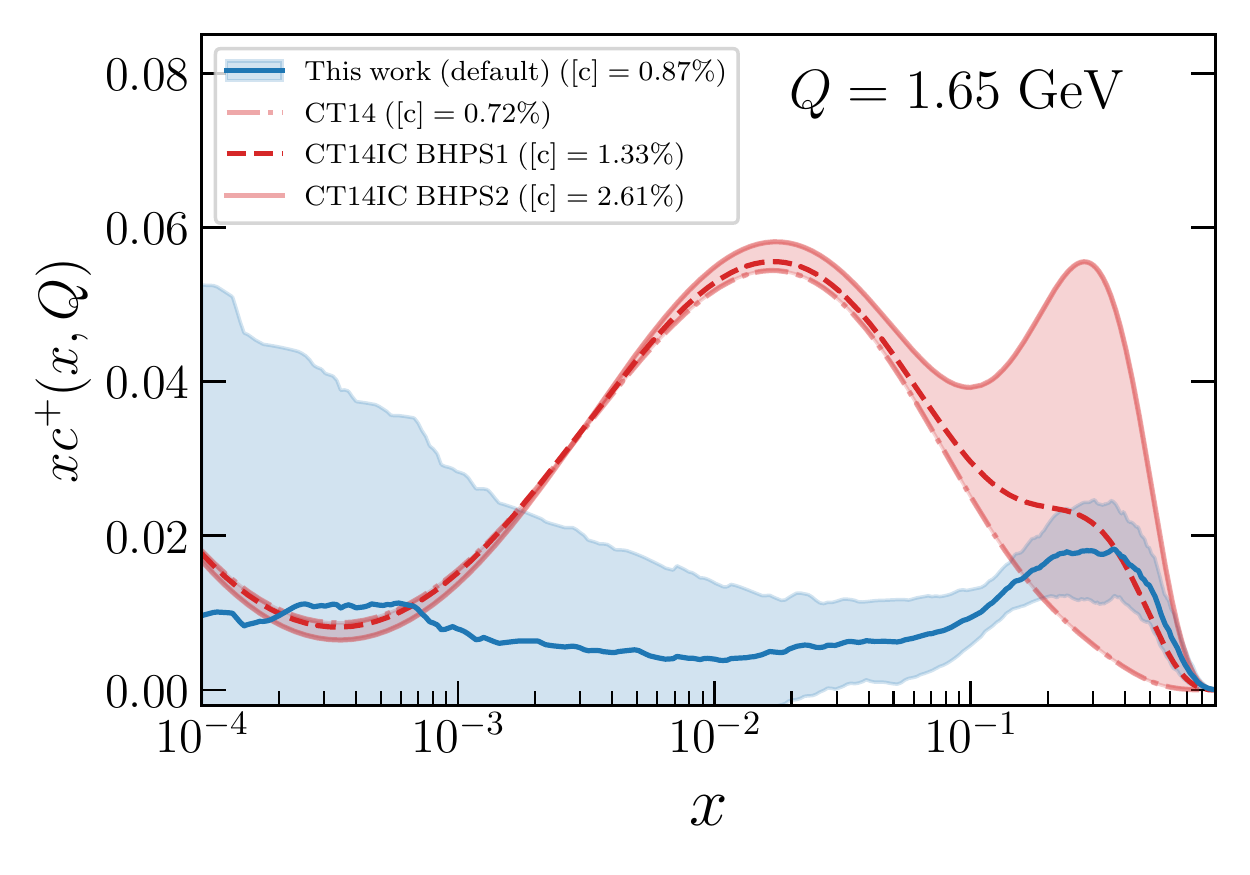}    \includegraphics[width=0.49\linewidth]{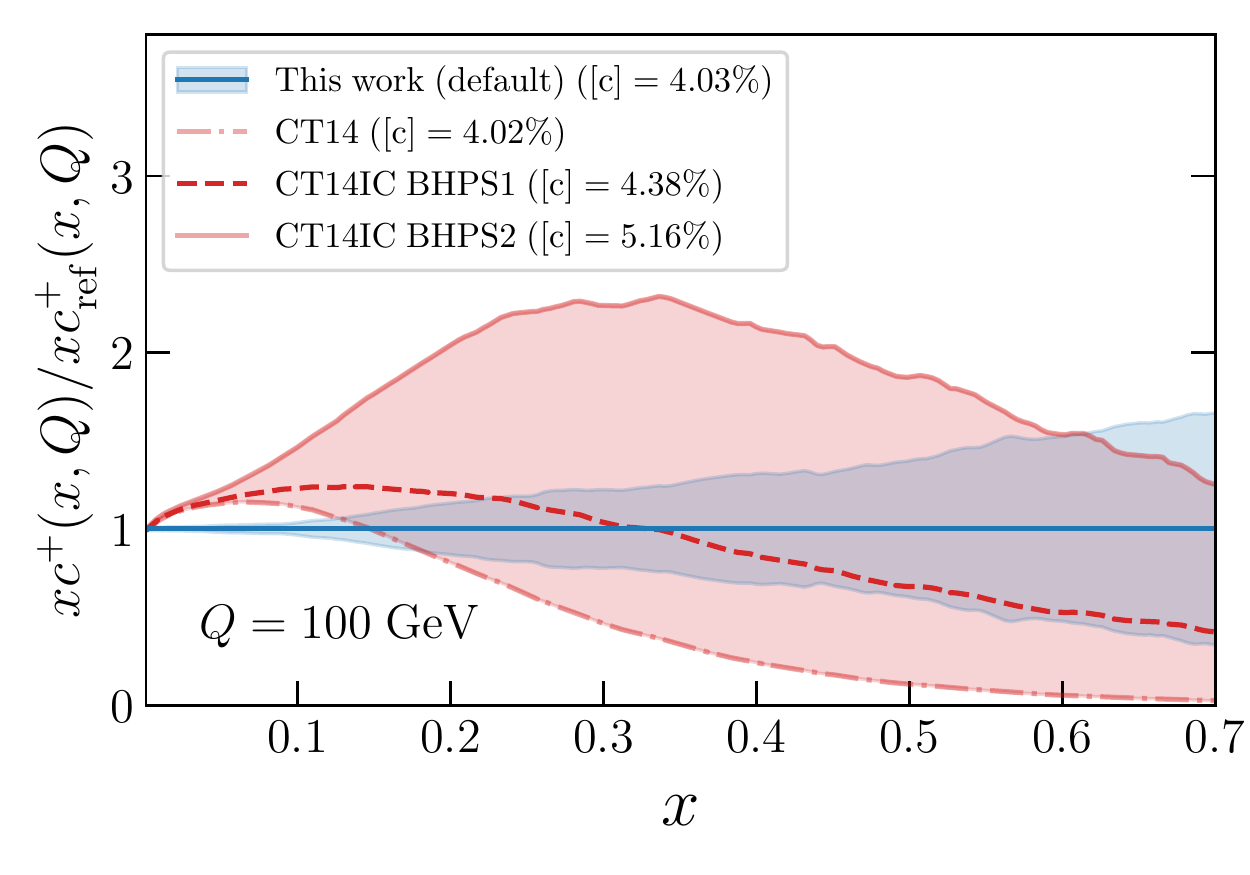}	
    \includegraphics[width=0.49\linewidth]{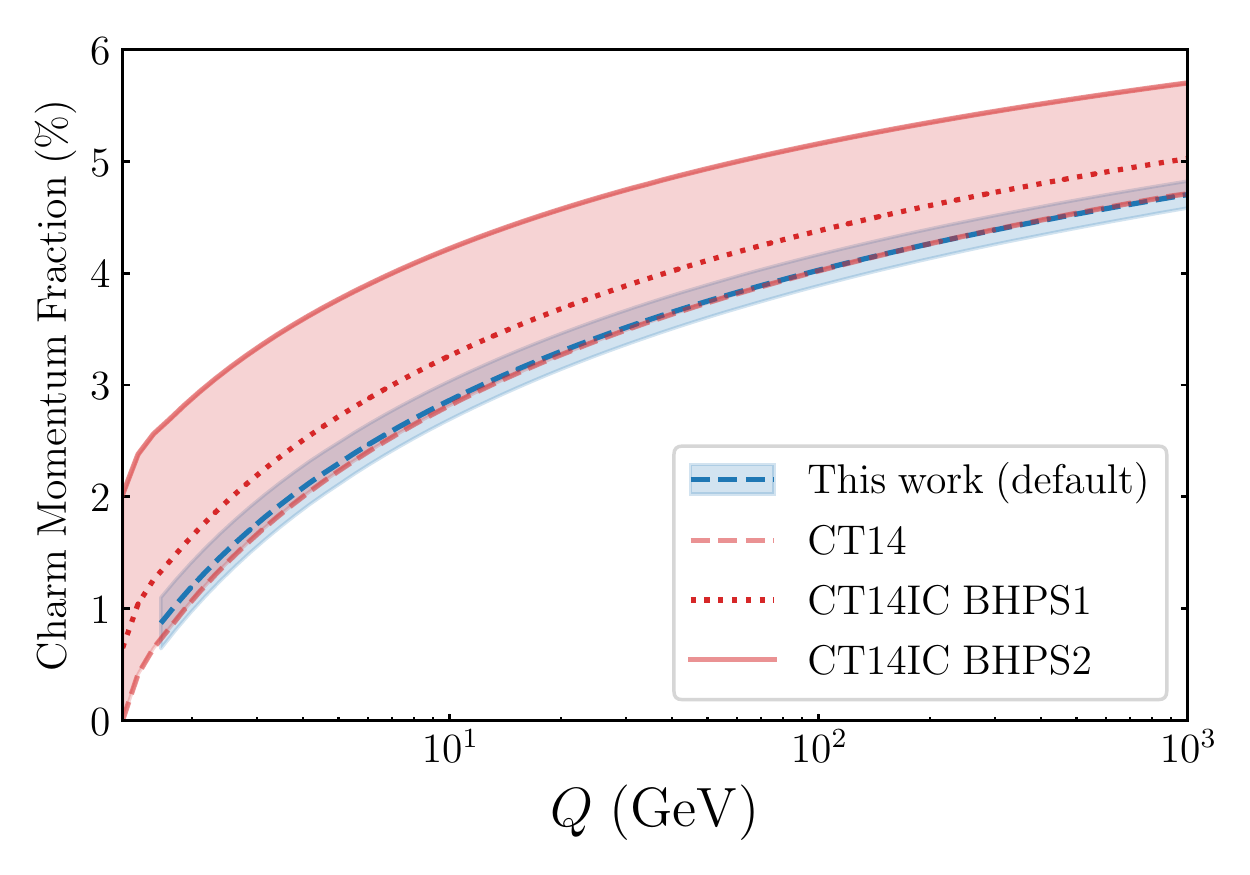}
    \caption{\small The 4FNS charm PDF
    from~\cite{Hou:2017khm} compared to our result (also in the 4FNS) 
     at $Q=1.65$~GeV on
    a linear (top left) and logarithmic (top right) scale in $x$, and
    at  $Q=100$~GeV on a linear scale in $x$ and as a ratio to our result
    (bottom left).
    The momentum fraction corresponding to either case
    is also shown as a function of $Q$ (bottom right). Note that for
    our result the uncertainty band is the 68\%CL PDF uncertainty,
    while for~\cite{Hou:2017khm} the central curve (labeled
    CT14IC BHPS1) corresponds to the BHPS model with best-fit
    normalization, the lower curve (labeled
    CT14) corresponds to the default CT14 perturbative charm PDF and
    the upper curve (labeled
    CT14IC BHPS2) corresponds to the BHPS model with normalization at
    the upper 90\% CL (see text). The value of the momentum fractions
    are also provided in each case.
  \label{fig:comparison_CT14} }
\end{center}
\end{figure}

The CT14IC charm PDF is compared to our result in
Fig.~\ref{fig:comparison_CT14}, at  $Q=1.65$ GeV and $Q=100$ GeV, in
the former case on both a logarithmic and linear scale in $x$ and in
the latter case on a linear scale only, as a ratio to our default
result.
Note that the uncertainty band has a different interpretation in the
two curves shown: for our result it is the 68\%~CL PDF uncertainty,
while for~\cite{Hou:2017khm}  it corresponds to the model
uncertainty estimated as described above.
In Fig.~\ref{fig:comparison_CT14} we also quote
the charm momentum fraction in each case, at the corresponding scale $Q$. 

As shown in Fig.~\ref{fig:charm_content_3fns} (right), our result for
the charm PDF is in good agreement with the BHPS model at large
$x$. Correspondingly, for $x\gsim 0.3$
we find reasonably good agreement between our
result and the central curve of~\cite{Hou:2017khm}, which
corresponds to a momentum fraction and thus a normalization of the charm
PDF not too different from our result (see Table~\ref{tab:momfrac_lowQ}).
Both the upper and lower curve from~\cite{Hou:2017khm} instead
do not agree with our result within uncertainties: indeed the
lower edge corresponds to the absence of intrinsic charm (which we
exclude) and the upper edge to a momentum fraction which we exclude
at more than the $5\sigma$ level (see Table~\ref{tab:momfrac_lowQ}). 

For intermediate values $3\cdot10^{-3}\lsim x\lsim 0.3$  our result disagrees
with that of 
~\cite{Hou:2017khm}, while at very small $x$ all results agree,
the intrinsic charm being compatible with zero.
The disagreement at intermediate $x$ is mostly due
to the fact that in ~\cite{Hou:2017khm} charm is assumed to take
the BHPS form, which vanishes for $x\lsim 0.1$,
in the 4FNS at the low scale $Q=1.3$~GeV.
Due to perturbative evolution from  $Q=1.3$~GeV to $Q=1.65$~GeV the charm 
PDF then develops the large bump that is
seen in Fig.~\ref{fig:comparison_CT14}, where we instead find that 
the 4FNS charm PDF is quite small.
This difference persists at large scales as seen  in
Fig.~\ref{fig:charm_content_3fns} (bottom left).

In terms of momentum fractions, shown in Fig.~\ref{fig:charm_content_3fns} (bottom right),
as already mentioned our result is
compatible with the central value of~\cite{Hou:2017khm} within
uncertainties; and also with the lower edge of ~\cite{Hou:2017khm}
that corresponds to perturbative charm.
The upper edge  of the prediction from~\cite{Hou:2017khm} is instead
ruled out at more than $5\sigma$. 

\clearpage
\section{$Z$+charm production in the forward region}
\label{sec:zcharm}

Here we provide full details on our computation of  $Z$+charm
production and on the inclusion of the LHCb data for this
process in the determination of the charm PDF shown in
Fig.~\ref{fig:Zc}. 

\paragraph{Computational settings.}
Theoretical predictions for
the $Z$+charm measurements in the forward region 
by LHCb~\cite{LHCb:2021stx} follow the 
 settings described in~\cite{Boettcher:2015sqn}.
$Z$+jet events at NLO QCD theory are generated for $\sqrt{s}= 13$ TeV  using the $Zj$ package of the
{\sc\small POWHEG-BOX}~\cite{Alioli:2010xd}.
The parton-level events produced by {\sc\small POWHEG}
are then interfaced to {\sc\small Pythia8}~\cite{Sjostrand:2007gs}
with the Monash 2013 tune~\citesupp{Skands:2014pea} for showering,
hadronization, and simulation of the underlying event and multiple
parton interactions.
Long-lived hadrons, including charmed hadrons,
are assumed stable and not decayed.

Selection criteria on these particle-level events are imposed
to match the LHCb acceptance~\cite{LHCb:2021stx}.
$Z$ bosons are reconstructed in the dimuon final state by
requiring $60~{\rm GeV}\le m_{\mu\mu} \le 120~{\rm GeV}$,
and
only events where these muons satisfy
    $p_T^\mu \ge 20~{\rm GeV}$ and $2.0 \le \eta_{\mu}\le 4.5$
    are retained.
Stable visible hadrons within the LHCb acceptance of
$2.0 \le \eta \le 4.5$ are clustered with
the anti-$k_T$ algorithm with radius parameter
of $R=0.5$~\citesupp{Cacciari:2008gp}.
Only events with a hardest jet satisfying
  $ 20~{\rm GeV} \le p_T^{\rm jet} \le 100~{\rm GeV}$
and $2.2 \le \eta_{\rm jet}\le 4.2$ are retained.
Charm jets are defined as jets containing
a charmed hadron, specifically  jets satisfying
$\Delta R(j, c{\rm-hadron})\le 0.5$ for a charmed
hadron with $p_T(c{\rm-hadron})\ge 5~{\rm GeV}$.
Jets and muons are required to be separated
in rapidity and azimuthal angle, so
we require $\Delta R(j, \mu)\ge 0.5$.
The resulting events
are then binned in the $Z$ bosom rapidity $y_Z = y_{\mu \mu}$.

The physical observable measured by LHCb is the ratio of the fraction of $Z$+jet
    events with and without a charm tag,
    \begin{equation}
    \label{eq:Rcj}
        \mathcal{R}_j^c \equiv \frac{\sigma(pp\to Z+{\rm charm~ jet})}{\sigma(pp \to Z+{\rm jet})}=
         \frac{N(c{\rm -tag})}{ 
        N({\rm jets})} \, .
    \end{equation}
 Here  $N(c{\rm -tag})$ and $N({\rm jets})$ are, respectively, the number
    of charm-tagged and un-tagged jets, for a  $Z$ boson rapidity interval
    that satisfies the selection and acceptance criteria.
    The denominator of Eq.~(\ref{eq:Rcj}) includes all jets, even those
    containing heavy hadrons.
The charm tagging efficiency is already accounted for at the level
of the experimental measurement, so it is not required in the theory
simulations.

Predictions for Eq.~(\ref{eq:Rcj}) are produced using our default PDF
determination (NNPDF4.0 NNLO), as well as the corresponding PDF set
with perturbative charm (see SI Sect.~\ref{app:consistency}).
We have
explicitly checked that our results are essentially independent of the
value of the charm mass.
We have evaluated MHOUs and PDF uncertainties using the
output of the {\sc\small POWHEG+Pythia8} calculations.
We have checked that MHOUs, evaluated with the standard
seven-point prescription, essentially cancel in the ratio
Eq.~(\ref{eq:Rcj}). Note that 
this is not the case for  PDF uncertainties, because the dominant
partonic subchannels in the numerator and denominator are not the same.

\paragraph{Inclusion of the LHCb data.}

\begin{table}[h]
  \small
    \renewcommand{\arraystretch}{1.45}
\begin{tabularx}{\textwidth}{C{3.5cm}C{2.5cm}C{2.5cm}C{2.5cm}C{2.5cm}}
  \toprule
 \multirow{2}{*}{ $\chi^2/N_{\rm dat}$}  &   \multicolumn{2}{c}{ default charm}   &\multicolumn{2}{c}{perturbative charm} \\
                       &  $\rho_{\rm sys}=0$   & $\rho_{\rm sys}=1$ &  $\rho_{\rm sys}=0$ &   $\rho_{\rm sys}=1$ \\
  \midrule
 Prior        &  1.85   &  3.33      &   3.54  & 3.85      \\
 \midrule
 Reweighted   &  1.81   &  3.14      &   $-$   &  $-$     \\
\bottomrule
\end{tabularx}
\vspace{0.3cm}
\caption{\label{tab:chi2_zcharm} The values of $\chi^2/N_{\rm dat}$
 for the LHCb $Z$+charm data before (prior) and after (reweighted)
 their inclusion in the PDF fit. Results are given for two 
 experimental correlation models, denoted as
 $\rho_{\rm sys}=0$ and $\rho_{\rm sys}=1$. We also report values
 before inclusion for the perturbative charm PDFs.
}
\end{table}

We first compare the quality of the description of the LHCb data
before their inclusion. In Table~\ref{tab:chi2_zcharm} we show the
values of $\chi^2/N_{\rm dat}$ for the LHCb $Z$+charm data
both with default and perturbative charm.
Since the experimental covariance matrix is not available for the LHCb
data we determine the $\chi^2$ values assuming two limiting scenarios
for the correlation of experimental systematic uncertainties.
Namely, 
we either add in quadrature statistical and systematic errors ($\rho_{\rm sys}=0$),
or alternatively we assume that the total systematic uncertainty
is fully correlated between $y_Z$ bins ($\rho_{\rm sys}=1$). Fit
quality is always significantly better in our default intrinsic charm
scenario than with perturbative charm.
As is clear from
Fig.~\ref{fig:Zc} (top left), the somewhat poor fit quality is mostly due to the first
rapidity bin, which is essentially uncorrelated to the amount of
intrinsic charm (see
Fig.~\ref{fig:Zc}, top right).

The LHCb $Z$+charm data are then included in the PDF determination
through
Bayesian reweighting~\citesupp{Ball:2010gb,Ball:2011gg}. The
$\chi^2/N_{\rm dat}$ values obtained using the PDFs found after their
inclusion are given in
Table~\ref{tab:chi2_zcharm}.
They are computed by combining the PDF and
experimental covariance matrix so both sources of uncertainty are
included --- as mentioned above, MHOUs are negligible.
The fit quality is seen to improve only
mildly, and the effective number of
replicas~\citesupp{Ball:2010gb,Ball:2011gg} after reweighting
is only moderately reduced, from the prior $N_{\rm rep}=100$ to $N_{\rm
eff}=92$ or $N_{\rm eff}=84$ in the
$\rho_{\rm sys}=0$ and $\rho_{\rm sys}=1$ scenarios respectively.
This
demonstrates that the inclusion of the LHCb $Z$+charm measurements  affects
the PDFs only weakly. This agrees with the results shown in 
Figs.~\ref{fig:Zc}~(center) in
the main manuscript, where it is seen that the inclusion of the LHCb
data has essentially no impact on the shape of the charm PDF, but
it moderately reduces its uncertainty in the region of the valence peak.

\clearpage
\section{Parton luminosities}
\label{sec:lumis}

The impact of intrinsic charm on hadron collider observables can be
assessed by studying  parton luminosities. Indeed, the
cross-section for hadronic processes at leading order is typically
proportional to an individual parton luminosity or linear combination
of parton luminosities.
Comparing parton luminosities determined
using our default PDF set to those obtained imposing perturbative
charm (see SI Sect.~\ref{app:consistency}) provides a qualitative estimate of the
measurable impact of intrinsic charm. Of course this is then modified
by higher-order
perturbative corrections, which generally depend on more partonic
subchannels and thus on more luminosities.
In this section we illustrate this by considering the parton
luminosities that are relevant for the computation of the
$Z$+charm process in the LHCb kinematics, see SI Sect.~\ref{sec:zcharm}.

The parton luminosity without any restriction on the rapidity $y_X$ of the final state is
\be
\label{eq:lumi1D}
\mathcal{L}_{ab}(m_X)= \frac{1}{s}\int_{\tau}^1 \frac{dx}{x}f_a \lp x,m_X^2\rp
f_b\lp \tau/x,m_X^2\rp \, ,\qquad
\tau=\frac{m_X^2}{s} \, ,
\ee
where $a,b$ label the species of incoming partons, $\sqrt{s}$ is the center-of-mass energy of
the hadronic collision, and $m_X$ is the final state invariant mass.
For the more realistic situation where the final state rapidity
is restricted, $y_{\rm min}\le y_X\le y_{\rm max}$,
Eq.~(\ref{eq:lumi1D}) is modified as
\be
\label{eq:lumi1D_restricted}
\mathcal{L}_{ab}(m_X)= \frac{1}{s}\int_{\tau}^1 \frac{dx}{x}f_a \lp x,m_X^2\rp
f_b\lp \tau/x,m_X^2\rp \theta\lp y_X-y_{\rm min}  \rp
\theta\lp y_{\rm max}-y_X  \rp\, , 
\ee
where $y_X = \lp \ln x^2/\tau \rp /2$.

We consider in particular the quark-gluon and the charm-gluon luminosities, defined as
\be
\label{eq:lumis}
\mathcal{L}_{qg}(m_X)\equiv \sum_{i=1}^{n_f} \lp \mathcal{L}_{q_ig}(m_X)+
\mathcal{L}_{\bar{q}_ig}(m_X) \rp\, , \quad
\mathcal{L}_{cg}(m_X)\equiv  \lp \mathcal{L}_{cg}(m_X)+
\mathcal{L}_{\bar{c}g}(m_X) \rp\, , 
\ee
where $n_f$ is the number of active quark flavors for a given value of $Q=m_X$
with a maximum value of $n_f=5$.
These are the combinations that provide the leading contributions
respectively to the numerator ($\mathcal{L}_{cg}$) and the 
denominator
($\mathcal{L}_{qg}$) of  $\mathcal{R}_j^c$ in Eq.~(\ref{eq:Rcj}). 

\begin{figure}[htbp]
  \begin{center}
    \includegraphics[width=0.99\linewidth]{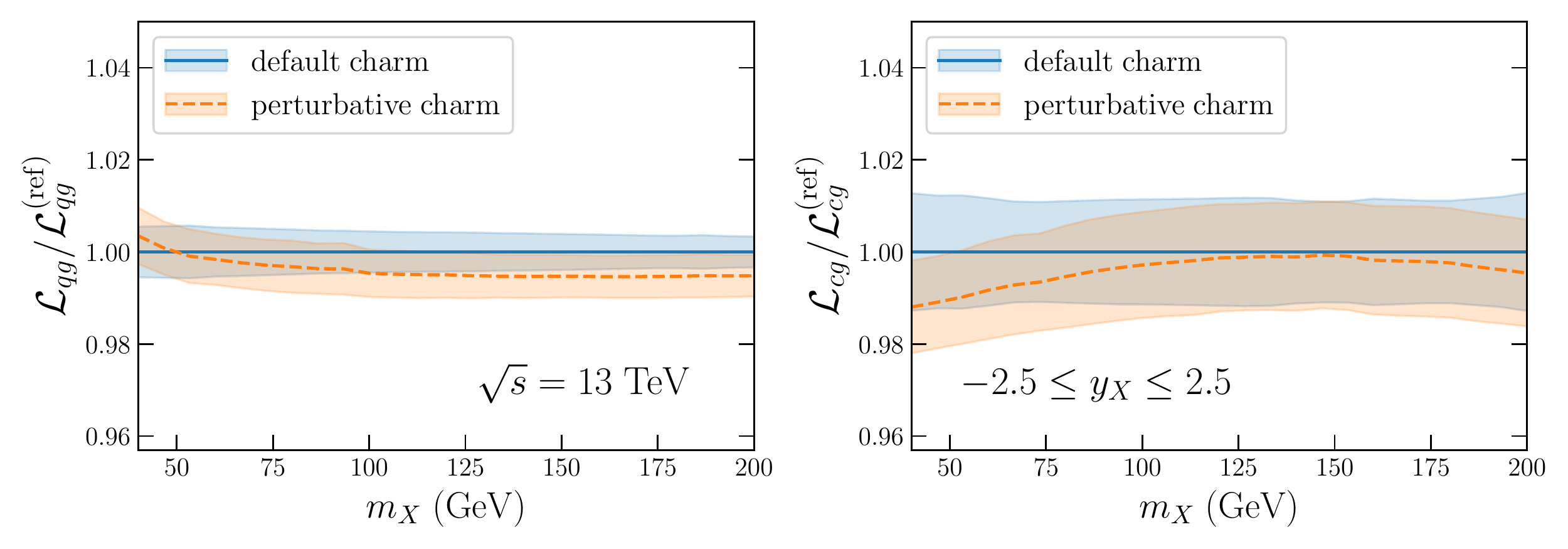}
    \includegraphics[width=0.99\linewidth]{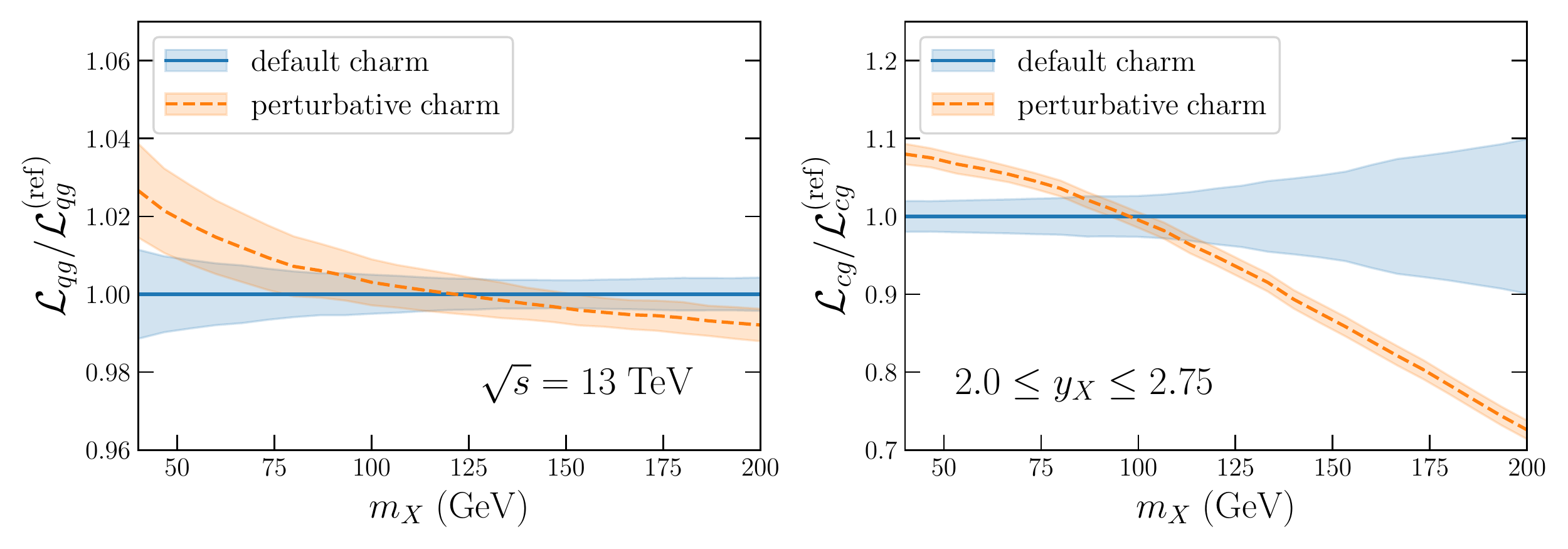}
    \includegraphics[width=0.99\linewidth]{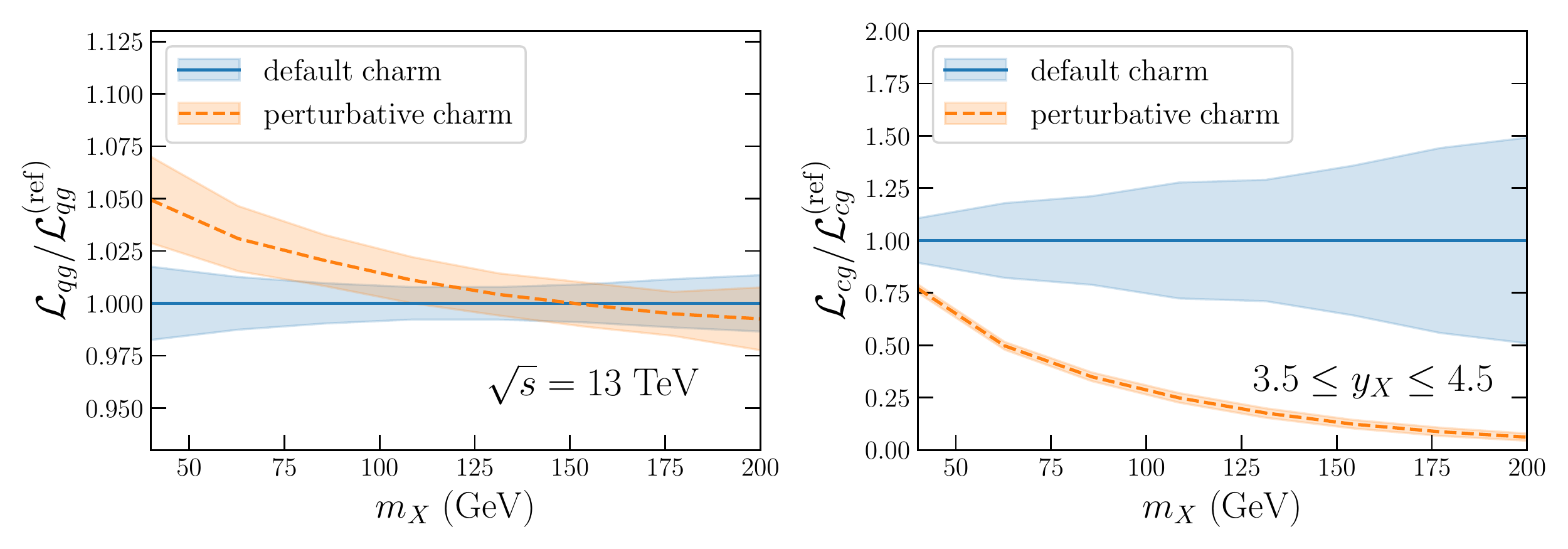}
    \caption{\small The quark-gluon (left) and charm-gluon (right)
      parton luminosities in the  $m_X$ region
      relevant for $Z$+charm production and  three different
      rapidity bins (see text). Results are shown both for our default charm
    PDFs and for the variant with perturbative charm. 
  \label{fig:charm_luminosities} }
\end{center}
\end{figure}

The luminosities are displayed in Fig.~\ref{fig:charm_luminosities},
in the  invariant mass region,
$40~{\rm GeV}\le m_X \le 200~{\rm GeV}$ which is most relevant for
$Z$+charm production.
Results are shown 
for three different
rapidity bins, $-2.5 \le y_X \le 2.5$ (central production in ATLAS and CMS),
$2.0 \le y_X \le 2.75$ (forward production, corresponding to the
central bin in LHCb),
and $3.5 \le y_X \le 4.5$ (highly boosted production, corresponding to
the most forward bin in the LHCb selection), as a ratio to our default case.

For central production it is clear that both the quark-gluon and
charm-gluon luminosities with our without intrinsic charm are very similar.
This means that central $Z$+charm production in this invariant mass
range is insensitive to intrinsic charm.
For forward production, corresponding to the  central LHCb rapidity
bin, $2.0 \le y_X \le 2.75$, in the invariant mass region  $m_X\simeq
100$~GeV again there is little difference between results with or
without intrinsic charm, but as the invariant mass increases the
charm-gluon luminosity with intrinsic charm is significantly enhanced.
For very forward production, such as the highest rapidity bin of LHCb,
$3.5 \le y_X \le 4.5$, the charm-gluon luminosity 
at $m_X \simeq 100$ GeV is enhanced  by a factor of about 4 in our
default result in comparison to the perturbative charm case, corresponding
to a $\simeq 3\sigma$ difference in units of the PDF uncertainty,
consistently with the behavior observed for the 
$\mathcal{R}_j^c$ observable in Fig.~\ref{fig:Zc}~(top left) in the
most forward rapidity  bin.
This observation provides a qualitative explanation of
the results of SI Sect.~\ref{sec:zcharm}.

\clearpage

\providecommand{\href}[2]{#2}\begingroup\raggedright\endgroup


\end{document}